\documentclass[12pt,english,onecolumn]{IEEEtran}
\usepackage[T3,T1]{fontenc}
\usepackage[utf8]{inputenc}
\usepackage{color}
\usepackage{babel}
\usepackage{array}
\usepackage{float}
\usepackage{mathtools}
\usepackage{multirow}
\usepackage{amsmath}
\usepackage{amsthm}
\usepackage{amssymb}
\usepackage{graphicx}
\usepackage[unicode=true,pdfusetitle,
 bookmarks=true,bookmarksnumbered=true,bookmarksopen=true,bookmarksopenlevel=3,
 breaklinks=false,pdfborder={0 0 1},backref=false,colorlinks=true]
 {hyperref}
\hypersetup{
 pdfborderstyle=,pdfborderstyle={},pdfborderstyle={},pdfborderstyle={},pdfborderstyle={},pdfborderstyle={},pdfborderstyle={},pdfborderstyle={},pdfborderstyle={},pdfborderstyle={},pdfborderstyle={},pdfborderstyle={},pdfborderstyle={},pdfborderstyle={},pdfborderstyle={},pdfborderstyle={},pdfborderstyle={},pdfborderstyle={},pdfborderstyle={},pdfborderstyle={},pdfpagelayout=OneColumn,pdfnewwindow=true,pdfstartview=XYZ,plainpages=false,linkcolor=blue,urlcolor=blue,citecolor=red,anchorcolor=blue,linkcolor=blue,urlcolor=blue,citecolor=red,anchorcolor=blue}

\makeatletter

\providecommand{\tabularnewline}{\\}

\theoremstyle{plain}
\newtheorem{thm}{\protect\theoremname}
\theoremstyle{definition}
\newtheorem{example}{\protect\examplename}
\theoremstyle{plain}
\newtheorem{prop}{\protect\propositionname}
\theoremstyle{plain}
\newtheorem{cor}{\protect\corollaryname}
\theoremstyle{remark}
\newtheorem{rem}{\protect\remarkname}
\theoremstyle{plain}
\newtheorem{fact}{\protect\factname}
\theoremstyle{plain}
\newtheorem{conjecture}{\protect\conjecturename}
\theoremstyle{plain}
\newtheorem{lem}{\protect\lemmaname}

\usepackage{babel}
\usepackage[mathscr]{eucal}
\usepackage{epsfig,epsf,psfrag}
\usepackage{amssymb,amsmath,amsthm,latexsym}
\usepackage{amsmath,graphicx,xcolor,url}
\usepackage[caption=false]{subfig} 
\usepackage{fixltx2e}%ordering of single and double column floats
\usepackage{array}%array and tabular environments
\usepackage{verbatim}
\usepackage{bm}
\usepackage{algorithmic}
\usepackage{algorithm}
\usepackage{verbatim}
\usepackage{textcomp}
\usepackage{mathrsfs,overpic}
\usepackage{epstopdf}
\usepackage{amsfonts}
\usepackage[numbers]{natbib}
\mathtoolsset{showonlyrefs} 

\usepackage{tikz}
\usepackage{float}
\usepackage{tabularx}
\usepackage{multirow}
\usetikzlibrary{patterns}

\def\1{\mathbf{1}}

\catcode`~=11 \def\UrlSpecials{\do\~{\kern -.15em\lower .7ex\hbox{~}\kern .04em}} \catcode`~=13 

\allowdisplaybreaks[1]

\DeclareMathAlphabet{\mathbsf}{OT1}{cmss}{bx}{n}
\DeclareMathAlphabet{\mathssf}{OT1}{cmss}{m}{sl}% slanted sans serif

\DeclareSymbolFont{bsfletters}{OT1}{cmss}{bx}{n}  
\DeclareSymbolFont{ssfletters}{OT1}{cmss}{m}{n}
\DeclareMathSymbol{\bsfGamma}{0}{bsfletters}{'000}
\DeclareMathSymbol{\ssfGamma}{0}{ssfletters}{'000}
\DeclareMathSymbol{\bsfDelta}{0}{bsfletters}{'001}
\DeclareMathSymbol{\ssfDelta}{0}{ssfletters}{'001}
\DeclareMathSymbol{\bsfTheta}{0}{bsfletters}{'002}
\DeclareMathSymbol{\ssfTheta}{0}{ssfletters}{'002}
\DeclareMathSymbol{\bsfLambda}{0}{bsfletters}{'003}
\DeclareMathSymbol{\ssfLambda}{0}{ssfletters}{'003}
\DeclareMathSymbol{\bsfXi}{0}{bsfletters}{'004}
\DeclareMathSymbol{\ssfXi}{0}{ssfletters}{'004}
\DeclareMathSymbol{\bsfPi}{0}{bsfletters}{'005}
\DeclareMathSymbol{\ssfPi}{0}{ssfletters}{'005}
\DeclareMathSymbol{\bsfSigma}{0}{bsfletters}{'006}
\DeclareMathSymbol{\ssfSigma}{0}{ssfletters}{'006}
\DeclareMathSymbol{\bsfUpsilon}{0}{bsfletters}{'007}
\DeclareMathSymbol{\ssfUpsilon}{0}{ssfletters}{'007}
\DeclareMathSymbol{\bsfPhi}{0}{bsfletters}{'010}
\DeclareMathSymbol{\ssfPhi}{0}{ssfletters}{'010}
\DeclareMathSymbol{\bsfPsi}{0}{bsfletters}{'011}
\DeclareMathSymbol{\ssfPsi}{0}{ssfletters}{'011}
\DeclareMathSymbol{\bsfOmega}{0}{bsfletters}{'012}
\DeclareMathSymbol{\ssfOmega}{0}{ssfletters}{'012}

\newcommand{\qednew}{\nobreak \ifvmode \relax \else
      \ifdim\lastskip<1.5em \hskip-\lastskip
      \hskip1.5em plus0em minus0.5em \fi \nobreak
      \vrule height0.75em width0.5em depth0.25em\fi}

\usepackage{bm,bbm}

\usepackage{lineno}
\usepackage{makecell}

\allowdisplaybreaks

\providecommand{\corollaryname}{Corollary}

\providecommand{\factname}{Fact}
\providecommand{\lemmaname}{Lemma}
\providecommand{\propositionname}{Proposition}
\providecommand{\remarkname}{Remark}
\providecommand{\theoremname}{Theorem}

\providecommand{\conjecturename}{Conjecture}

\providecommand{\examplename}{Example}
\usepackage{array}
\usepackage{varwidth}

\makeatother

\providecommand{\conjecturename}{Conjecture}
\providecommand{\corollaryname}{Corollary}
\providecommand{\examplename}{Example}
\providecommand{\factname}{Fact}
\providecommand{\lemmaname}{Lemma}
\providecommand{\propositionname}{Proposition}
\providecommand{\remarkname}{Remark}
\providecommand{\theoremname}{Theorem}

\begin{document}
\title{On the Complete Monotonicity of Rényi Entropy}
\author{Hao Wu, Lei Yu, and Laigang Guo\thanks{H. Wu and L. Yu are with the School of Statistics and Data Science,
LPMC, KLMDASR, and LEBPS, Nankai University, Tianjin 300071, China
(e-mails: haowu@mail.nankai.edu.cn and leiyu@nankai.edu.cn). L. Guo
is with the School of Mathematical Sciences, Beijing Normal University,
Beijing 100875, China (e-mail: lgguo@bnu.edu.cn). This work was supported
by the NSFC grant 62101286 and the Fundamental Research Funds for
the Central Universities of China (Nankai University). The corresponding
author is Lei Yu. }}
\maketitle
\begin{abstract}
In this paper, we investigate the complete monotonicity of Rényi entropy
along the heat flow. We confirm this property for the order of derivative
up to $4$, when the order of Rényi entropy is in certain regimes.
We also investigate concavity of Rényi entropy power and the complete
monotonicity of Tsallis entropy. We recover and slightly extend Hung's
result on the fourth-order derivative of the Tsallis entropy, and
observe that the complete monotonicity holds for Tsallis entropy of
order $2$, which is equivalent to that the noise stability with respect
to the heat semigroup is completely monotone. Based on this observation,
we conjecture that the complete monotonicity holds for Tsallis entropy
of all orders $\alpha\in(1,2)$. Our proofs in this paper are based
on the techniques of integration-by-parts, sum-of-squares, and curve-fitting. 
\end{abstract}


\section{Introduction }

Let $X$ be a random variable on $\mathbb{R}$ with density $f$,
and $Z\sim\mathcal{N}(0,1)$ be a standard Gaussian random variable
independent of $X$. For $t\ge0$, let 
\begin{equation}
X_{t}:=X+\sqrt{t}Z.\label{eq:-15}
\end{equation}
The density of $X_{t}$ is 
\begin{equation}
p(x,t)=\int_{-\infty}^{+\infty}f(y)\frac{1}{\sqrt{2\pi t}}e^{-\frac{(x-y)^{2}}{2t}}dy.\label{eq:-16}
\end{equation}
In information-theoretic language, $X_{t}$ is the output of an additive
white Gaussian noise (AWGN) channel with noise power $t$ when the
input is $X\sim f$. This kind of channel is closely related to thermodynamics
and is well investigated in literature. In thermodynamics, the function
$p$ is the solution to the following heat equation: 
\begin{equation}
\frac{\partial}{\partial t}p(x,t)=\frac{1}{2}\frac{\partial^{2}}{\partial x^{2}}p(x,t).\label{eq:heat equation}
\end{equation}
For brevity, throughout this paper, we write $p_{t}(x,t):=\frac{\partial}{\partial t}p(x,t)$
and $p_{k}(x,t):=\frac{\partial^{k}}{\partial x^{k}}p(x,t)$, and
the arguments $(x,t)$ of functions $p,p_{t},p_{k}$ are sometimes
omitted. So, the above heat equation can be shortly rewritten as 
\[
p_{t}=\frac{1}{2}p_{2}.
\]
The theory of the heat equation was first developed by Joseph Fourier
in 1822 for the purpose of modeling how a quantity such as heat diffuses
through a given region.

The entropy of a random variable is important in describing the amount
of information contained in the random variable. The differential
Shannon entropy (or simply, entropy) of a random variable $Z$ with
density $g$ is defined by 
\[
h(Z):=h(g):=-\int g(x)\log g(x)dx.
\]
The base of $\log$ is $e$. How does the entropy of $X_{t}$ evolve
as $t$ increases? By the technique of integration-by-parts, in 1966
McKean \cite{McKean1966} derived the first and second order derivatives
of $h(X_{t})$ as follows. For $t>0$, 
\begin{align}
\frac{\partial h(p)}{\partial t} & =\frac{1}{2}J(X_{t}),\label{eq:1st derivative-2}\\
\frac{\partial^{2}h(p)}{\partial t^{2}} & =-\frac{1}{2}\int p\left(\frac{p_{2}}{p}-(\frac{p_{1}}{p})^{2}\right)^{2}dx,
\end{align}
where 
\[
J(X_{t}):=\int\frac{p_{1}^{2}}{p}dx
\]
is known as the Fisher information of $X_{t}$. McKean \cite{McKean1966}
conjectured that given the variance of $X$, $(-1)^{k}\frac{\partial^{k}h(p)}{\partial t^{k}},k\ge1$
is maximized by Gaussian $X$, and confirmed it for $k=1,2$. A weaker
version of McKean's conjecture is that the signs of the derivatives
change alternately as the order of the derivatives increases. This
conjecture, known as the completely monotone conjecture, was explicitly
stated by Cheng and Geng in \cite{Cheng2015Higher}. Cheng and Geng
investigated the completely monotone conjecture for the higher order
derivatives of $h(p)$. They showed that the completely monotone conjecture
holds for the third and fourth order derivatives. Furthermore, the
Gaussian extremality conjecture and the completely monotone conjecture
in higher dimensional Euclidean spaces were investigated in \cite{zhang2018gaussian,guo2022lower}.

There are many other related works in the literature. The derivatives
of the Shannon entropy were used by Costa to prove a certain class
of entropy power inequalities \cite{costa1985new}. Dembo provided
a simple proof of Costa's entropy power inequalities, equivalently,
the concavity of the entropy power \cite{dembo1989simple}, which
was further simplified by Villani in \cite{villani2000short}. The
signs of higher derivatives of the entropy power were determined by
Toscani \cite{toscani2015concavity} under the log-concave measure
assumption. Courtade \cite{courtade2017concavity} generalizes Costa’s
entropy power inequalities to non-Gaussian additive perturbations.
Costa's entropy power inequalities were extended to the nonlinear
heat equation in \cite{savare2014concavity,li2020renyi,guo2021generalization}.
The complete monotonicity of the Tsallis entropy was investigated
by Hung \cite{hung2022generalization}, determining the signs of derivatives
of order up to $5$ when the order of the Tsallis entropy is in certain
regions. A related conjecture on the log-convexity of the Fisher information
was posed in \cite{Cheng2015Higher} and resolved in \cite{ledoux2021log}.
See reviews on related topics \cite{ledoux2022differentials,nair2020signs}.

In this paper, we focus on the Rényi entropy of $X_{t}.$ The Rényi
entropy \cite{Renyi1961OnMO} is a natural generalization of Shannon
entropy, which provides a finer characterization of information amount
contained in a random variable. For a probability density $g$, its
Rényi entropy of order $\alpha$ is defined by 
\[
h_{\alpha}(g):=\frac{1}{1-\alpha}\log\int g^{\alpha}(x)dx,
\]
for $\alpha\in(0,+\infty)\backslash\{1\}$. For $\alpha\in\{0,1,\infty\}$,
the Rényi entropy is defined by continuous extension. Specifically,
for $\alpha=1$, the Rényi entropy of order $1$ is 
\[
h_{1}(g):=h(g),
\]
i.e., the Shannon entropy. For $\alpha=\infty$, the Rényi entropy
is 
\[
h_{\infty}(g):=-\log\mathrm{ess}\sup_{x}g(x),
\]
which is also known as the min-entropy. For $\alpha=0$, the Rényi
entropy is 
\[
h_{0}(g):=\log\mu(g>0),
\]
where $\mu$ is the Lebesgue measure. In our setting, $h_{0}(p(\cdot,t))=\infty$
for any $t>0$ if $g>0$ a.s.

The Rényi entropy is closely related to the Tsallis entropy. For a
probability density $g$, the Tsallis entropy of order $\alpha$ is
defined by 
\[
\hat{h}_{\alpha}(g):=\frac{1}{1-\alpha}\left(\int g^{\alpha}(x)dx-1\right),
\]
for $\alpha\in(0,+\infty)\backslash\{1\}$. For $\alpha=1$, the Tsallis
entropy is defined by continuous extension. Specifically, $\hat{h}_{1}(g):=h(g)$.
The Rényi entropy and Tsallis entropy admit the following relation:
\begin{equation}
h_{\alpha}(g):=\frac{1}{1-\alpha}\log\left(1+(1-\alpha)\hat{h}_{\alpha}(g)\right).\label{eq:-12}
\end{equation}

In this paper, we investigate the evolution of the Rényi entropy
along a heat flow. We only focus on the range of time at which the
Rényi entropy is finite. More precisely, we focus on the case $t\in(0,t_{\alpha})$,
where $t_{\alpha}=\sup\{t\in(0,\infty):|h_{\alpha}(p(\cdot,t))|<\infty\}$.
By basic properties of the Rényi entropy, 
\[
t_{\alpha}=\begin{cases}
\infty, & \alpha\in(1,\infty)\\
\sup\{t\in(0,\infty):h_{\alpha}(p(\cdot,t))<\infty\}, & \alpha\in(0,1]
\end{cases}.
\]
For brevity, we assume the following convention throughout this paper. 

\textbf{Convention: }Throughout this paper, when we say $t>0$, we
means $t\in(0,t_{\alpha})$. 

As an extension of the completely monotone conjecture for Shannon
entropy, the complete monotonicity of Tsallis entropy was investigated
by Hung \cite{hung2022generalization}. In the present paper, we focus
on another extension, i.e., the complete monotonicity of Rényi entropy.
Here are two natural questions that we try to answer. Consider the
following inequality: 
\begin{equation}
(-1)^{k-1}\,\frac{\partial^{k}h_{\alpha}(p(\cdot,t))}{\partial t^{k}}\ge0,\qquad\forall t>0.\label{eq:derivatives}
\end{equation}
Denote 
\begin{align}
a & :=\inf\{s:\eqref{eq:derivatives}\textrm{ holds for all }k\ge1,\alpha\in[s,1]\},\nonumber \\
b & :=\sup\{s:\eqref{eq:derivatives}\textrm{ holds for all }k\ge1,\alpha\in[1,s]\}.\label{eq:-11}
\end{align}
\textbf{Question 1.}  \textit{Is it true that the complete monotonicity}\footnote{Rigorously speaking, the completely monotone conjecture is about the
complete monotonicity of the first-order derivative of the entropy,
instead of the entropy itself. Throughout this paper, we abuse the
terminology ``the complete monotonicity for the entropy'' to indicate
the complete monotonicity of its first-order derivative. }\textit{{} holds for Rényi entropy $h_{\alpha}$ if and only if $\alpha\in[a,b]$?
That is, is it true that \eqref{eq:derivatives} holds for all $k\ge1$,
if and only if $\alpha\in[a,b]$?}

\vspace{1em}

{} 

Given $k\ge1$, denote 
\begin{align*}
a_{k} & :=\inf\{s:\eqref{eq:derivatives}\textrm{ holds for all }\alpha\in[s,1]\},\\
b_{k} & :=\sup\{s:\eqref{eq:derivatives}\textrm{ holds for all }\alpha\in[1,s]\}.
\end{align*}
 \textbf{Question 2.}  \textit{Is it true that given $k\ge1$, \eqref{eq:derivatives}
holds if and only if $\alpha\in[a_{k},b_{k}]$? }

\vspace{1em}

In this paper, we investigate the values of $a_{k}$ and $b_{k}$.

\subsection{\label{subsec:Why-is-R=00003D0000E9nyi}Why is Rényi Entropy Important? }

Before introducing our main contributions, we first illustrate our
motivation on investigating the Rényi entropy.  
\begin{enumerate}
\item The Rényi entropy has explicit mathematical/physical meanings.  
\begin{enumerate}
\item The Rényi entropy is a measure of uncertainty or disorders. Rényi
proposed the Rényi entropy in \cite{Renyi1961OnMO} by modifing one
of Shannon's axioms for Shannon entropy. So, the Rényi entropy is
a natural extension of the Shannon entropy. It is well-known that
the Shannon entropy is a measure of uncertainty or disorders. So is
the Rényi entropy.  
\item The Rényi entropy is ``equivalent'' to the $L^{\alpha}$-norm. By
definition, the $\alpha$-Rényi entropy and the $L^{\alpha}$-norm
admit the following intimate relation: 
\begin{equation}
h_{\alpha}(f)=\frac{\alpha}{1-\alpha}\log\|f\|_{\alpha},\label{eq:renyi}
\end{equation}
where $\|f\|_{\alpha}:=[\int f^{\alpha}(x)dx]^{1/\alpha}$ is the
$L^{\alpha}$-norm of $f$. That is, the $\alpha$-Rényi entropy and
the $L^{\alpha}$-norm are mutually determined by each other.  
\item For different orders, the Rényi entropy admits different meanings.
For $\alpha=0$, the $\alpha$-Rényi entropy measures the support
size of $f$. For $\alpha=1$, the $\alpha$-Rényi entropy is just
the Shannon entropy. For $\alpha=2$, the $\alpha$-Rényi entropy
measures the ``energy'' of $f$. For $\alpha=\infty$, the $\alpha$-Rényi
entropy measures the maximum value of $f$.  
\end{enumerate}
\item The Rényi entropy is important in studying the contractivity/hypercontractivity
of a diffusion process. The contraction/hypercontraction of a diffusion
process is characterized by the following kind of inequalities: 
\begin{equation}
\|T_{t}f\|_{\beta}\le\|f\|_{\alpha},\;\forall\textrm{ nonnegative functions }f,\label{eq:-14}
\end{equation}
where $T_{t}$ is a semigroup operator. This kind of inequalities
can be written as inequalities on Rényi entropies. To see this point,
for simplicity, consider $T_{t}$ as the heat semigroup, and $f$
as a density function w.r.t. the Lebesgue measure. Then, 
\[
T_{t}f(y)=\int_{-\infty}^{+\infty}f(x)\frac{1}{\sqrt{2\pi t}}e^{-\frac{(x-y)^{2}}{2t}}dx=p(y,t),
\]
which is exactly the density given in \eqref{eq:-16}. So, the Rényi
entropies of $X$ and $X_{t}$ connect to the norms of $f$ and $T_{t}f$
in the following way: 
\begin{align}
h_{\beta}(X_{t}) & =h_{\beta}(T_{t}f)=\frac{\beta}{1-\beta}\log\|T_{t}f\|_{\beta},\label{eq:renyi-1}\\
h_{\alpha}(X) & =h_{\alpha}(f)=\frac{\alpha}{1-\alpha}\log\|f\|_{\alpha}.
\end{align}
By these relations, \eqref{eq:-14} can be equivalently expressed
as 
\begin{equation}
\frac{1-\beta}{\beta}h_{\beta}(X_{t})\le\frac{1-\alpha}{\alpha}h_{\alpha}(X),\;\forall\textrm{ inputs }X.\label{eq:-1-3}
\end{equation}
See more details on this equivalence in \cite{raginsky2013logarithmic}.
When $\alpha=\beta\ge1$, \eqref{eq:-14} is known as the contractivity
of a diffusion process, which, by \eqref{eq:-1-3}, is equivalent
to the monotonicity of $h_{\alpha}(X_{t})$ in $t$, i.e., $\frac{\partial h_{\alpha}(X_{t})}{\partial t}\ge0$.
This is in fact one of our results in this paper. Motivated by this
point, it is interesting to investigate the higher order derivatives
of the Rényi entropy. In fact, Gross is the first to investigate the
derivative of the Rényi entropy (more precisely, the derivative of
$\|T_{t}f\|_{\alpha}$) \cite{gross1975logarithmic} during his discovery
of the logarithmic Sobolev inequality, although he did not explicitly
mention the concept ``Rényi entropy'' in his paper. 
\item The Rényi entropy is important in information theory and statistics.
The log-likelihood ratio is a key quantity in hypothesis testing,
which is also known as the information density in information theory.
The Shannon entropy (or the relative entropy) is the expectation of
the information density, and the Rényi entropy (or Rényi divergence)
is essentially the logarithmic moment generating function of the information
density. The Rényi entropy has applications in characterizing the
convergence exponents of optimal hypothesis testing and also has applications
in characterizing the convergence exponents of optimal source and
channel coding; see e.g., \cite{gallagerIT,yu2023theentro}. 
\end{enumerate}

\subsection{Our Contributions}

Our contributions in this paper are as follows.
\begin{enumerate}
\item We provide bounds on $a_{k}$ and $b_{k}$ for $1\le k\le4$. In particular,
we show the following. 
\begin{enumerate}
\item For $t>0$, $\frac{\partial h_{\alpha}(p)}{\partial t}\ge0$, whenever
$\alpha\in(0,+\infty]$. 
\item For $t>0$, $\frac{\partial^{2}h_{\alpha}(p)}{\partial t^{2}}\le0$,
whenever $\alpha\in(0,3]$. 
\item For $t>0$, $\frac{\partial^{3}h_{\alpha}(p)}{\partial t^{3}}\ge0$,
whenever $\alpha\in[\beta_{0},1.76)$, where $\beta_{0}\approx0.389$. 
\item For $t>0$, $\frac{\partial^{4}h_{\alpha}(p)}{\partial t^{4}}\le0$,
whenever $\alpha\in(0.93,1.76)$.
\end{enumerate}
These generalize the classic results on the first two order derivatives
of the Shannon entropy in \cite{McKean1966,costa1985new} and the
third and fourth-order derivatives of the Shannon entropy in \cite{Cheng2015Higher}
to the Rényi entropy. 
\item The same technique is then used by us to investigate the concavity
of Rényi entropy power $N_{\alpha}^{\beta}(X_{t})$ with respect to
$t$, where $N_{\alpha}(X_{t}):=\exp[2h_{\alpha}(p)]$ and $\beta>0$.
In particular, we show that $N_{\alpha}^{\frac{1}{2}}(X_{t})$ is
concave in $t$, whenever $\alpha\in(0,\frac{3}{2}+\sqrt{2}]$; and
$N_{\alpha}^{\frac{\alpha+1}{2}}(X_{t})$ is concave in $t$, whenever
$\alpha\in(0,1)$. We can recover Costa's entropy power inequality
by letting $\alpha\to1$ in the latter result. 
\item We also investigate the complete monotonicity for Tsallis entropy.
This setting was first investigated in \cite{hung2022generalization}.
For the Tsallis entropy, we recover and slightly extend Hung's result
for the fourth-order derivative. Also, we observe that the complete
monotonicity holds for order $2$. As far as we know, this is the
first example for the complete monotonicity to hold. 
\item In the aspect of proof techniques, our proofs in this paper are based
on the techniques of integration-by-parts, curve-fitting, and sum-of-squares/semidefinite
optimization. The techniques of integration-by-parts and sum-of-squares/semidefinite
optimization are consistent with the ones used in \cite{McKean1966,costa1985new,villani2000short,Cheng2015Higher},
but the curve-fitting technique is newly introduced into this kind
of problems. Specifically, there are two main differences between
ours and existing proofs. Firstly, different from the existing works,
there are additional product integral terms in the expressions of
derivatives should be treated specifically since they cannot be written
into the sum of squares. This makes our proofs more complicated. Secondly,
a key step in both our proofs and existing ones, e.g., in \cite{hung2022generalization}
is to verify the positive semidefiniteness of the coefficient matrix.
Note that the Rényi entropy and Tsallis entropy involve the order
$\alpha$. So, the coefficient matrix involves the indeterminate $\alpha$
as well, which make it difficult to verify the positive semidefiniteness
using computers, since computers are more convenient to verify the
positive semidefiniteness for a numerical matrix. Hung \cite{hung2022generalization}
used lemmas from \cite{jungel2006algorithmic} to determine the positive
semidefiniteness for the order $3$. In contrast, we adopt a curve-fitting
strategy to determine the positive semidefiniteness.  Compared with
the strategy in \cite{jungel2006algorithmic,hung2022generalization},
our strategy is more capable to deal with matrices of sizes $4\times4$
or larger. Hence, our strategy is more powerful. In theory, it can
be applied to determining signs of higher order derivatives, although
this is not easy to conduct in practice since the sizes of coefficient
matrices might be very large. 
\end{enumerate}
Our results are summarized in Table \ref{tab:Summary-of-results}.

\begin{table}[H]
\centering{}\caption{\label{tab:Summary-of-results}Summary of results on Rényi/Tsallis
entropy along heat flow.}
\begin{tabular}{|c|>{\centering}p{0.5\textwidth}|}
\hline 
\multicolumn{1}{|c|}{\textbf{Cases}} & \textbf{Sign of $k$-th derivative}\tabularnewline
\hline 
\hline 
\multirow{5}{*}{Rényi entropy} & $k=1,2$ and $\alpha=1$: \cite{McKean1966,costa1985new}; $k=3,4$
and $\alpha=1$: \cite{Cheng2015Higher}\tabularnewline
\cline{2-2} 
 & $k=1$ and $\alpha\in(0,+\infty]$: Theorem \ref{thm:sign of 1st derivative}\tabularnewline
\cline{2-2} 
 & $k=2$ and $\alpha\in(0,3]$: Theorem \ref{thm:sign of 2nd derivative}\tabularnewline
\cline{2-2} 
 & $k=3$ and $\alpha\in[\beta_{0},1.76],\beta_{0}\approx0.389$: Theorem
\ref{thm:sign of 3rd derivative}\tabularnewline
\cline{2-2} 
 & $k=4$ and $\alpha\in[0.93,1.76]$: Theorem \ref{thm:sign of 4th derivative}\tabularnewline
\hline 
\multirow{1}{*}{Rényi entropy power} & Second derivatives of $N_{\alpha}^{\frac{1}{2}}(p)$ for $\alpha\in(0,\frac{3}{2}+\sqrt{2}]$
and $N_{\alpha}^{\frac{\alpha+1}{2}}(p)$ for $\alpha\in(0,1)$: Theorem
\ref{thm:REP-1}\tabularnewline
\hline 
\multirow{4}{*}{Tsallis entropy} & $k=1$ and $\alpha\in(0,+\infty]$; $k=2$ and $\alpha=(0,3]$; $k=3$
and $\alpha\in[\beta_{0},2]$: \cite{hung2022generalization}\tabularnewline
\cline{2-2} 
 & Numerical verification for $k=4$ and $\alpha\in(1,2),$ as well as,
for $k=5$ and $\alpha\in(\beta_{1},2)$ with $\beta_{1}\in(1.54,1.55)$:
\cite{hung2022generalization}\tabularnewline
\cline{2-2} 
 & Rigorous proof for $k=4$ and $\alpha\in[0.93,2]$: Theorem \ref{thm:4th Tsallis entropy}\tabularnewline
\cline{2-2} 
 & $k\in\mathbb{N}$ and $\alpha=2$: Fact \ref{thm:Tsallis2} (complete
monotonicity)\tabularnewline
\hline 
\end{tabular}
\end{table}

\subsection{Organization}

The organization of this paper is as follows. We state our main result
on signs of up to fourth order derivatives of the Rényi entropy in
Section \ref{sec:Main-Results}. In the same section, we also investigate
the derivatives of the Rényi entropy power and the derivatives of
the Tsallis entropy. In Section \ref{subsec:Notations-and-Preliminaries},
we provide notations and preliminaries for our proofs. In Section
\ref{sec:First-order-derivative}, we focus on the first-order derivative,
and prove the first statement of Theorem \ref{thm:main}. In Section
\ref{sec:Second-order-derivative}, we focus on the second-order derivative,
and prove the second statement of Theorem \ref{thm:main}. In Section
\ref{sec:Third-order--derivative}, we focus on the third-order derivative,
and prove the third statement of Theorem \ref{thm:main}. In Section
\ref{sec:Fourth-order-Derivative}, we focus on the fourth-order derivative,
and prove the fourth statement of Theorem \ref{thm:main}. The remaining
proofs are provided in Sections \ref{sec:Proof of sign of 2nd derivative}-\ref{sec:conjecture-1},
as well as in the appendix.

\section{\label{sec:Main-Results}Main Results}

\subsection{Rényi entropy }

Our main result is the following which provides bounds on $a_{k}$
and $b_{k}$. 
\begin{thm}
\label{thm:main}The following hold. 
\begin{enumerate}
\item For $t>0$, $\frac{\partial h_{\alpha}(p)}{\partial t}\ge0$, whenever
$\alpha\in(0,+\infty]$. 
\item For $t>0$, $\frac{\partial^{2}h_{\alpha}(p)}{\partial t^{2}}\le0$,
whenever $\alpha\in(0,3]$. 
\item For $t>0$, $\frac{\partial^{3}h_{\alpha}(p)}{\partial t^{3}}\ge0$,
whenever $\alpha\in[\beta_{0},1.76]$, where $\beta_{0}\approx0.389$.
\item For $t>0$, $\frac{\partial^{4}h_{\alpha}(p)}{\partial t^{4}}\le0$,
whenever $\alpha\in[0.93,1.76]$.
\end{enumerate}
\end{thm}
Statement 1 is in fact equivalent to the contractivity of the heat
semigroup; see the second point in Section \ref{subsec:Why-is-R=00003D0000E9nyi}.
So, essentially speaking, this is not new. Moreover, Statements 2-4
in Theorem \ref{thm:main} for $\alpha=1$ are neither new. Statements
1 and 2 for $\alpha=1$ are classic results (see e.g., \cite{McKean1966,costa1985new}),
Statements 3 and 4 for $\alpha=1$ were proven in \cite{Cheng2015Higher}.
Statement 2 for $\alpha\in(0,3]\backslash\{1\}$, Statement 3 for
$\alpha\in[\beta_{0},1.76]\backslash\{1\}$ and Statement 4 for $[0.93,1.76]\backslash\{1\}$
are new. 

The results in Theorem \ref{thm:main} look similar to Hung's results
in \cite{hung2022generalization} on the complete monotonicity for
Tsallis entropy. By using techniques of integration-by-parts and sum-of-squares,
Hung showed that  
\begin{enumerate}
\item For $t>0$, $\frac{\partial\hat{h}_{\alpha}(p)}{\partial t}\ge0$,
whenever $\alpha\in(0,+\infty)$. 
\item For $t>0$, $\frac{\partial^{2}\hat{h}_{\alpha}(p)}{\partial t^{2}}\le0$,
whenever $\alpha\in(0,3]$. 
\item For $t>0$, $\frac{\partial^{3}\hat{h}_{\alpha}(p)}{\partial t^{3}}\ge0$,
whenever $\alpha\in[\beta_{0},2]$, where $\beta_{0}$ is the same
as the one given in Theorem \ref{thm:main}. 
\item For $t>0$, $\frac{\partial^{4}\hat{h}_{\alpha}(p)}{\partial t^{4}}\le0$,
whenever $\alpha\in(1,2)$ (only numerical verification).
\end{enumerate}
The main differences are that for the third order derivative, our
range of $\alpha$ is $[\beta_{0},1.76)$ but his is $[\beta_{0},2]$,
and for the fourth order derivative, our range of $\alpha$ is $[0.93,1.76]$
but his is $(1,2)$. This difference comes from the different definitions
of Rényi entropy and Tsallis entropy. The specific reason is given
as follows. By differentiating the two sides of \eqref{eq:-12} with
respect to $t$, we observe that 
\begin{equation}
\frac{\partial h_{\alpha}(p)}{\partial t}=\frac{1}{\int p^{\alpha}dx}\frac{\partial\hat{h}_{\alpha}(p)}{\partial t}.\label{eq:-12-1}
\end{equation}
For the $k$-th order derivative, 
\begin{equation}
\frac{\partial^{k}h_{\alpha}(p)}{\partial t^{k}}=\frac{1}{\int p^{\alpha}dx}\frac{\partial^{k}\hat{h}_{\alpha}(p)}{\partial t^{k}}+\Sigma,\label{eq:-13}
\end{equation}
where $\Sigma$ is a linear combination of the products of at least
two{} integrals with each integral in the form $\mathbb{E}_{\alpha}[p_{k_{1}}p_{k_{2}}...p_{k_{m}}]$.
Here, $\mathbb{E}_{\alpha}[g]:=\int g\tilde{p}_{\alpha}dx$ with $\tilde{p}_{\alpha}:=\frac{p^{\alpha}}{\int p^{\alpha}dx}$.
For example, in the third order derivative expression of Rényi entropy,
$\Sigma$ is a linear combination of $\mathbb{E}_{\alpha}[\bar{p}_{1}^{4}]\mathbb{E}_{\alpha}[\bar{p}_{1}^{2}]$,
$\mathbb{E}_{\alpha}[\bar{p}_{1}^{2}]\mathbb{E}_{\alpha}[\bar{p}_{2}^{2}]$,
and $\mathbb{E}_{\alpha}[\bar{p}_{1}^{2}]^{3}$. If we remove these
product terms, then $\frac{\partial^{k}h_{\alpha}(p)}{\partial t^{k}}$
wound turn into $\frac{\partial^{k}\hat{h}_{\alpha}(p)}{\partial t^{k}}$
times a factor $\frac{1}{\int p^{\alpha}dx}$. This intimate relation
shows that if we can write $\frac{\partial^{k}h_{\alpha}(p)}{\partial t^{k}}$
as a sum-of-squares, then by setting product terms to zero, we obtain
a form in sum-of-squares for $\frac{\partial^{k}\hat{h}_{\alpha}(p)}{\partial t^{k}}$.
This is the reason why the ranges of $\alpha$ obtained in Theorem
\ref{thm:main} are included in the ranges of $\alpha$ obtained by
Hung. Compared to the Tsallis entropy, for certain values of $(k,\alpha)$
(e.g., $(k,\alpha)=(3,\beta_{0})$), adding the product terms (i.e.,
$\Sigma$) in \eqref{eq:-13} essentially does not change the sum-of-squares
for $\frac{\partial^{k}\hat{h}_{\alpha}(p)}{\partial t^{k}}$, but
for other certain values of $(k,\alpha)$ (e.g., $(k,\alpha)=(3,2)$),
adding the product terms in \eqref{eq:-13} indeed changes the sum-of-squares. 

Theorem \ref{thm:main} implies that $a_{1}=0,b_{1}=\infty$, $a_{2}=0,b_{2}\ge3$,
and $a_{3}\le\beta_{0},b_{3}\ge1$. The following example shows that
$b_{2}<8$, $b_{3}<5$, $b_{4}<4$, $b_{k}<3$ with $5\le k\le8$,
and $b_{9}<2$. Hence, the complete monotonicity fails even for $\alpha=2$.
That is, $b<2$ where $b$ is defined in \eqref{eq:-11}. 
\begin{example}
\label{exa:Suppose-the-distribution}Suppose the distribution of $X$
is $\frac{1}{2}(\delta_{1}+\delta_{-1})$, where $\delta_{i}$ represents
the Dirac measure (i.e., the single-point distribution) at $i$.
Then, for this case, 
\[
p(x,t)=\frac{1}{2\sqrt{2\pi t}}e^{-\frac{(x-1)^{2}}{2t}}+\frac{1}{2\sqrt{2\pi t}}e^{-\frac{(x+1)^{2}}{2t}}.
\]
We use Mathematica to verify that the inequality in \eqref{eq:derivatives}
does not hold for 
\begin{equation}
(k,\alpha)\in\{(2,8),(3,5),(4,4),(5,3),(9,2)\}.\label{eq:-5}
\end{equation}
Specifically, when $\alpha$ is set to an integer, it is easy to
use Mathematica to obtain the closed form expressions for Rényi entropy
$h_{\alpha}(p)$ and its derivatives $\frac{\partial^{k}h_{\alpha}(p)}{\partial t^{k}}$.
Given $(k,\alpha)\in\mathbb{N}^{+}\times\mathbb{N}^{+}$, we plot
$\frac{\partial^{k}h_{\alpha}(p)}{\partial t^{k}}$ as functions of
the time $t$, and then observe whether the sign of $\frac{\partial^{k}h_{\alpha}(p)}{\partial t^{k}}$
changes around some point. For example, we find that when $(k,\alpha)=(5,3)$,
$\frac{\partial^{5}h_{3}(p)}{\partial t^{5}}$ is not always non-negative,
since for some range of $t$, it takes negative values. Therefore,
$(k,\alpha)=(5,3)$ is a pair of parameters that violates complete
monotonicity. Other pairs of $(k,\alpha)$ are verified similarly. 
\end{example}
We next look for other pairs of $(k,\alpha)$ besides those listed
in \eqref{eq:-5} that violate complete monotonicity. To this end,
we need the following proposition, whose proof is given in Section
\ref{sec:Proof-of-Proposition}. 
\begin{prop}
\label{prop:sign}For any $(k,\alpha)\in\mathbb{N}^{+}\times(\frac{1}{3},+\infty)$,
if $\frac{\partial^{k+1}}{\partial t^{k+1}}h_{\alpha}(X_{t})$ is
non-positive (resp. non-negative) for all $t>0$, then $\frac{\partial^{k}}{\partial t^{k}}h_{\alpha}(X_{t})$
is non-negative (resp. non-positive) for all $t>0$. 
\end{prop}
The proposition above shows that if $\frac{\partial^{k}h_{\alpha}(p)}{\partial t^{k}}$
is not always non-negative (resp. not always non-positive), then $\frac{\partial^{k+1}}{\partial t^{k+1}}h_{\alpha}(X_{t})$
is not always non-positive (resp. not always non-negative). Combining
this result with the example above, one can find that all pairs of
$(k,\alpha)$ such that $k\ge2$ for $\alpha=8$, $k\ge3$ for $\alpha=5$,
$k\ge4$ for $\alpha=4$, $k\ge5$ for $\alpha=3$, and $k\ge9$ for
$\alpha=2$, violate complete monotonicity. 

\subsection{Rényi entropy power}

For a random variable $X$ with probability density $g$, its Rényi
entropy power of order $\alpha$ is defined by 
\[
N_{\alpha}(X):=N_{\alpha}(g):=\exp[2h_{\alpha}(g)]=\left(\int g(x)^{\alpha}dx\right)^{-\frac{2}{\alpha-1}}.
\]
Specifically, for $\alpha=1$, the Rényi entropy power of order $1$
is defined by 
\[
N_{1}(X):=N_{1}(g):=\exp[2h(g)]=\exp[-2\int g(x)\log g(x)dx],
\]
which is also known as the Shannon entropy power. We consider the
concavity of $N_{\alpha}^{\beta}(X_{t})$ with respect to $t$, where
$\beta>0$. 
\begin{thm}
\label{thm:REP-1} The following statements hold. 
\begin{enumerate}
\item $N_{\alpha}^{\frac{1}{2}}(X_{t})$ is concave in $t>0$, whenever
$\alpha\in(0,\frac{3}{2}+\sqrt{2}]$. 
\item $N_{\alpha}^{\frac{\alpha+1}{2}}(X_{t})$ is concave in $t>0$, whenever
$\alpha\in(0,1)$. 
\end{enumerate}
\end{thm}
The proof of Theorem \ref{thm:REP-1} is deferred to Section \ref{sec:Proof of concavity of REP-1}.
From the above, we immediately obtain a new version of Rényi entropy
power inequality in Costa's sense. 
\begin{cor}
\label{cor:REP-2}The following statements hold. 
\begin{enumerate}
\item For $\alpha\in(0,\frac{3}{2}+\sqrt{2}]$, 
\[
e^{h_{\alpha}(X+\sqrt{t}Z)}\ge(1-t)e^{h_{\alpha}(X)}+te^{h_{\alpha}(X+Z)},\qquad t\in[0,1].
\]
\item For $\alpha\in(0,1)$, 
\begin{equation}
e^{(\alpha+1)h_{\alpha}(X+\sqrt{t}Z)}\ge(1-t)e^{(\alpha+1)h_{\alpha}(X)}+te^{(\alpha+1)h_{\alpha}(X+Z)},\qquad t\in[0,1].\label{eq:}
\end{equation}
\end{enumerate}
\end{cor}
It is known that the Rényi entropy is continuous in its order at which
the Rényi entropy is finite \cite{Erven} \cite[Theorem 2]{yu2023theentro}.
So, letting $\alpha\uparrow1$ in \eqref{eq:} yields 
\begin{equation}
e^{2h(X+\sqrt{t}Z)}\ge(1-t)e^{2h(X)}+te^{2h(X+Z)},\qquad t\in[0,1],\label{eq:-10}
\end{equation}
which is exactly Costa's entropy power inequality \cite{costa1985new}. 

The complete monotonicity was extended to the Shannon entropy power
in \cite{ledoux2022differentials}. It has been shown recently \cite{wang2024entropy}
that the complete monotonicity of the Shannon entropy power implies
the McKean conjecture, and hence also implies the complete monotonicity
of the Shannon entropy. We next prove an analogous result in the Rényi
setting, i.e., the complete monotonicity of the Rényi entropy power
implies the complete monotonicity of the Rényi entropy of the same
order.  The proof is given in Section \ref{sec:conjecture-1-1}.
\begin{prop}
\label{prop:For-any-}For any $K\in\mathbb{N}^{+}$ and $\alpha\in(0,+\infty)$,
$(-1)^{k-1}\frac{d^{k}}{dt^{k}}N_{\alpha}(p)\ge0$ for all $1\le k\le K$
implies $(-1)^{k-1}\frac{d^{k}}{dt^{k}}h_{\alpha}(p)\ge0$ for all
$1\le k\le K$.
\end{prop}
\begin{rem}
This result also holds for the $n$-dimensional case, for which the
Rényi entropy power is defined as $N_{\alpha}(p):=e^{\frac{2}{n}h_{\alpha}(p)}$. 
\end{rem}

\subsection{Tsallis entropy}

We next consider the complete monotonicity for the Tsallis entropy.
Using our curve-fitting method, we could recover (prove rigorously)
and extend slightly Hung's result for the fourth order derivative
of the Tsallis entropy.
\begin{thm}
\label{thm:4th Tsallis entropy}For $t>0$, when $\alpha\in[0.93,2]$,
$\frac{\partial^{4}\hat{h}_{\alpha}(p)}{\partial t^{4}}\le0$.
\end{thm}
The proof of this theorem is deferred to Section \ref{sec:Proof of 4th Tsallis entropy}.
Numerical results show that the left end point $0.93$ could be improved
to $0.92$. 

Interestingly, the complete monotonicity can be verified easily for
Tsallis entropy of order $2$. This is the first example for the complete
monotonicity to hold. Note that in contrast, Example \ref{exa:Suppose-the-distribution}
shows that the complete monotonicity for Rényi entropy of order $2$
does not hold. 
\begin{fact}[Complete Monotonicity for Tsallis Entropy of Order $2$]
\label{thm:Tsallis2} For $t>0$, 
\[
\frac{\partial^{k}\hat{h}_{2}(p)}{\partial t^{k}}=(-1)^{k-1}\int p_{k}^{2}dx,
\]
whenever $k\in\mathbb{N}^{+}$. As a consequence, for any $t>0$,
\[
(-1)^{k-1}\frac{\partial^{k}\hat{h}_{2}(p)}{\partial t^{k}}\ge0,
\]
whenever $k\in\mathbb{N}^{+}$. 
\end{fact}
This fact is probably already known to experts, but we have not found
it explicitly stated in the literature. So, a proof of this fact is
given in Section \ref{sec:conjecture-1} which is based on the integration-by-parts
and quite simple. Besides, the complete monotonicity for Tsallis entropy
of order $2$ can be also proven by Fourier analysis.

It is natural to make the following conjecture. 
\begin{conjecture}
\label{conj:The-completely-monotone} The complete monotonicity holds
for Tsallis entropy $\hat{h}_{\alpha}$ with $\alpha\in(1,2)$, i.e.,
for such $\alpha$'s, 
\begin{equation}
(-1)^{k-1}\frac{\partial^{k}\hat{h}_{\alpha}(p)}{\partial t^{k}}\ge0,\qquad\forall t>0.\label{eq:TsallisConj}
\end{equation}
\end{conjecture}
This conjecture implies the completely monotone conjecture for differential
entropy $h$.

The complete monotonicity of $\hat{h}_{\alpha}$ is equivalent to
complete monotonicity of $-\int p^{\alpha}(x,t)dx$. The theorem above
implies that $-\int p^{\alpha}(x,t)dx$ is completely monotone for
$\alpha=2$. On the other hand, for $\alpha=1$, $-\int p^{\alpha}(x,t)dx=-1$
holds trivially. Hence, $-\int p^{\alpha}(x,t)dx$ is also completely
monotone for $\alpha=1$. If one can show that all the signs of derivatives
of $t\mapsto-\int p^{\alpha}(x,t)dx$ do not change on $\alpha\in[1,2]$,
then the conjecture above follows directly.

The complete monotonicity for Tsallis entropy of other orders up to
$5$ was verified in \cite{hung2022generalization}. Specifically,
it was shown that \eqref{eq:TsallisConj} holds for the case of $k=1$
and $\alpha\in(0,1)\cup(1,\infty)$, the case of $k=2$ and $\alpha\in(0,1)\cup(1,3]$,
the case of $k=3$ and $\alpha\in[\beta_{0},1)\cup(1,2]$ with $\beta_{0}\approx0.389$,
the case of $k=4$ and $\alpha\in(1,2)$, and the case of $k=5$ and
$\alpha\in(\beta_{1},2)$ with some $\beta_{1}\in(1.54,1.55)$.

We next investigate when the complete monotonicity for Tsallis entropy
does not hold. 
\begin{example}
We consider the distribution $\frac{1}{2}(\delta_{1}+\delta_{-1})$
of $X$ again. For this case, we use Mathematica to verify that the
inequality in \eqref{eq:TsallisConj} does not hold for 
\begin{equation}
(k,\alpha)\in\{(3,33),(4,8),(5,5),(6,4),(9,3)\}.\label{eq:-19}
\end{equation}
In particular, $\frac{\partial^{k}\hat{h}_{\alpha}(p)}{\partial t^{k}}<0$
for $(k,\alpha)=(9,3)$ and some $t>0$. So, the complete monotonicity
holds for Tsallis entropy $\hat{h}_{\alpha}$ with $\alpha\in(1,b]$
only if $b<3$.  From this observation, the case of $\alpha\in(0,b]$
for some $b<3$ is of most interest to us. 
\end{example}
We next look for other pairs of $(k,\alpha)$ besides those listed
in \eqref{eq:-19} that violate complete monotonicity.  Similarly
to Proposition \ref{prop:sign}, the following proposition holds.
The proof is almost same as the one of Proposition \ref{prop:sign},
and hence, omitted here. 
\begin{prop}
\label{prop:sign-1}For any $(k,\alpha)\in\mathbb{N}^{+}\times(\frac{1}{3},+\infty)$,
if $\frac{\partial^{k+1}}{\partial t^{k+1}}\hat{h}_{\alpha}(X_{t})$
is non-positive (resp. non-negative) for all $t>0$, then $\frac{\partial^{k}}{\partial t^{k}}\hat{h}_{\alpha}(X_{t})$
is non-negative (resp. non-positive) for all $t>0$. 
\end{prop}
Combining this result with the example above, one can find that all
pairs of $(k,\alpha)$ such that $k\ge3$ for $\alpha=33$, $k\ge4$
for $\alpha=8$, $k\ge5$ for $\alpha=5$, and $k\ge9$ for $\alpha=3$,
violate complete monotonicity for Tsallis entropy.  Comparing the
analyses around Examples 1 and 2, the complete monotonicity of Tsallis
entropy is more likely to hold, compared to the complete monotonicity
of Rényi entropy for the same order $\alpha$.

\subsection{Connections to $\alpha$-Stability}

Denote the heat semigroup as $T_{t}$ which is given by\footnote{Here the function $f$ is an arbitrary Lebesgue integrable function
whose integral is not necessarily restricted to be $1$. However,
there is no significant difference from density functions, since one
can always normalize a Lebesgue integrable function to a density function. } 
\[
T_{t}f(x)=\mathbb{E}[f(x+\sqrt{t}Z)]=\int f(y)\frac{1}{\sqrt{2\pi t}}e^{-\frac{(x-y)^{2}}{2t}}dy.
\]
That is, 
\[
p(x,t)=T_{t}f(x).
\]
The $\alpha$-stability of $f$ with respect to the heat semigroup
$T_{t}$ is defined as 
\[
S_{t}^{\alpha}(f):=\int(T_{t}f(x))^{\alpha}dx=\int p(x,t)^{\alpha}dx.
\]
In particular, for $\alpha=2$, 
\[
S_{t}^{2}(f):=\int(T_{t}f(x))^{2}dx=\int f(x)T_{2t}f(x)dx=\int f(x)p(x,2t)dx,
\]
which is known as the noise stability of $f$.

It is easy to observe that the $\alpha$-Rényi entropy and $\alpha$-Tsallis
entropy of $f$ can be expressed in terms of $\alpha$-stability as
follows: 
\begin{align*}
h_{\alpha}(p) & =\frac{1}{1-\alpha}\log S_{t}^{\alpha}(f),\\
\hat{h}_{\alpha}(p) & =\frac{1}{1-\alpha}\left(S_{t}^{\alpha}(f)-1\right).
\end{align*}
So, our results given above imply how the $\alpha$-stability $S_{t}^{\alpha}(f)$
evolves as the time $t$ increases. In particular, for $\alpha=2$,
Fact \ref{thm:Tsallis2} implies the complete monotonicity of $S_{t}^{2}(f)$
in $t$. 
\begin{cor}[Complete Monotonicity of Noise Stability]
\label{thm:Tsallis2-1} For $t>0$, 
\[
\frac{\partial^{k}S_{t}^{2}(f)}{\partial t^{k}}=(-1)^{k}\int\left(\frac{\partial^{k}}{\partial x^{k}}T_{t}f(x)\right)^{2}dx,
\]
whenever $k\in\mathbb{N}^{+}$. 
\end{cor}
The $\alpha$-stability or noise stability is an important and general
concept, since it can be used to recover several classic results,
e.g., the Euclidean/Gaussian isoperimetric inequality and the hypercontractivity
inequality. Consider the $n$-dimensional Euclidean space. The noise
stability was used by Ledoux \cite{ledoux1994semigroup} to recover
the Euclidean isoperimetric inequality, i.e., the Euclidean balls
have the minimum surface area when the volume is given. The key ingredient
in his proof is the following inequalities: for any $t\ge0$, 
\[
a-\sqrt{\frac{2t}{\pi}}\mathrm{Vol}_{n-1}(\partial A)\le S_{t}^{2}(1_{A})\le S_{t}^{2}(1_{B}),
\]
where an arbitrary set $A$ and an Euclidean ball $B$ satisfy $\mathrm{Vol}_{n}(A)=\mathrm{Vol}_{n}(B)=a$,
and $\mathrm{Vol}_{n-1}(\partial A)$ denotes the surface area of
$A$. By differentiating all terms in $t$ and letting $t\downarrow0$
yields 
\[
-\lim_{t\downarrow0}\sqrt{2\pi t}\frac{\partial S_{t}^{2}(1_{B})}{\partial t}\le-\lim_{t\downarrow0}\sqrt{2\pi t}\frac{\partial S_{t}^{2}(1_{A})}{\partial t}\le\mathrm{Vol}_{n-1}(\partial A).
\]
The most left quantity above is exactly equal to $\mathrm{Vol}_{n-1}(\partial B)$
\cite{ledoux1994semigroup}. So, the Euclidean isoperimetric inequality
is recovered. The inequalities above connect the derivative of the
noise stability of $1_{A}$ and the surface area of $A$. This observation
inspires us to investigate whether $(-1)^{k}\frac{\partial^{k}S_{t}^{2}(1_{A})}{\partial t^{k}},k\ge1,t>0$
are still minimized by the Euclidean balls when the volume of $A$
is given.

The noise stability with respect to the Ornstein--Uhlenbeck semigroup
was also well studied in the literature; see, e.g., \cite{borell1985geometric,ledoux1994semigroup,ledoux2014remarks,mossel2015robust,eldan2015two}.
The noise stability in this setting was used to recover the Gaussian
isoperimetric inequality in \cite{ledoux1994semigroup} and used to
recover the hypercontractivity inequality in \cite{ledoux2014remarks}.
The complete monotonicity of the noise stability for the Ornstein--Uhlenbeck
semigroup follows directly by harmonic analysis, specifically, by
expressing it in terms of Hermite polynomials; see e.g., \cite[Proposition 11.37]{ODonnell14analysisof}.

\section{\label{subsec:Notations-and-Preliminaries}Notations and Preliminaries
for Proofs}

Throughout this paper, we use the notation $\partial_{t}:=\frac{\partial}{\partial t}$.
For the derivatives, in addition to the usages of $p_{xx},p_{xt}$
and $\frac{\partial^{2}}{\partial x^{2}}p$, by $p^{(n)}$ we always
mean 
\[
p^{(n)}:=\frac{\partial^{n}}{\partial x^{n}}p.
\]
Sometimes, for ease of notation we also denote 
\[
p_{n}:=p^{(n)}=\frac{\partial^{n}}{\partial x^{n}}p.
\]
The integration interval, usually $(-\infty,+\infty)$, will be omitted,
unless it is not clear from the context. Denote a new probability
density function 
\[
\tilde{p}_{\alpha}:=\frac{p^{\alpha}}{\int p^{\alpha}dx},
\]
which is known as the $\alpha$-tilted version of $p$. Denote 
\[
\mathbb{E}_{\alpha}[g]:=\int g\tilde{p}_{\alpha}dx.
\]
That is, $\mathbb{E}_{\alpha}$ is the expectation operator with respect
to the density $\tilde{p}_{\alpha}$. Denote 
\[
\bar{p}_{i}=\frac{p_{i}}{p}.
\]

We will invoke the following lemmas in our proofs. If exchanging differentiation
and integration is feasible, then it will be seen that all the derivatives
of $p$ w.r.t. $t$ can be written as the integrals of linear combinations
of terms $p\Pi_{i=1}^{r}\left(\frac{p^{(n_{i})}}{p}\right)^{k_{i}}$.
The absolute integrability of this kind of functions was investigated
in the literature; see e.g., \cite[Proposition 2]{Cheng2015Higher},
which guarantees the existence of the integrals of these functions.
However, to guarantee the feasibility of exchanging differentiation
and integration, we need to show a stronger integrability condition---the
functions $p\Pi_{i=1}^{r}\left(\frac{p^{(n_{i})}}{p}\right)^{k_{i}}$
are uniformly dominated by an integrable function, which, combined
with the mean value theorem and the dominated convergence theorem,
implies the feasibility of exchanging differentiation and integration.
In fact, the feasibility of exchanging differentiation and integration
is rarely discussed in the literature except in a few works, e.g.,
\cite{barron1984monotonic}. 
\begin{lem}
\label{lem:k>=00003D00003D00003D1}Given $0<\alpha<+\infty$, $r,n_{i},k_{i}\in\mathbb{Z}_{+}$,
$0<a<b<+\infty$, and $\epsilon>0$, it holds that for any $t\in[a,b]$,
\begin{align*}
p^{\alpha}\prod_{i=1}^{r}\left|\frac{p^{(n_{i})}}{p}\right|^{k_{i}} & \le Cp^{\alpha}(x,b+\epsilon),
\end{align*}
where $C=C_{\alpha,a,b,\epsilon,(n_{i},k_{i})_{i=1}^{r}}$ is a constant.
In particular, this implies that 
\[
\lim_{|x|\to\infty}p^{\alpha}\prod_{i=1}^{r}\left(\frac{p^{(n_{i})}}{p}\right)^{k_{i}}=0.
\]
\end{lem}
The proof of Lemma \ref{lem:k>=00003D00003D00003D1} is provided in
Appendix \ref{sec:-Proof-of}. 
\begin{lem}
\label{lem:0<k<1}\cite[Lemma 2.1]{hung2022generalization}For any
$0<k<1$, 
\[
\lim_{|x|\to\infty}\frac{p_{1}}{p^{k}}=0,\quad t>0.
\]
\end{lem}
The following lemma characterizes the relation between the differential
and difference operations. 
\begin{lem}
\label{lem:differential-difference}For any $k$-th order differentiable
function $f$, 
\begin{align*}
f^{(k)}(t) & =\lim_{\Delta\to0}\frac{1}{\Delta^{k}}F_{k}(t,\Delta),
\end{align*}
where $F_{k}(t,\Delta):=\sum_{i=0}^{k}{k \choose i}(-1)^{i}f(t+(k-i)\Delta)$
is the $k$-th order difference of $f$. Moreover, $f^{(k)}(t)\ge0,\forall t$
if and only if $F_{k}(t,\Delta)\ge0,\forall t,\forall\Delta\ge0$. 
\end{lem}
This can be easily proven by Taylor's theorem.

Besides, in order to determine the sign of the third-order derivative
of $h_{\alpha}(p)$, we use the method of semidefinite optimization
(also known as semidefinite programming, SDP). Specifically, a general
SDP problem can be defined as any mathematical programming problem
of the form 
\begin{align}
\min_{x^{1},\dots,x^{n}\in\mathbb{R}^{n}} & \sum_{i,j\in[n]}c_{i,j}(x^{i}\cdot x^{j})\nonumber \\
\text{subject to} & \sum_{i,j\in[n]}p_{i,j,k}(x^{i}\cdot x^{j})\le q_{k},\quad k=1,\dots,m\label{eq:SDP}
\end{align}
where $[n]:=\left\{ 1,\dots,n\right\} $, the $c_{i,j},p_{i,j,k},$and
the $q_{k}$ are real numbers and $x^{i}\cdot x^{j}$ is the dot product
of $x^{i}$ and $x^{j}$.

Denote by $\mathbb{S}^{n}$ the space of all $n\times n$ real symmetric
matrices. The space is equipped with the inner product: 
\[
\langle P,Q\rangle:=\text{trace}(P^{T}Q):=\sum_{i,j=1}^{n}p_{ij}q_{ij},\;\text{for all }P,Q\in\mathbb{S}^{n},
\]
where $p_{ij}$ is the entry $i,j$ of $P$. Then we can rewrite \eqref{eq:SDP}
equivalently as 
\begin{align}
\min_{X\in\mathbb{S}^{n}} & \langle C,X\rangle\nonumber \\
\text{subject to } & \langle P_{k},X\rangle\le q_{k},\quad k=1,\dots,m\label{eq:SDP-1}\\
 & X\succeq0,\nonumber 
\end{align}
where $X\succeq0$ means $X$ is positive semidefinite, entry $i,j$
in $C$ is given by $\frac{c_{i,j}+c_{j,i}}{2}$ from $\eqref{eq:SDP}$,
and $P_{k}$ is a symmetric $n\times n$ matrix having $i,j$th entry
$\frac{p_{i,j,k}+p_{j,i,k}}{2}$ from \eqref{eq:SDP}. Thus, the matrices
$C$ and $P_{k}$ are symmetric and the above inner products are well
defined.

\section{\label{sec:First-order-derivative}First-order Derivative}

We compute the first-order derivative of $h_{\alpha}(p)$. 
\begin{thm}
\label{thm:1st derivative}For $t>0$, 
\begin{equation}
\frac{\partial h_{\alpha}(p)}{\partial t}=\frac{\alpha}{2}\frac{\int p(x,t)^{\alpha-2}p_{1}(x,t)^{2}dx}{\int p(x,t)^{\alpha}dx}.\label{eq:1st derivative}
\end{equation}
Or equivalently, 
\begin{equation}
\frac{\partial h_{\alpha}(p)}{\partial t}=\frac{\alpha}{2}\mathbb{E}_{\alpha}[\bar{p}_{1}^{2}].\label{eq:1st derivative-1}
\end{equation}
\end{thm}
Here, $\mathbb{E}_{\alpha}[\bar{p}_{1}^{2}]$ is called the $\alpha$-tilted
Fisher information, which reduces to the classic Fisher information
when $\alpha=1$. 
\begin{IEEEproof}
By definition of $h_{\alpha}(p)$ and by invoking Lemma \ref{lem:k>=00003D00003D00003D1},
the mean value theorem and the dominated convergence theorem (see
e.g., \cite[Corollary 2.8.7]{bogachev2007measure1}), it holds that
\begin{align*}
\frac{\partial h_{\alpha}(p)}{\partial t} & =\frac{1}{1-\alpha}\frac{\int\alpha p^{\alpha-1}p_{t}dx}{\int p^{\alpha}dx}\\
 & =\frac{1}{2}\frac{\alpha}{1-\alpha}\frac{\int p^{\alpha-1}p_{2}dx}{\int p^{\alpha}dx}\\
 & =\frac{1}{2}\frac{\alpha}{1-\alpha}\frac{\int p^{\alpha-1}dp_{1}}{\int p^{\alpha}dx}\\
 & =\frac{1}{2}\frac{\alpha}{1-\alpha}\frac{p^{\alpha-1}p_{1}\bigg|_{-\infty}^{+\infty}-\int(\alpha-1)p^{\alpha-2}p_{1}^{2}dx}{\int p^{\alpha}dx}\\
 & =\frac{\alpha}{2}\frac{\int p^{\alpha-2}p_{1}^{2}dx}{\int p^{\alpha}dx},
\end{align*}
where the second equality follows from heat equation \eqref{eq:heat equation},
the third equality follows from that given $t$, $dp_{1}=p_{2}dx$,
the forth equality follows from integration by parts, and the fifth
equality follows from $p^{\alpha-1}p_{1}\bigg|_{-\infty}^{+\infty}=0$,
no matter whether $0<\alpha<1$ or $\alpha>1$, by Lemmas \ref{lem:k>=00003D00003D00003D1}
and \ref{lem:0<k<1}. 
\end{IEEEproof}
In this paper, Lemmas \ref{lem:k>=00003D00003D00003D1} and \ref{lem:0<k<1}
will be used many times in the calculation of integral by parts, in
order to guarantee the feasibility of exchanging differentiation and
integration and the feasibility of the integral by parts, and also
guarantee that the first term of the integral by parts vanishes. For
brevity of presentation, we will not mention each time. As a corollary
to Theorem \ref{thm:1st derivative}, we have the following result. 
\begin{thm}
\label{thm:sign of 1st derivative}For $t>0$, $\frac{\partial h_{\alpha}(p)}{\partial t}\ge0$,
whenever $\alpha\in(0,+\infty]$. 
\end{thm}
\begin{IEEEproof}
By Theorem \ref{thm:1st derivative}, we immediately have that $\frac{\partial h_{\alpha}(p)}{\partial t}\ge0$
for $\alpha\in(0,1)\cup(1,+\infty)$. The order $\alpha$ can be extended
to $1$ and $+\infty$ by Lemma \ref{lem:differential-difference}. 
\end{IEEEproof}

\section{\label{sec:Second-order-derivative}Second-order Derivative}

The following lemma is instrumental in deriving $\frac{\partial^{2}h_{\alpha}(p)}{\partial t^{2}}$. 
\begin{lem}
\label{lem:(p^alpha)_t}For $t>0$, 
\begin{align}
\frac{\partial(\int p^{\alpha-2}p_{1}^{2}dx)}{\partial t} & =\frac{(\alpha-2)(\alpha-3)}{6}\int p^{\alpha-4}p_{1}^{4}dx-\int p^{\alpha-2}p_{2}^{2}dx,\label{eq:p^alpha-2p_1^2}\\
\frac{\partial(\int p^{\alpha}dx)}{\partial t} & =-\frac{\alpha(\alpha-1)}{2}\int p^{\alpha-2}p_{1}^{2}dx.\label{eq:p^alpha}
\end{align}
\end{lem}
The proof of Lemma \ref{lem:(p^alpha)_t} is presented in Appendix
\ref{sec:Appendix-(p^alpha)_t}. Then we give the second-order derivative
of $h_{\alpha}(p)$. 
\begin{thm}
\label{thm:2nd derivative}For $t>0$, 
\begin{equation}
\frac{\partial^{2}h_{\alpha}(p)}{\partial t^{2}}=\frac{\alpha}{12}\frac{\left((\alpha-2)(\alpha-3)\int p^{\alpha-4}p_{1}^{4}dx-6\int p^{\alpha-2}p_{2}^{2}dx\right)\int p^{\alpha}dx+3\alpha(\alpha-1)(\int p^{\alpha-2}p_{1}^{2}dx)^{2}}{(\int p^{\alpha}dx)^{2}}.\label{eq:2nd derivative}
\end{equation}
Or equivalently, 
\begin{equation}
\frac{\partial^{2}h_{\alpha}(p)}{\partial t^{2}}=\frac{\alpha}{12}\left((\alpha-2)(\alpha-3)\mathbb{E}_{\alpha}\left[\bar{p}_{1}^{4}\right]-6\mathbb{E}_{\alpha}\left[\bar{p}_{2}^{2}\right]+3\alpha(\alpha-1)(\mathbb{E}_{\alpha}\left[\bar{p}_{1}^{2}\right])^{2}\right).\label{eq:2nd derivative-1-1}
\end{equation}
\end{thm}
\begin{IEEEproof}
Taking the derivative of $t$ on both sides of \eqref{eq:1st derivative},
we have 
\begin{align*}
\frac{\partial^{2}h_{\alpha}(p)}{\partial t^{2}} & =\frac{\alpha}{2}\frac{(\int p^{\alpha-2}p_{1}^{2}dx)_{t}\cdot\int p^{\alpha}dx-\int p^{\alpha-2}p_{1}^{2}dx\cdot(\int p^{\alpha}dx)_{t}}{(\int p^{\alpha}dx)^{2}}\\
 & =\frac{\alpha}{2}\frac{\left(\frac{(\alpha-2)(\alpha-3)}{6}\int p^{\alpha-4}p_{1}^{4}dx-\int p^{\alpha-2}p_{2}^{2}dx\right)\int p^{\alpha}dx+\frac{\alpha(\alpha-1)}{2}\int p^{\alpha-2}p_{1}^{2}dx\cdot\int p^{\alpha-2}p_{1}^{2}dx}{(\int p^{\alpha}dx)^{2}}\\
 & =\textrm{RHS of }\eqref{eq:2nd derivative},
\end{align*}
where the second equality follows from Lemma \ref{lem:(p^alpha)_t}. 
\end{IEEEproof}
Our main result about the second-order derivative of $h_{\alpha}(p)$
is the following theorem. 
\begin{thm}
\label{thm:sign of 2nd derivative}When $0<\alpha<1$ or $1<\alpha\le3$,
$h_{\alpha}(p(x,t))$ is concave in $t$, where $t>0$. 
\end{thm}
The order $\alpha$ can be extended to $1$ by Lemma \ref{lem:differential-difference}.
The proof of this theorem is deferred to Section \ref{sec:Proof of sign of 2nd derivative}.

\section{\label{sec:Third-order--derivative}Third-order Derivative}

The following lemmas are instrumental in deriving $\frac{\partial^{3}h_{\alpha}(p)}{\partial t^{3}}$. 
\begin{lem}
\label{lem:E_alphap_1^4}For $t>0$, 
\begin{align*}
\frac{\partial\mathbb{E}_{\alpha}[\bar{p}_{1}^{4}]}{\partial t} & =\frac{3(\alpha-4)(\alpha-5)}{10}\mathbb{E}_{\alpha}[\bar{p}_{1}^{6}]-6\mathbb{E}_{\alpha}[\bar{p}_{1}^{2}\bar{p}_{2}^{2}]+\frac{\alpha(\alpha-1)}{2}\mathbb{E}_{\alpha}[\bar{p}_{1}^{4}]\mathbb{E}_{\alpha}[\bar{p}_{1}^{2}],\\
\frac{\partial\mathbb{E}_{\alpha}[\bar{p}_{2}^{2}]}{\partial t} & =\frac{(\alpha-2)(\alpha-3)}{2}\mathbb{E}_{\alpha}[\bar{p}_{1}^{2}\bar{p}_{2}^{2}]+(\alpha-2)\mathbb{E}_{\alpha}[\bar{p}_{2}^{3}]-\mathbb{E}_{\alpha}[\bar{p}_{3}^{2}]+\frac{\alpha(\alpha-1)}{2}\mathbb{E}_{\alpha}[\bar{p}_{1}^{2}]\mathbb{E}_{\alpha}[\bar{p}_{2}^{2}],\\
\frac{\partial\mathbb{E}_{\alpha}[\bar{p}_{1}^{2}]}{\partial t} & =\frac{(\alpha-2)(\alpha-3)}{6}\mathbb{E}_{\alpha}[\bar{p}_{1}^{4}]-\mathbb{E}_{\alpha}[\bar{p}_{2}^{2}]+\frac{\alpha(\alpha-1)}{2}(\mathbb{E}_{\alpha}[\bar{p}_{1}^{2}])^{2},\\
\frac{\partial(\mathbb{E}_{\alpha}[\bar{p}_{1}^{2}])^{2}}{\partial t} & =\mathbb{E}_{\alpha}[\bar{p}_{1}^{2}]\left(\frac{(\alpha-2)(\alpha-3)}{3}\mathbb{E}_{\alpha}[\bar{p}_{1}^{4}]-2\mathbb{E}_{\alpha}[\bar{p}_{2}^{2}]+\alpha(\alpha-1)(\mathbb{E}_{\alpha}[\bar{p}_{1}^{2}])^{2}\right).
\end{align*}
\end{lem}
The proof of Lemma \ref{lem:E_alphap_1^4} is presented in Appendix
\ref{sec:Appendix-E_alphap_1^4}. Then we compute the third-order
derivative of $h_{\alpha}(p)$. 
\begin{thm}
For $t>0$, 
\begin{align}
\frac{\partial^{3}h_{\alpha}(p)}{\partial t^{3}} & =\frac{\alpha}{12}\left(\frac{3(\alpha-2)(\alpha-3)(\alpha-4)(\alpha-5)}{10}\mathbb{E}_{\alpha}[\bar{p}_{1}^{6}]-9(\alpha-2)(\alpha-3)\mathbb{E}_{\alpha}[\bar{p}_{1}^{2}\bar{p}_{2}^{2}]\right.\nonumber \\
 & \qquad+\frac{3}{2}\alpha(\alpha-1)(\alpha-2)(\alpha-3)\mathbb{E}_{\alpha}[\bar{p}_{1}^{4}]\mathbb{E}_{\alpha}[\bar{p}_{1}^{2}]-6(\alpha-2)\mathbb{E}_{\alpha}[\bar{p}_{2}^{3}]\nonumber \\
 & \qquad+6\mathbb{E}_{\alpha}[\bar{p}_{3}^{2}]-9\alpha(\alpha-1)\mathbb{E}_{\alpha}[\bar{p}_{1}^{2}]\mathbb{E}_{\alpha}[\bar{p}_{2}^{2}]+3\alpha^{2}(\alpha-1)^{2}(\mathbb{E}_{\alpha}[\bar{p}_{1}^{2}])^{3}\bigg).\label{eq:3rd derivative}
\end{align}
\end{thm}
\begin{IEEEproof}
From \eqref{eq:2nd derivative-1-1}, we have 
\begin{align*}
\frac{\partial^{3}h_{\alpha}(p)}{\partial t^{3}} & =\frac{\alpha}{12}\left((\alpha-2)(\alpha-3)\frac{\partial\mathbb{E}_{\alpha}\left[\bar{p}_{1}^{4}\right]}{\partial t}-6\frac{\partial\mathbb{E}_{\alpha}\left[\bar{p}_{2}^{2}\right]}{\partial t}+3\alpha(\alpha-1)\frac{\partial(\mathbb{E}_{\alpha}\left[\bar{p}_{1}^{2}\right])^{2}}{\partial t}\right)\\
 & =\frac{\alpha}{12}\left((\alpha-2)(\alpha-3)(\frac{3(\alpha-4)(\alpha-5)}{10}\mathbb{E}_{\alpha}[\bar{p}_{1}^{6}]-6\mathbb{E}_{\alpha}[\bar{p}_{1}^{2}\bar{p}_{2}^{2}]+\frac{\alpha(\alpha-1)}{2}\mathbb{E}_{\alpha}[\bar{p}_{1}^{4}]\mathbb{E}_{\alpha}[\bar{p}_{1}^{2}])\right.\\
 & \qquad-6(\frac{(\alpha-2)(\alpha-3)}{2}\mathbb{E}_{\alpha}[\bar{p}_{1}^{2}\bar{p}_{2}^{2}]+(\alpha-2)\mathbb{E}_{\alpha}[\bar{p}_{2}^{3}]-\mathbb{E}_{\alpha}[\bar{p}_{3}^{2}]+\frac{\alpha(\alpha-1)}{2}\mathbb{E}_{\alpha}[\bar{p}_{1}^{2}]\mathbb{E}_{\alpha}[\bar{p}_{2}^{2}])\\
 & \qquad\left.+3\alpha(\alpha-1)\mathbb{E}_{\alpha}[\bar{p}_{1}^{2}](\frac{(\alpha-2)(\alpha-3)}{3}\mathbb{E}_{\alpha}[\bar{p}_{1}^{4}]-2\mathbb{E}_{\alpha}[\bar{p}_{2}^{2}]+\alpha(\alpha-1)(\mathbb{E}_{\alpha}[\bar{p}_{1}^{2}])^{2})\right)\\
 & =\textrm{RHS of }\eqref{eq:3rd derivative},
\end{align*}
where the second equality follows from Lemma \ref{lem:E_alphap_1^4}. 
\end{IEEEproof}
\begin{thm}
\label{thm:sign of 3rd derivative}For $t>0$, when $\alpha\in[\beta_{0},1.76]$,
$\frac{\partial^{3}h_{\alpha}(t)}{\partial t^{3}}\ge0$. 
\end{thm}
The proof of this theorem is deferred to Section \ref{sec:Proof of sign of 3rd derivative}.

\section{\label{sec:Fourth-order-Derivative}Fourth-order Derivative}

The following lemmas are instrumental in deriving $\frac{\partial^{4}h_{\alpha}(p)}{\partial t^{4}}$. 
\begin{lem}
\label{lem:E_alphap_1^6}For $t>0$, 
\begin{align*}
\frac{\partial\mathbb{E}_{\alpha}[\bar{p}_{1}^{6}]}{\partial t} & =\frac{5}{14}(\alpha-6)(\alpha-7)\mathbb{E}_{\alpha}[\bar{p}_{1}^{8}]-15\mathbb{E}_{\alpha}[\bar{p}_{1}^{4}\bar{p}_{2}^{2}]+\frac{\alpha(\alpha-1)}{2}\mathbb{E}_{\alpha}[\bar{p}_{1}^{6}]\mathbb{E}_{\alpha}[\bar{p}_{1}^{2}],\\
\frac{\partial\mathbb{E}_{\alpha}[\bar{p}_{1}^{2}\bar{p}_{2}^{2}]}{\partial t} & =\frac{(\alpha-4)(\alpha-5)}{2}\mathbb{E}_{\alpha}[\bar{p}_{1}^{4}\bar{p}_{2}^{2}]+\frac{7}{3}(\alpha-4)\mathbb{E}_{\alpha}[\bar{p}_{1}^{2}\bar{p}_{2}^{3}]+\frac{1}{3}\mathbb{E}_{\alpha}[\bar{p}_{2}^{4}]-\mathbb{E}_{\alpha}[\bar{p}_{1}^{2}\bar{p}_{3}^{2}]+\frac{\alpha(\alpha-1)}{2}\mathbb{E}_{\alpha}[\bar{p}_{1}^{2}]\mathbb{E}_{\alpha}[\bar{p}_{1}^{2}\bar{p}_{2}^{2}],\\
\frac{\partial\mathbb{E}_{\alpha}[\bar{p}_{2}^{3}]}{\partial t} & =\frac{(\alpha-3)(\alpha-4)}{2}\mathbb{E}_{\alpha}[\bar{p}_{1}^{2}\bar{p}_{2}^{3}]+(\alpha-3)\mathbb{E}_{\alpha}[\bar{p}_{2}^{4}]-3\mathbb{E}_{\alpha}[\bar{p}_{2}\bar{p}_{3}^{2}]+\frac{\alpha(\alpha-1)}{2}\mathbb{E}_{\alpha}[\bar{p}_{2}^{3}]\mathbb{E}_{\alpha}[\bar{p}_{1}^{2}],\\
\frac{\partial\mathbb{E}_{\alpha}[\bar{p}_{3}^{2}]}{\partial t} & =\frac{(\alpha-2)(\alpha-3)}{2}\mathbb{E}_{\alpha}[\bar{p}_{1}^{2}\bar{p}_{3}^{2}]+(\alpha-2)\mathbb{E}_{\alpha}[\bar{p}_{2}\bar{p}_{3}^{2}]-\mathbb{E}_{\alpha}[\bar{p}_{4}^{2}]+\frac{\alpha(\alpha-1)}{2}\mathbb{E}_{\alpha}[\bar{p}_{1}^{2}]\mathbb{E}_{\alpha}[\bar{p}_{3}^{2}].
\end{align*}
\end{lem}
The proof of Lemma \ref{lem:E_alphap_1^6} is presented in Appendix
\ref{sec:Appendix-F_alphap_1^6}. Then we compute the fourth-order
derivative of $h_{\alpha}(p)$.
\begin{thm}
For $t>0$,
\begin{align}
\frac{\partial^{4}h_{\alpha}(p)}{\partial t^{4}} & =\frac{\alpha}{12}\left(\frac{3}{28}(\alpha-2)(\alpha-3)(\alpha-4)(\alpha-5)(\alpha-6)(\alpha-7)\mathbb{E}_{\alpha}[\bar{p}_{1}^{8}]-9(\alpha-2)(\alpha-3)(\alpha-4)(\alpha-5)\mathbb{E}_{\alpha}[\bar{p}_{1}^{4}\bar{p}_{2}^{2}]\right.\nonumber \\
 & +\frac{3}{5}\alpha(\alpha-1)(\alpha-2)(\alpha-3)(\alpha-4)(\alpha-5)\mathbb{E}_{\alpha}[\bar{p}_{1}^{6}]\mathbb{E}_{\alpha}[\bar{p}_{1}^{2}]-24(\alpha-2)(\alpha-3)(\alpha-4)\mathbb{E}_{\alpha}[\bar{p}_{1}^{2}\bar{p}_{2}^{3}]\nonumber \\
 & -9(\alpha-2)(\alpha-3)\mathbb{E}_{\alpha}[\bar{p}_{2}^{4}]+12(\alpha-2)(\alpha-3)\mathbb{E}_{\alpha}[\bar{p}_{1}^{2}\bar{p}_{3}^{2}]-18\alpha(\alpha-1)(\alpha-2)(\alpha-3)\mathbb{E}_{\alpha}[\bar{p}_{1}^{2}]\mathbb{E}_{\alpha}[\bar{p}_{1}^{2}\bar{p}_{2}^{2}]\nonumber \\
 & +3\alpha^{2}(\alpha-1)^{2}(\alpha-2)(\alpha-3)(\mathbb{E}_{\alpha}[\bar{p}_{1}^{2}])^{2}\mathbb{E}_{\alpha}[\bar{p}_{1}^{4}]+\frac{1}{4}\alpha(\alpha-1)(\alpha-2)^{2}(\alpha-3)^{2}(\mathbb{E}_{\alpha}[\bar{p}_{1}^{4}])^{2}\nonumber \\
 & -3\alpha(\alpha-1)(\alpha-2)(\alpha-3)\mathbb{E}_{\alpha}[\bar{p}_{1}^{4}]\mathbb{E}_{\alpha}[\bar{p}_{2}^{2}]+24(\alpha-2)\mathbb{E}_{\alpha}[\bar{p}_{2}\bar{p}_{3}^{2}]-12\alpha(\alpha-1)(\alpha-2)\mathbb{E}_{\alpha}[\bar{p}_{2}^{3}]\mathbb{E}_{\alpha}[\bar{p}_{1}^{2}]\label{eq:4th derivative}\\
 & -6\mathbb{E}_{\alpha}[\bar{p}_{4}^{2}]+12\alpha(\alpha-1)\mathbb{E}_{\alpha}[\bar{p}_{1}^{2}]\mathbb{E}_{\alpha}[\bar{p}_{3}^{2}]+9\alpha(\alpha-1)(\mathbb{E}_{\alpha}[\bar{p}_{2}^{2}])^{2}\nonumber \\
 & \left.-18\alpha^{2}(\alpha-1)^{2}(\mathbb{E}_{\alpha}[\bar{p}_{1}^{2}])^{2}\mathbb{E}_{\alpha}[\bar{p}_{2}^{2}]+\frac{9}{2}\alpha^{3}(\alpha-1)^{3}(\mathbb{E}_{\alpha}[\bar{p}_{1}^{2}])^{4}\right).\nonumber 
\end{align}
\end{thm}
\begin{IEEEproof}
From \eqref{eq:3rd derivative}, we have
\begin{align*}
\frac{\partial^{4}h_{\alpha}(p)}{\partial t^{4}} & =\frac{\alpha}{12}\left(\frac{3(\alpha-2)(\alpha-3)(\alpha-4)(\alpha-5)}{10}\frac{\partial\mathbb{E}_{\alpha}[\bar{p}_{1}^{6}]}{\partial t}-9(\alpha-2)(\alpha-3)\frac{\partial\mathbb{E}_{\alpha}[\bar{p}_{1}^{2}\bar{p}_{2}^{2}]}{\partial t}\right.\\
 & \qquad+\frac{3}{2}\alpha(\alpha-1)(\alpha-2)(\alpha-3)\frac{\partial(\mathbb{E}_{\alpha}[\bar{p}_{1}^{4}]\mathbb{E}_{\alpha}[\bar{p}_{1}^{2}])}{\partial t}-6(\alpha-2)\frac{\partial\mathbb{E}_{\alpha}[\bar{p}_{2}^{3}]}{\partial t}\\
 & \qquad+6\frac{\partial\mathbb{E}_{\alpha}[\bar{p}_{3}^{2}]}{\partial t}-9\alpha(\alpha-1)\frac{\partial(\mathbb{E}_{\alpha}[\bar{p}_{1}^{2}]\mathbb{E}_{\alpha}[\bar{p}_{2}^{2}])}{\partial t}+3\alpha^{2}(\alpha-1)^{2}\frac{\partial(\mathbb{E}_{\alpha}[\bar{p}_{1}^{2}])^{3}}{\partial t}\bigg)\\
 & =\textrm{RHS of }\eqref{eq:4th derivative},
\end{align*}
where the second equality follows from Lemma \ref{lem:E_alphap_1^4}
and Lemma \ref{lem:E_alphap_1^6}.
\end{IEEEproof}
\begin{thm}
\label{thm:sign of 4th derivative}For $t>0$, when $\alpha\in[0.93,1.76]$,
$\frac{\partial^{4}h_{\alpha}(t)}{\partial t^{4}}\le0$. 
\end{thm}
The proof of this theorem is deferred to Section \ref{sec:Proof of sign of 4th derivative}.
As mentioned before, numerical results show that the left end point
$0.93$ could be improved to $0.92$. However, this will require polynomials
of higher orders in our curve-fitting argumentwhich makes the proof
more complicated. So, we prefer not doing this. 

\section{\label{sec:Proof of sign of 2nd derivative}Proof of Theorem \ref{thm:sign of 2nd derivative}}

\subsection{Case of $0<\alpha<1$}

For $0<\alpha<1$, to show $\frac{\partial^{2}h_{\alpha}(p)}{\partial t^{2}}\le0$,
it suffices to express the expression $(\alpha-2)(\alpha-3)\int p^{\alpha-4}p_{1}^{4}dx-6\int p^{\alpha-2}p_{2}^{2}dx$
in \eqref{eq:2nd derivative} in the following form: 
\begin{equation}
-\int p^{\alpha}\left(a\frac{p_{2}}{p}-b\frac{p_{1}^{2}}{p^{2}}\right)^{2}dx,\label{eq:-8}
\end{equation}
where $a$ and $b$ are coefficients. To this end, we expand \eqref{eq:-8}
as follows: 
\begin{align}
-\int p^{\alpha}\left(a\frac{p_{2}}{p}-b\frac{p_{1}^{2}}{p^{2}}\right)^{2}dx & =-\int(a^{2}p_{2}^{2}p^{\alpha-2}-2abp_{2}p_{1}^{2}p^{\alpha-3}+b^{2}p_{1}^{4}p^{\alpha-4})dx\nonumber \\
 & =-a^{2}\int p^{\alpha-2}p_{2}^{2}dx-\left(2ab\cdot\frac{\alpha-3}{3}+b^{2}\right)\int p^{\alpha-4}p_{1}^{4}dx,\label{eq:-a^2p_2^2p^alpha-2}
\end{align}
where the second equality follows from \eqref{eq:p^alpha-3p_1^2p_2}.
To make the coefficients matched, we need to solve the following equations:
\[
\begin{cases}
-a^{2}=-6\\
-\frac{2ab}{3}(\alpha-3)-b^{2}=(\alpha-2)(\alpha-3)
\end{cases}.
\]
In order to ensure that these equations have at least one solution,
it is required that 
\[
\begin{cases}
a=\pm\sqrt{6}\\
\Delta=(\alpha-3)\left(\frac{4a^{2}}{9}(\alpha-3)-4(\alpha-2)\right)\ge0
\end{cases},
\]
that is, 
\[
\begin{cases}
a=\pm\sqrt{6}\\
0\le\alpha\le3
\end{cases}.
\]
Without loss of generality, we take $a=\sqrt{6}$, yielding $b=\frac{\frac{2\sqrt{6}(\alpha-3)}{3}\pm\sqrt{-\frac{4}{3}\alpha(\alpha-3)}}{-2}$.

\subsection{Case of $1<\alpha\le3$}

For $1<\alpha\le3$, in order to show $\frac{\partial^{2}h_{\alpha}(p)}{\partial t^{2}}\le0$,
we would like to write the expression on the right hand side of \eqref{eq:2nd derivative-1-1}
in the following form: 
\[
L:=-\frac{\alpha}{12}\mathbb{E}_{\alpha}\left[\left(a\bar{p}_{2}+b\bar{p}_{1}^{2}+c\mathbb{E}_{\alpha}\left[\bar{p}_{1}^{2}\right]\right)^{2}\right].
\]
Expanding this expression yields that 
\begin{align}
L & =\frac{\alpha}{12}\left(-\mathbb{E}_{\alpha}\left[\left(a\bar{p}_{2}+b\bar{p}_{1}^{2}\right)^{2}\right]-c^{2}\left(\mathbb{E}_{\alpha}\left[\bar{p}_{1}^{2}\right]\right)^{2}-2c\mathbb{E}_{\alpha}\left[\bar{p}_{1}^{2}\right]\mathbb{E}_{\alpha}\left[\left(a\bar{p}_{2}+b\bar{p}_{1}^{2}\right)\right]\right)\nonumber \\
 & =\frac{\alpha}{12}\left(-a^{2}\mathbb{E}_{\alpha}\left[\bar{p}_{2}^{2}\right]-\left(b^{2}-2ab\cdot\frac{\alpha-3}{3}\right)\mathbb{E}_{\alpha}\left[\bar{p}_{1}^{4}\right]-\left(c^{2}+2c(b-(\alpha-1)a)\right)\left(\mathbb{E}_{\alpha}\left[\bar{p}_{1}^{2}\right]\right)^{2}\right),\label{eq:-1}
\end{align}
where \eqref{eq:-1} follows from \eqref{eq:-a^2p_2^2p^alpha-2} and
the fact that $\mathbb{E}_{\alpha}\left[\bar{p}_{2}\right]=-(\alpha-1)\mathbb{E}_{\alpha}\left[\bar{p}_{1}^{2}\right]$,
which is contained in the proof of Theorem \ref{thm:1st derivative}.

Comparing the coefficients in \eqref{eq:2nd derivative-1-1} and the
ones in \eqref{eq:-1}, we set $a,b,c$ such that 
\begin{equation}
\begin{cases}
-a^{2}=-6\\
-\left(b^{2}-2ab\cdot\frac{\alpha-3}{3}\right)=(\alpha-2)(\alpha-3)\\
-\left(c^{2}+2c(b-(\alpha-1)a)\right)=3\alpha(\alpha-1)
\end{cases}.\label{eq:-2}
\end{equation}
To find a solution to these equations, we let 
\[
\begin{cases}
a=-\sqrt{6}\\
b=\frac{2a\frac{\alpha-3}{3}+\sqrt{\left(2a\frac{\alpha-3}{3}\right)^{2}-4(\alpha-2)(\alpha-3)}}{2}\\
\qquad=\sqrt{6}\left(1-\frac{\alpha}{3}\right)+\sqrt{\alpha\left(1-\frac{\alpha}{3}\right)}
\end{cases},
\]
where $b$ is real as long as $0\le\alpha\le3$. This set of $(a,b)$
satisfies the first two equations of \eqref{eq:-2}. A real number
$c$ that together with $a,b$ forms a solution to \eqref{eq:-2}
exists if and only if 
\[
\left(2(b-(\alpha-1)a)\right)^{2}\ge12\alpha(\alpha-1).
\]
Equivalently, 
\[
\sqrt{6}\left(1-\frac{\alpha}{3}\right)+\sqrt{\alpha\left(1-\frac{\alpha}{3}\right)}+(\alpha-1)\sqrt{6}\ge\sqrt{3\alpha(\alpha-1)}
\]
for $1<\alpha\le3$. That is, 
\[
\frac{2}{3}\sqrt{6\alpha}+\sqrt{1-\frac{\alpha}{3}}\ge\sqrt{3(\alpha-1)}.
\]
Taking square, it is easy to see that this inequality always holds
for $1<\alpha\le3$. Therefore, for $1<\alpha\le3$, there always
exists a solution $(a,b,c)$ to \eqref{eq:-2}, which implies $\frac{\partial^{2}h_{\alpha}(p)}{\partial t^{2}}\le0$.

Based on the above, we have obtained $\frac{\partial^{2}h_{\alpha}(p)}{\partial t^{2}}\le0$
for $\alpha\in(0,1)\cup(1,3]$. This result can be extended to $\alpha=1$
by Lemma \ref{lem:differential-difference}.

\section{\label{sec:Proof of sign of 3rd derivative}Proof of Theorem \ref{thm:sign of 3rd derivative}}

It follows directly from Theorem \ref{thm:sign of 4th derivative}
and Proposition \ref{prop:sign} that $\frac{\partial^{3}h_{\alpha}(p)}{\partial t^{3}}\ge0$
whenever $\alpha\in[0.93,1.76]$. Thus, it suffices to consider the
case where $\alpha\le1$. We will divide the proof of Theorem \ref{thm:sign of 3rd derivative}
into three cases: $\alpha\in[\beta_{0},\frac{5-2\sqrt{2}}{3}]$, $\alpha\in(0.5,0.84)$,
and $\alpha\in(0.83,1]$, where $\frac{5-2\sqrt{2}}{3}\approx0.72385763$.
The union of these throughout intervals is obviously $[\beta_{0},1]$.

\subsection{Case 1: $\alpha\in[\beta_{0},\frac{5-2\sqrt{2}}{3}]$}

In order to determine the sign, we next would like to write \eqref{eq:3rd derivative}
in the following form:\footnote{Note that here $e$ denotes an arbitrary real number to be determined,
not Euler's number.} 
\[
K:=\frac{\alpha}{12}\left(\mathbb{E}_{\alpha}[(a\bar{p}_{3}+b\bar{p}_{1}\bar{p}_{2}+c\bar{p}_{1}^{3}+d\bar{p}_{1}\mathbb{E}_{\alpha}[\bar{p}_{1}^{2}])^{2}]+e^{2}\mathbb{E}_{\alpha}[\bar{p}_{1}^{6}]+f^{2}\mathbb{E}_{\alpha}[\bar{p}_{1}^{4}]\mathbb{E}_{\alpha}[\bar{p}_{1}^{2}]+g^{2}(\mathbb{E}_{\alpha}[\bar{p}_{1}^{2}])^{3}+h^{2}\mathbb{E}_{\alpha}[\bar{p}_{1}^{2}]\mathbb{E}_{\alpha}[\bar{p}_{2}^{2}]\right).
\]

Expanding this expression yields that 
\begin{align}
K & =\frac{\alpha}{12}\left(a^{2}\mathbb{E}_{\alpha}[\bar{p}_{3}^{2}]+(b^{2}-ab(\alpha-3)-6ac)\mathbb{E}_{\alpha}[\bar{p}_{1}^{2}\bar{p}_{2}^{2}]+(c^{2}+\frac{2}{5}ac(\alpha-4)(\alpha-5)-\frac{2}{5}bc(\alpha-5)+e^{2})\mathbb{E}_{\alpha}[\bar{p}_{1}^{6}]\right.\nonumber \\
 & -ab\mathbb{E}_{\alpha}[\bar{p}_{2}^{3}]+(d^{2}+g^{2})(\mathbb{E}_{\alpha}[\bar{p}_{1}^{2}])^{3}+(-2ad+h^{2})\mathbb{E}_{\alpha}[\bar{p}_{1}^{2}]\mathbb{E}_{\alpha}[\bar{p}_{2}^{2}]\label{eq:expandK}\\
 & \left.+(2d(\frac{1}{3}a(\alpha-2)(\alpha-3)-\frac{1}{3}b(\alpha-3)+c)+f^{2})\mathbb{E}_{\alpha}[\bar{p}_{1}^{2}]\mathbb{E}_{\alpha}[\bar{p}_{1}^{4}]\right),\nonumber 
\end{align}
where \eqref{eq:expandK} follows from \eqref{eq:p^alpha-3p_1p_2p_3},
\eqref{eq:p^alpha-4p_1^3p_3}, \eqref{eq:p^alpha-2p_1p_3}, \eqref{eq:p^alpha-5p_1^4p_2}
and \eqref{eq:p^alpha-3p_1^2p_2}.

Comparing the coefficients in \eqref{eq:3rd derivative} and the ones
in \eqref{eq:expandK}, we choose $a,b,c,d,e,f,g,h$ such that 
\begin{equation}
\begin{cases}
a^{2}=6\\
b^{2}-ab(\alpha-3)-6ac=-9(\alpha-2)(\alpha-3)\\
c^{2}+\frac{2}{5}ac(\alpha-4)(\alpha-5)-\frac{2}{5}bc(\alpha-5)+e^{2}=\frac{3(\alpha-2)(\alpha-3)(\alpha-4)(\alpha-5)}{10}\\
-ab=-6(\alpha-2)\\
d^{2}+g^{2}=3\alpha^{2}(\alpha-1)^{2}\\
-2ad+h^{2}=-9\alpha(\alpha-1)\\
2d(\frac{1}{3}a(\alpha-2)(\alpha-3)-\frac{1}{3}b(\alpha-3)+c)+f^{2}=\frac{3}{2}\alpha(\alpha-1)(\alpha-2)(\alpha-3)
\end{cases}.\label{eq:equations-2}
\end{equation}
The above equations \eqref{eq:equations-2} have a solution: 
\[
\begin{cases}
a=\sqrt{6}\\
b=\sqrt{6}\alpha-2\sqrt{6}\\
c=\frac{1}{12}(3\sqrt{6}\alpha^{2}-13\sqrt{6}\alpha+14\sqrt{6})\\
e=\frac{\sqrt{-9\alpha^{4}+30\alpha^{3}-53\alpha^{2}+68\alpha-20}}{2\sqrt{30}}\\
f=\frac{\sqrt{9\alpha^{4}-54\alpha^{3}+99\alpha^{2}-54\alpha-3\sqrt{6}\alpha^{2}d+13\sqrt{6}\alpha d-14\sqrt{6}d}}{\sqrt{6}}\\
g=\sqrt{3\alpha^{4}-6\alpha^{3}+3\alpha^{2}-d^{2}}\\
h=\sqrt{-9\alpha^{2}+9\alpha+2\sqrt{6}d}
\end{cases}
\]
which is real if and only if 
\begin{equation}
\begin{cases}
-9\alpha^{4}+30\alpha^{3}-53\alpha^{2}+68\alpha-20\ge0\\
9\alpha^{4}-54\alpha^{3}+99\alpha^{2}-54\alpha-3\sqrt{6}\alpha^{2}d+13\sqrt{6}\alpha d-14\sqrt{6}d\ge0\\
3\alpha^{4}-6\alpha^{3}+3\alpha^{2}-d^{2}\ge0\\
-9\alpha^{2}+9\alpha+2\sqrt{6}d\ge0
\end{cases}.\label{eq:equations-3}
\end{equation}
It is not hard to check that we can find a real number $d$ that satisfies
\eqref{eq:equations-3} when $\beta_{0}\le\alpha\le\frac{5-2\sqrt{2}}{3}$,
where $\beta_{0}$ is the unique real root of $9x^{3}-12x^{2}+29x-10=0$,
and $\beta_{0}\approx0.38921378$. Therefore, for $\beta_{0}\le\alpha\le\frac{5-2\sqrt{2}}{3}$,
there always exists a solution $(a,b,c,d,e,f,g,h)$ to \eqref{eq:equations-2},
which implies $\frac{\partial^{3}h_{\alpha}(p)}{\partial t^{3}}\ge0$.

\subsection{Case 2: $\alpha\in(0.5,0.84)$}

For $\alpha\in(0.5,0.84)$, we would like to write \eqref{eq:3rd derivative}
in the following more general form: 
\[
M:=\frac{\alpha}{12}\left(\mathbb{E}_{\alpha}[N]+d\mathbb{E}_{\alpha}[\bar{p}_{1}^{4}]\mathbb{E}_{\alpha}[\bar{p}_{1}^{2}]+e\mathbb{E}_{\alpha}[\bar{p}_{1}^{2}]\mathbb{E}_{\alpha}[\bar{p}_{2}^{2}]\right),
\]
where $d\ge0,e\ge0$, and $N=\overrightarrow{z}A\overrightarrow{z}^{\top}$
with a positive semidefinite matrix $A$ and the vector $\overrightarrow{z}:=\left(\bar{p}_{3},\bar{p}_{1}\bar{p}_{2},\bar{p}_{1}^{3},\bar{p}_{1}\mathbb{E}_{\alpha}[\bar{p}_{1}^{2}]\right)^{\top}$.
We choose $A$ as 
\[
A:=\begin{bmatrix}6 & 6(\alpha-2) & b & \frac{e}{2}+\frac{9\alpha(\alpha-1)}{2}\\
\\
6(\alpha-2) & -3(\alpha-2)(\alpha-3)+6b & c & a\\
\\
b & c & \begin{array}{cc}
\frac{3(\alpha-2)(\alpha-3)(\alpha-4)(\alpha-5)}{10}\\
-\frac{2(\alpha-4)(\alpha-5)}{5}b+\frac{2(\alpha-5)}{5}c
\end{array} & \begin{array}{cc}
-\frac{3\alpha(\alpha-1)(\alpha-2)(\alpha-3)}{4}\\
+\frac{(\alpha-3)a}{3}-\frac{d}{2}-\frac{(\alpha-2)(\alpha-3)}{6}e
\end{array}\\
\\
\frac{e}{2}+\frac{9\alpha(\alpha-1)}{2} & a & \begin{array}{cc}
-\frac{3\alpha(\alpha-1)(\alpha-2)(\alpha-3)}{4}\\
+\frac{(\alpha-3)a}{3}-\frac{d}{2}-\frac{(\alpha-2)(\alpha-3)}{6}e
\end{array} & 3\alpha^{2}(\alpha-1)^{2}
\end{bmatrix}.
\]
Using Lemma \ref{lem:3rd derivative-1}, it is straightforward to
check that $M$ is equal to \eqref{eq:3rd derivative}, when $A$
is taken as the above expression. Hence, to establish the non-negativity
of $M$, it is suffice to ensure the positive semidefinitemess of
$A$ by choosing suitable values of $(a,b,c,d,e)=(a(\alpha),b(\alpha),c(\alpha),d(\alpha),e(\alpha))$,
depending on the value of $\alpha$.

It is in fact difficult to guess what are the best choice of the functions
$(a(\alpha),b(\alpha),c(\alpha),d(\alpha),e(\alpha))$. Instead, we
use the curve-fitting method to find the desired choice of $(a(\alpha),b(\alpha),c(\alpha),d(\alpha),e(\alpha))$.
To this end, we use optimization function ``SemidefiniteOptimization''
in Mathematica to obtain numerical points $(\alpha,a,b,c,d,e)$ that
meet the requirements that $A$ is positive semidefinite and $d\ge0,e\ge0$.
Indeed, recall \eqref{eq:SDP-1}, the problem of semidefinite optimization
we concerned is 
\begin{align}
\min_{A\in\mathbb{S}^{4}} & \,0\nonumber \\
\text{subject to } & d\ge0,e\ge0\label{eq:SDP-2}\\
 & A\succeq0.\nonumber 
\end{align}
These numerical points $(\alpha,a,b,c,d,e)$ are obtained when running
the SDP \eqref{eq:SDP-2} in Mathematica for $\alpha$ taking values
of $0.4,0.5,0.6,0.7,0.8,0.9$ and $1$, respectively. Then we fit
these points using polynomial functions.

For example, upon rounding off the numerical values to an appropriate
decimal precision, the points $(\alpha,a)$ obtained by the optimization
process are 
\begin{align*}
L_{1} & :=\{(0.4,1.58125),(0.5,1.53958),(0.6,1.38319),(0.7,1.12931),\\
 & \qquad\qquad(0.8,0.800784),(0.9,0.414932),(1,0.000135571)\}.
\end{align*}
Then we use a fourth-order polynomial function to fit $L_{1}$ and
obtain the best curve-fitting function 
\[
\hat{a}(\alpha)=0.3+6.4\alpha-8.7\alpha^{2}+1.1\alpha^{3}+0.8\alpha^{4}.
\]
Similarly, for the points of $(\alpha,b)$, $(\alpha,c)$, $(\alpha,d)$,
and $(\alpha,e)$, the best curve-fitting functions found are 
\begin{align*}
\hat{b}(\alpha) & =9.4-24.9\alpha+49.1\alpha^{2}-48.6\alpha^{3}+17.7\alpha^{4},\\
\hat{c}(\alpha) & =-25.7+114.6\alpha-262.6\alpha^{2}+262.2\alpha^{3}-94.2\alpha^{4},\\
\hat{d}(\alpha) & =-0.2\alpha+0.6\alpha^{2}-0.4\alpha^{3},\\
\hat{e}(\alpha) & =0.5-2.3\alpha+5.6\alpha^{2}-3.7\alpha^{3}.
\end{align*}
The graphs of $\hat{a}(\alpha),\hat{b}(\alpha),\hat{c}(\alpha),\hat{d}(\alpha),\hat{e}(\alpha)$
are plotted in Fig. \ref{fig:Curve-fitting}. 
\begin{center}
\begin{figure}
\begin{centering}
\includegraphics[width=0.45\textwidth]{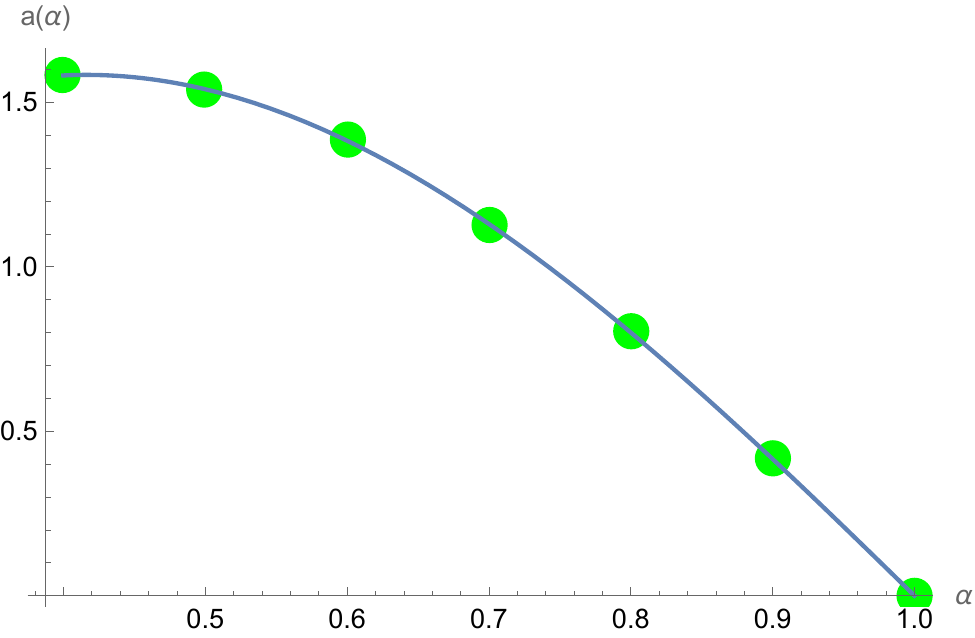}\includegraphics[width=0.45\textwidth]{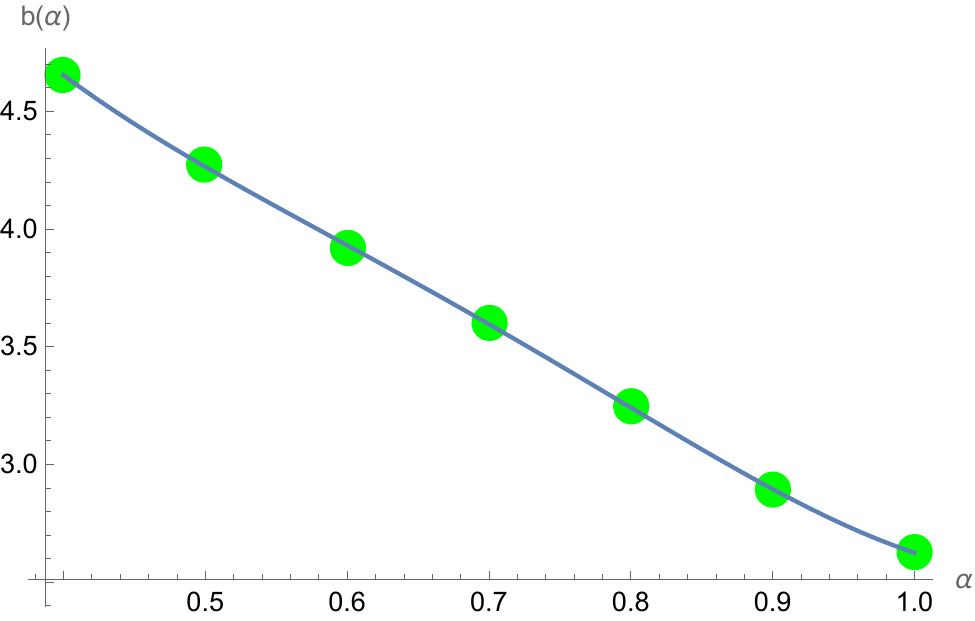} 
\par\end{centering}
\begin{centering}
\includegraphics[width=0.45\textwidth]{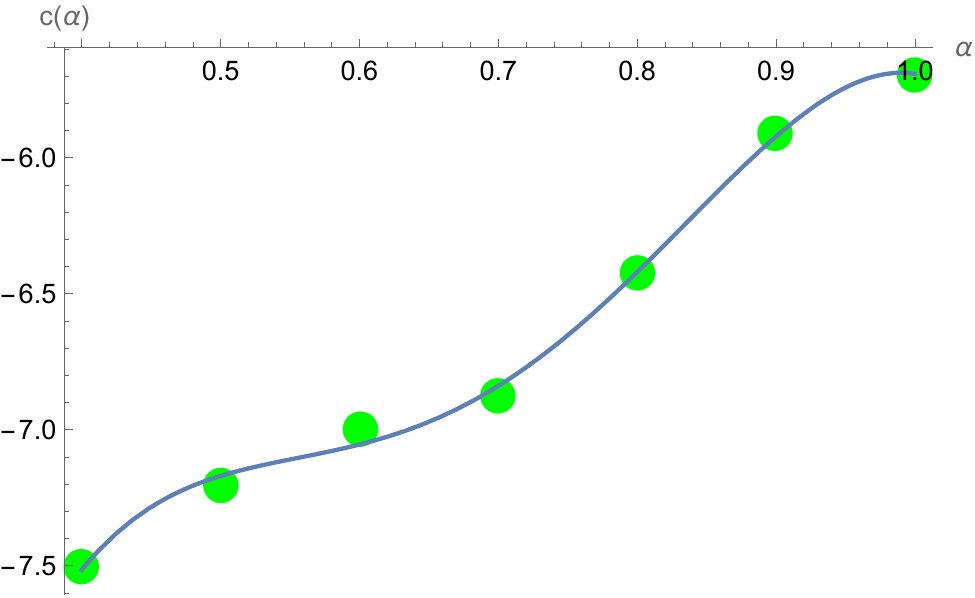}\includegraphics[width=0.45\textwidth]{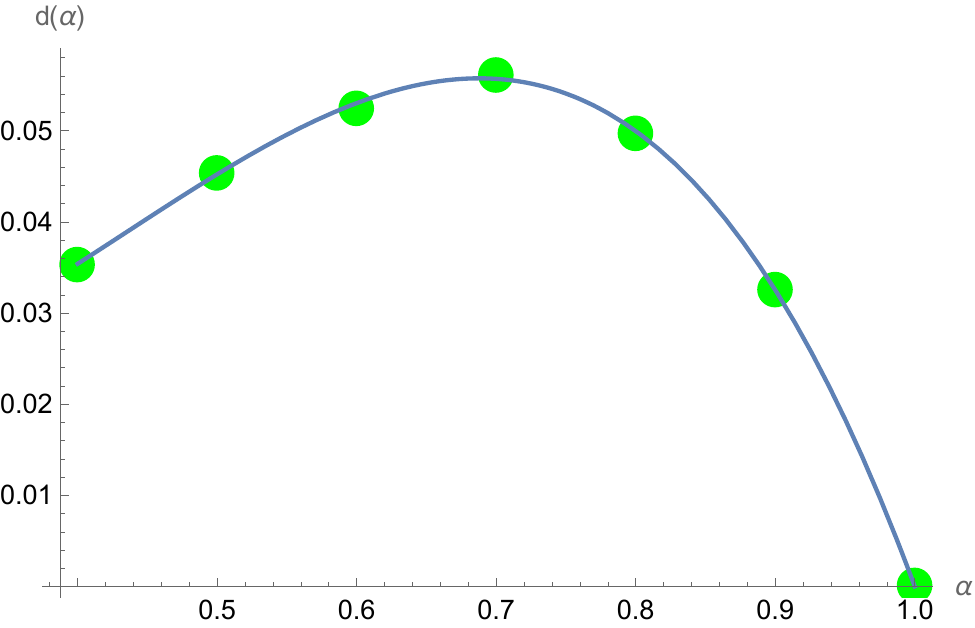} 
\par\end{centering}
\begin{centering}
\includegraphics[width=0.45\textwidth]{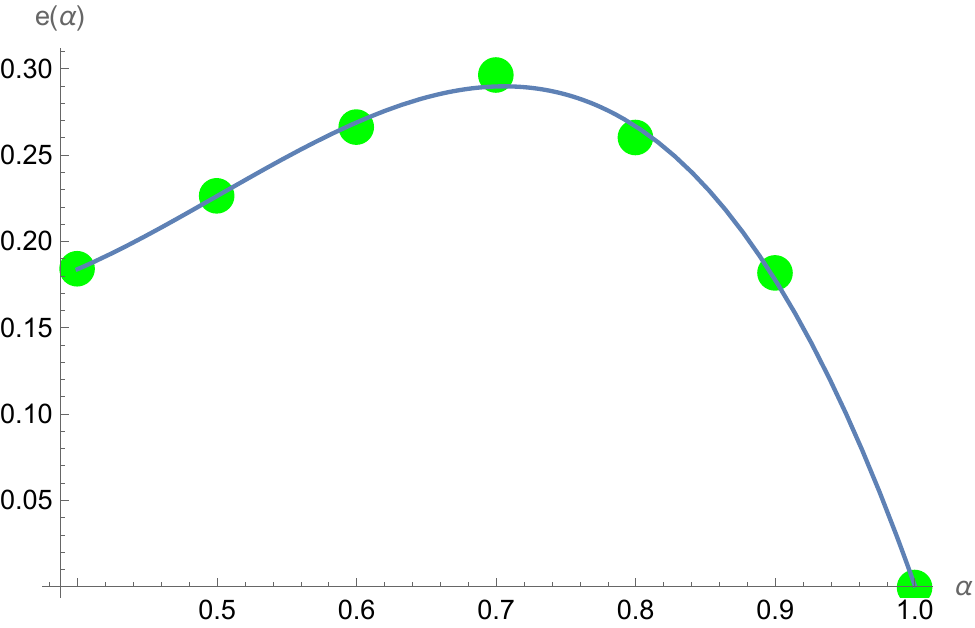} 
\par\end{centering}
\caption{\label{fig:Curve-fitting}Curve-fitting for the best $(a(\alpha),b(\alpha),c(\alpha),d(\alpha),e(\alpha))$. }
\end{figure}
\par\end{center}

Replacing $a,b,c,d,e$ with $\hat{a}(\alpha),\hat{b}(\alpha),\hat{c}(\alpha),\hat{d}(\alpha),\hat{e}(\alpha)$
respectively, the matrix $A$ then turns into {\small{}{} 
\[
\hat{A}=\begin{bmatrix}6 & 6(-2+\alpha) & \begin{array}{cc}
\frac{1}{10}(94-249\alpha+491\alpha^{2}\\
-486\alpha^{3}+177\alpha^{4})
\end{array} & \begin{array}{cc}
\frac{1}{20}(5-113\alpha+\\
146\alpha^{2}-37\alpha^{3})
\end{array}\\
\\
6(-2+\alpha) & \begin{array}{cc}
\frac{3}{5}(64-224\alpha+486\alpha^{2}\\
-486\alpha^{3}+177\alpha^{4})
\end{array} & \begin{array}{cc}
\frac{1}{10}(-257+1146\alpha\\
-2626\alpha^{2}+2622\alpha^{3}\\
-942\alpha^{4})
\end{array} & \begin{array}{cc}
\frac{1}{10}(3+64\alpha-87\alpha^{2}\\
+11\alpha^{3}+8\alpha^{4})
\end{array}\\
\\
\begin{array}{cc}
\frac{1}{10}(94-249\alpha+491\alpha^{2}\\
-486\alpha^{3}+177\alpha^{4})
\end{array} & \begin{array}{cc}
\frac{1}{10}(-257+1146\alpha\\
-2626\alpha^{2}+2622\alpha^{3}\\
-942\alpha^{4})
\end{array} & \begin{array}{cc}
\frac{1}{50}(610-2632\alpha+5307\alpha^{2}\\
-2906\alpha^{3}-2131\alpha^{4}\\
+2274\alpha^{5}-354\alpha^{6})
\end{array} & \begin{array}{cc}
\frac{1}{60}(-48+61\alpha-319\alpha^{2}\\
+567\alpha^{3}-312\alpha^{4}\\
+53\alpha^{5})
\end{array}\\
\\
\begin{array}{cc}
\frac{1}{20}(5-113\alpha\\
+146\alpha^{2}-37\alpha^{3})
\end{array} & \begin{array}{cc}
\frac{1}{10}(3+64\alpha-87\alpha^{2}\\
+11\alpha^{3}+8\alpha^{4})
\end{array} & \begin{array}{cc}
\frac{1}{60}(-48+61\alpha-319\alpha^{2}\\
+567\alpha^{3}-312\alpha^{4}\\
+53\alpha^{5})
\end{array} & 3(-1+\alpha)^{2}\alpha^{2}
\end{bmatrix}.
\]
} To show that $\hat{A}$ is a desired choice, it is sufficient to
show that $\hat{A}$ is positive definite, or equivalently, its all
sequential principal minors are positive. Let the sequential principal
minors of matrix $\hat{A}$ of orders one to four be respectively
$\hat{\mathsf{A}}_{1},\hat{\mathsf{A}}_{2},\hat{\mathsf{A}}_{3}$
and $\hat{\mathsf{A}}_{4}$. Obviously, $\hat{\mathsf{A}}_{1}=6>0$.
As for $\hat{\mathsf{A}}_{2}$, we have 
\begin{align*}
\hat{\mathsf{A}}_{2} & =\begin{vmatrix}6 & 6(-2+\alpha)\\
6(-2+\alpha) & \frac{3}{5}(64-224\alpha+486\alpha^{2}-486\alpha^{3}+177\alpha^{4})
\end{vmatrix}\\
 & =\frac{1}{5}(432-3312\alpha+8568\alpha^{2}-8748\alpha^{3}+3186\alpha^{4}).
\end{align*}
This is a polynomial function of $\alpha$, whose graph is plotted
in Fig. \ref{fig:The-sequential-principal}. It is easy to see that
$\hat{\mathsf{A}}_{2}$ is positive whenever $\alpha\in(0.5,1)$.
This point can be proven rigorously by standard but tedious techniques,
e.g., taking derivatives and determining the signs of its derivatives.
We omit the details.

Similarly, 
\begin{align*}
\hat{\mathsf{A}}_{3} & =-\frac{125991}{250}+\frac{1056702}{125}\alpha-\frac{8019621}{125}\alpha^{2}+\frac{7341099}{25}\alpha^{3}-\frac{227464371}{250}\alpha^{4}\\
 & \qquad+\frac{505782483}{250}\alpha^{5}-\frac{1660856481}{500}\alpha^{6}+\frac{507425778}{125}\alpha^{7}-\frac{365740659}{100}\alpha^{8}\\
 & \qquad+\frac{118059363}{50}\alpha^{9}-\frac{129122208}{125}\alpha^{10}+\frac{68516523}{250}\alpha^{11}-\frac{16635699}{500}\alpha^{12}\\
 & >0,
\end{align*}
whenever $\alpha\in(0.5,1)$, and

\begin{align*}
\hat{\mathsf{A}}_{4} & =-\frac{1152239}{40000}+\frac{15486703}{20000}\alpha-\frac{396809831}{40000}\alpha^{2}+\frac{7993634641}{100000}\alpha^{3}\\
 & \qquad-\frac{45104308341}{100000}\alpha^{4}+\frac{23516351671}{12500}\alpha^{5}-\frac{239125432197}{40000}\alpha^{6}\\
 & \qquad+\frac{147169345899}{10000}\alpha^{7}-\frac{707339063119}{25000}\alpha^{8}+\frac{1063361232613}{25000}\alpha^{9}\\
 & \qquad-\frac{4976176278313}{100000}\alpha^{10}+\frac{2239391676329}{50000}\alpha^{11}-\frac{3038852817539}{100000}\alpha^{12}\\
 & \qquad+\frac{150252323103}{10000}\alpha^{13}-\frac{510688609929}{100000}\alpha^{14}+\frac{10665829017}{10000}\alpha^{15}\\
 & \qquad-\frac{2579598531}{25000}\alpha^{16}\\
 & >0,
\end{align*}
whenever $\alpha\in(0.5,0.84).$ The graphs of these functions are
also plotted in Fig. \ref{fig:The-sequential-principal}. 
\begin{center}
\begin{figure}
\begin{centering}
\includegraphics[width=0.45\textwidth]{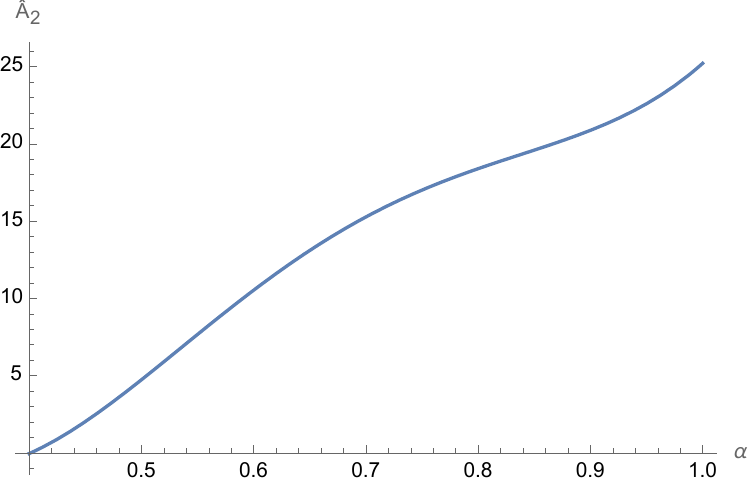}\includegraphics[width=0.45\textwidth]{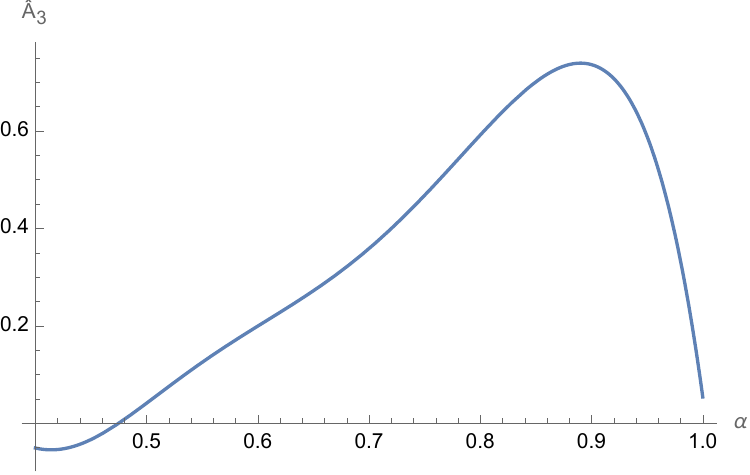} 
\par\end{centering}
\begin{centering}
\includegraphics[width=0.45\textwidth]{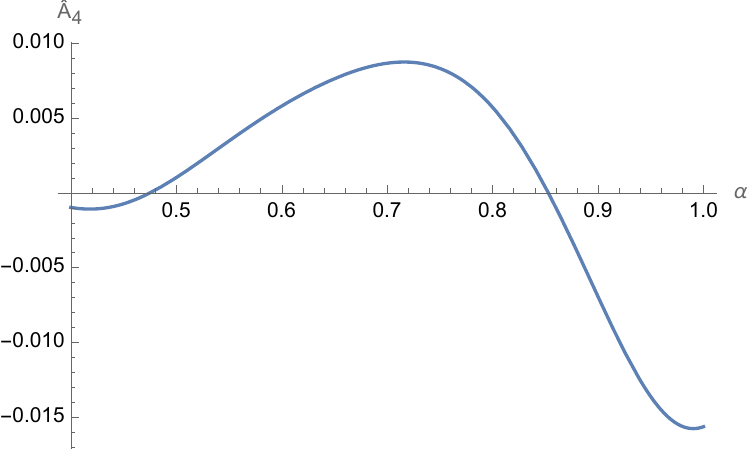} 
\par\end{centering}
\caption{\label{fig:The-sequential-principal}The sequential principal minors
of matrix $\hat{A}$.}
\end{figure}
\par\end{center}

Therefore, for $\alpha\in(0.5,0.84)$, $\hat{A}$ is positive definite,
completing the proof for the case of $\alpha\in(0.5,0.84)$.

\subsection{Case 3: $\alpha\in(0.83,1]$}

For $\alpha\in(0.83,1]$, we choose 
\begin{align*}
\tilde{a}(\alpha) & =4-4\alpha,\\
\tilde{b}(\alpha) & =6-3.5\alpha,\\
\tilde{c}(\alpha) & =-12+7\alpha,\\
\tilde{d}(\alpha) & =0,\\
\tilde{e}(\alpha) & =1.5-1.5\alpha.
\end{align*}
This choice of $(a,b,c,d,e)$ is obtained by the same method used
for Case 2.

Substituting $\tilde{a}(\alpha),\tilde{b}(\alpha),\tilde{c}(\alpha),\tilde{d}(\alpha),\tilde{e}(\alpha)$
into the matrix $A$ yields a new matrix 
\[
\tilde{A}=\begin{bmatrix}6 & 6(-2+\alpha) & 6-\frac{7}{2}\alpha & \frac{3}{4}(1-7\alpha+6\alpha^{2})\\
\\
6(-2+\alpha) & -3(-6+2\alpha+\alpha^{2}) & -12+7\alpha & 4-4\alpha\\
\\
6-\frac{7}{2}\alpha & -12+7\alpha & \begin{array}{cc}
\frac{1}{10}(120-154\alpha+91\alpha^{2}\\
-28\alpha^{3}+3\alpha^{4})
\end{array} & \begin{array}{cc}
\frac{1}{12}(-66+151\alpha-133\alpha^{2}\\
+57\alpha^{3}-9\alpha^{4})
\end{array}\\
\\
\frac{3}{4}(1-7\alpha+6\alpha^{2}) & 4-4\alpha & \begin{array}{cc}
\frac{1}{12}(-66+151\alpha-133\alpha^{2}\\
+57\alpha^{3}-9\alpha^{4})
\end{array} & 3(-1+\alpha)^{2}\alpha^{2}
\end{bmatrix}.
\]
Similarly, it is sufficient to show that for $\alpha\in(0.83,1]$,
$\tilde{A}$ is positive semidefinite. To this end, we consider all
sequential principal minors of matrix $\tilde{A}$. Let $\tilde{\mathsf{A}}_{1},\tilde{\mathsf{A}}_{2},\tilde{\mathsf{A}}_{3}$
and $\tilde{\mathsf{A}}_{4}$ denote the sequential principal minors
of matrix $\tilde{A}$ of orders one to four. Obviously, $\tilde{\mathsf{A}}_{1}=6>0$.
As for $\tilde{\mathsf{A}}_{2}$, we have 
\begin{align*}
\tilde{\mathsf{A}}_{2} & =\begin{vmatrix}6 & 6(-2+\alpha)\\
6(-2+\alpha) & -3(-6+2\alpha+\alpha^{2})
\end{vmatrix}\\
 & =-36+108\alpha-54\alpha^{2},
\end{align*}
which is positive whenever $\alpha\in(0.5,1)$. Similarly, 
\[
\tilde{\mathsf{A}}_{3}=-216+\frac{4752}{5}\alpha-\frac{17013}{10}\alpha^{2}+\frac{15687}{10}\alpha^{3}-\frac{15357}{20}\alpha^{4}+\frac{918}{5}\alpha^{5}-\frac{81}{5}\alpha^{6}>0,
\]
whenever $\alpha\in(0.8,1)$, and 
\begin{align*}
\tilde{\mathsf{A}}_{4} & =\frac{675}{2}-\frac{22197}{8}\alpha+\frac{395399}{40}\alpha^{2}-\frac{200303}{10}\alpha^{3}+\frac{4064497}{160}\alpha^{4}\\
 & \qquad-\frac{833569}{40}\alpha^{5}+\frac{43959}{4}\alpha^{6}-\frac{142191}{40}\alpha^{7}+\frac{100503}{160}\alpha^{8}-\frac{891}{20}\alpha^{9}\\
 & >0,
\end{align*}
whenever $\alpha\in(0.83,1)$. 
\begin{center}
\begin{figure}
\begin{centering}
\includegraphics[width=0.45\textwidth]{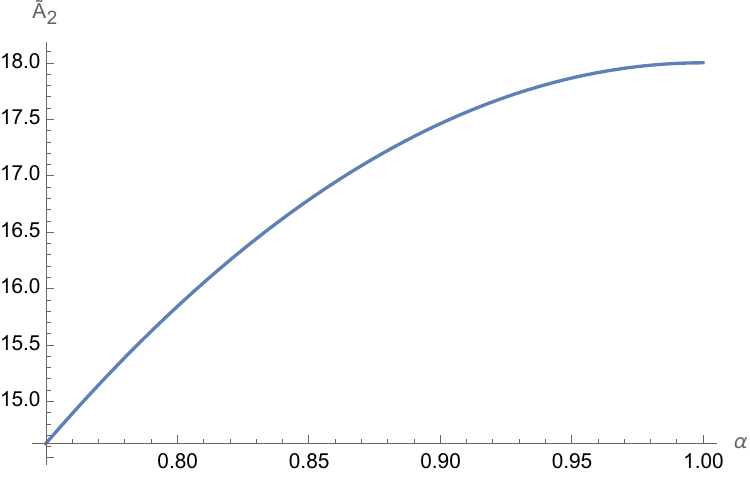}\includegraphics[width=0.45\textwidth]{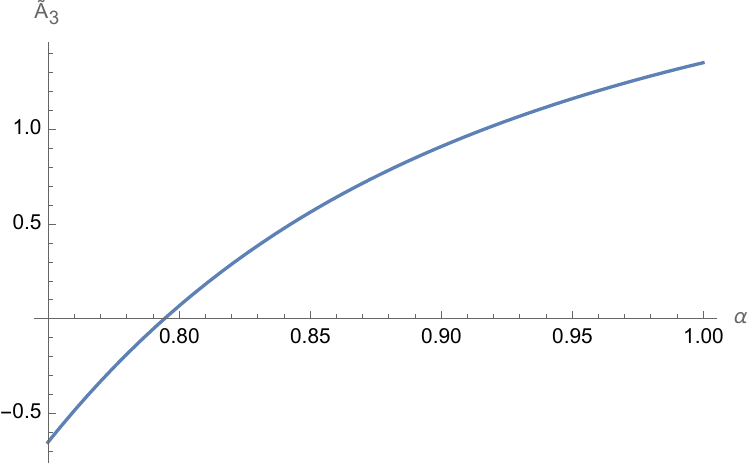} 
\par\end{centering}
\begin{centering}
\includegraphics[width=0.45\textwidth]{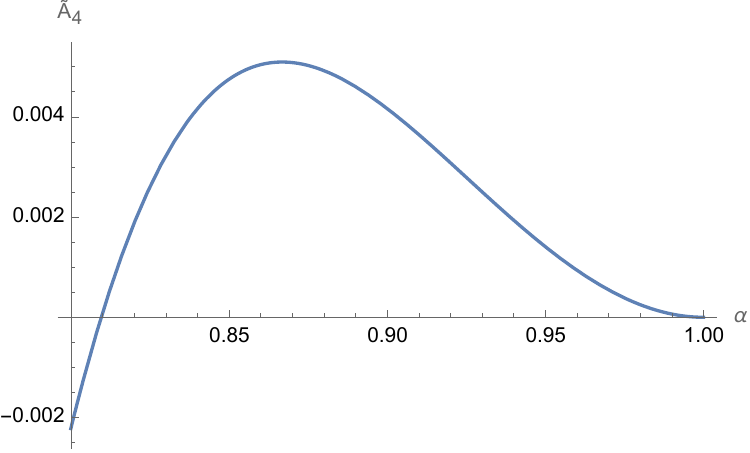} 
\par\end{centering}
\caption{\label{fig:The-sequential-principal-1}The sequential principal minors
of matrix $\tilde{A}$.}
\end{figure}
\par\end{center}

Therefore, for $\alpha\in(0.83,1)$, $\tilde{A}$ is positive definite.
Moreover, for $\alpha=1$, $\tilde{A}$ turns to be 
\[
\begin{pmatrix}6 & -6 & 2.5 & 0\\
-6 & 9 & -5 & 0\\
2.5 & -5 & 3.2 & 0\\
0 & 0 & 0 & 0
\end{pmatrix},
\]
which is obviously positive semidefinite. We complete the proof for
the case of $\alpha\in(0.83,1]$. 
\begin{rem}
In Case 1 of the proof above, we express the derivative $\frac{\partial^{3}h_{\alpha}(p)}{\partial t^{3}}$
as the integral of a specific form of sum of squares (plus some nonnegative
terms), while in Cases 2 and 3, we express the derivative as the integral
of a positive semidefinite quadratic form (plus some nonnegative terms),
which equivalently runs over all possible forms of sum of squares.
The former method is more straightforward, but the latter is more
powerful which can recover the former. 
\end{rem}

\section{\label{sec:Proof of sign of 4th derivative}Proof of Theorem \ref{thm:sign of 4th derivative}}

Similarly to the proof of Theorem \ref{thm:sign of 3rd derivative},
we would like to write \eqref{eq:4th derivative} in the following
more general form: 
\[
P:=-\frac{\alpha}{12}\left(\mathbb{E}_{\alpha}[Q]+c_{1}\mathbb{E}_{\alpha}[\bar{p}_{1}^{6}]\mathbb{E}_{\alpha}[\bar{p}_{1}^{2}]+c_{2}\mathbb{E}_{\alpha}[\bar{p}_{1}^{2}]\mathbb{E}_{\alpha}[\bar{p}_{1}^{2}\bar{p}_{2}^{2}]+c_{3}\mathbb{E}_{\alpha}[\bar{p}_{1}^{4}]\mathbb{E}_{\alpha}[\bar{p}_{2}^{2}]+c_{4}\mathbb{E}_{\alpha}[\bar{p}_{1}^{2}]\mathbb{E}_{\alpha}[\bar{p}_{3}^{2}]\right),
\]
where $c_{1},c_{2},c_{3},c_{4}\ge0$, and $Q=\overrightarrow{y}B\overrightarrow{y}^{\top}$
with a positive semidefinite matrix $B$ and the vector 
\[
\overrightarrow{y}:=\left(\bar{p}_{4},\bar{p}_{1}\bar{p}_{3},\bar{p}_{2}^{2},\bar{p}_{1}^{2}\bar{p}_{2},\bar{p}_{1}^{4},\mathbb{E}_{\alpha}[\bar{p}_{1}^{4}],(\mathbb{E}_{\alpha}[\bar{p}_{1}^{2}])\bar{p}_{2},\mathbb{E}_{\alpha}[\bar{p}_{2}^{2}],(\mathbb{E}_{\alpha}[\bar{p}_{1}^{2}])^{2},(\mathbb{E}_{\alpha}[\bar{p}_{1}^{2}])\bar{p}_{1}^{2}\right)^{\top}.
\]
We choose symmetric $B$ as $\left(b_{i,j}\right)_{1\le i,j\le10}$,
such that
\[
b_{11}=6,\;b_{12}=24(-2+\alpha)-4b_{13},\;b_{17}=\frac{1}{2}(-12\alpha+12\alpha^{2}+c_{4}),\;b_{22}=12(6-5\alpha+\alpha^{2})-4(\alpha-3)b_{13}+2b_{14},
\]
\begin{align*}
b_{33}=9(6-5\alpha+\alpha^{2})-\frac{2}{3}(\alpha-3)b_{13}-\frac{4}{3}b_{14}+\frac{2}{3}b_{23},b_{34} & =12(-24+26\alpha-9\alpha^{2}+\alpha^{3})-\frac{1}{3}(12-7\alpha+\alpha^{2})b_{13}\\
 & \qquad-\frac{13}{6}(\alpha-4)b_{14}-6b_{15}-\frac{1}{3}(4-\alpha)b_{23}+\frac{3}{2}b_{24},
\end{align*}
\[
b_{37}=\frac{b_{27}}{2}-b_{1,10}+\frac{1}{4}(24\alpha-36\alpha^{2}+12\alpha^{3}+2c_{4}-\alpha c_{4}),
\]
\[
b_{44}=9(\alpha-5)(\alpha-4)(\alpha-3)(\alpha-2)-(20-9\alpha+\alpha^{2})b_{14}-14(\alpha-5)b_{15}-(5-\alpha)b_{24}+10b_{25}-2b_{35},
\]
\[
b_{47}=-\frac{1}{2}(3-\alpha)b_{27}-4(\alpha-3)b_{1,10}+3b_{2,10}-b_{3,10}+\frac{1}{4}(-144\alpha+264\alpha^{2}-144\alpha^{3}+24\alpha^{4}-2c_{2}-6c_{4}+5\alpha c_{4}-\alpha^{2}c_{4}),
\]
\[
b_{55}=-\frac{3}{28}(\alpha-7)(\alpha-6)(\alpha-5)(\alpha-4)(\alpha-3)(\alpha-2)+\frac{2}{7}(\alpha-7)(\alpha-6)(\alpha-5)b_{15}-\frac{2}{7}(\alpha-7)(\alpha-6)b_{25}+\frac{2}{7}(\alpha-7)b_{45},
\]
\begin{align*}
b_{5,10} & =-\frac{1}{5}(5-\alpha)b_{57}-\frac{1}{5}(60-47\alpha+12\alpha^{2}-\alpha^{3})b_{1,10}-\frac{1}{5}(20-9\alpha+\alpha^{2})b_{2,10}-\frac{1}{5}(5-\alpha)b_{4,10}\\
 & \qquad+\frac{1}{10}(360\alpha-822\alpha^{2}+675\alpha^{3}-255\alpha^{4}+45\alpha^{5}-3\alpha^{6}-5c_{1}),
\end{align*}
\[
b_{66}=-\frac{1}{4}(\alpha-3)^{2}(\alpha-2)^{2}(\alpha-1)\alpha+\frac{2}{3}(\alpha-3)(\alpha-2)(\alpha-1)b_{16}-\frac{2}{3}(\alpha-3)(\alpha-2)b_{26}+\frac{2}{3}(\alpha-3)b_{46}-2b_{56},
\]
\begin{align*}
b_{68} & =(1-\alpha)b_{16}-\frac{1}{3}(6-11\alpha+6\alpha^{2}-\alpha^{3})b_{18}+b_{26}-\frac{1}{3}(6-5\alpha+\alpha^{2})b_{28}-b_{36}-\frac{1}{3}(3-\alpha)b_{48}-b_{58}\\
 & \qquad+\frac{1}{2}(-18\alpha+33\alpha^{2}-18\alpha^{3}+3\alpha^{4}-c_{3}),
\end{align*}
\[
b_{88}=9(\alpha-\alpha^{2})-2(\alpha-1)b_{18}+2b_{28}-2b_{38},
\]
\[
b_{8,10}=9(\alpha^{2}-2\alpha^{3}+\alpha^{4})-(\alpha-1)b_{19}+b_{29}-b_{39}-\frac{b_{77}}{2}-(1-\alpha)b_{78}-b_{89},
\]
\[
b_{9,10}=-\frac{9}{4}(-\alpha^{3}+3\alpha^{4}-3\alpha^{5}+\alpha^{6})-(1-\alpha)b_{79}-\frac{b_{99}}{2},
\]
\begin{align*}
b_{10,10} & =-3(\alpha-3)(\alpha-2)(\alpha-1)^{2}\alpha^{2}+\frac{2}{3}(\alpha-3)(\alpha-2)(\alpha-1)b_{19}-\frac{2}{3}(\alpha-3)(\alpha-2)b_{29}+\frac{2}{3}(\alpha-3)b_{49}\\
 & \qquad-2b_{59}+2(\alpha-1)b_{67}-2b_{69}-2b_{6,10}+\frac{2}{3}(\alpha-3)b_{7,10}.
\end{align*}
Using Lemma \ref{lem:3rd derivative-1} and Lemma \ref{lem:4th derivative-1},
it is straightforward to check that $P$ with this choice is equal
to \eqref{eq:4th derivative}. Hence, to establish the non-positivity
of $P$, it is suffice to ensure the positive semidefiniteness of
$B$ by choosing suitable values of 
\begin{align*}
 & (b_{13}(\alpha),\dots,b_{16}(\alpha),b_{18}(\alpha),\dots,b_{1,10}(\alpha),b_{23}(\alpha),\dots,b_{2,10}(\alpha),\\
 & b_{35}(\alpha),b_{36}(\alpha),b_{38}(\alpha),\dots,b_{3,10}(\alpha),b_{45}(\alpha),b_{46}(\alpha),b_{48}(\alpha),\dots,b_{4,10}(\alpha),\\
 & b_{56}(\alpha),\dots,b_{59}(\alpha),b_{67}(\alpha),b_{69}(\alpha),b_{6,10}(\alpha),b_{77}(\alpha),\dots,b_{7,10}(\alpha),\\
 & b_{89}(\alpha),b_{99}(\alpha),c_{1}(\alpha),\dots,c_{4}(\alpha))
\end{align*}
for each $\alpha$.

Like the proof in the case of third-order derivative, we use the curve-fitting
method to find the desired choice of $(b_{13}(\alpha),\dotsm c_{4}(\alpha))$.
To this end, we use optimization function ``SemidefiniteOptimization''
in Mathematica to obtain numerical points $(b_{13},\dots,c_{4})$
that meet the requirements that $B$ is positive semidefinite and
$c_{1}\ge0,c_{2}\ge0,c_{3}\ge0,c_{4}\ge0$. Indeed, recall \eqref{eq:SDP-1},
the problem of semidefinite optimization we concerned is 
\begin{align}
\min_{B\in\mathbb{S}^{10}} & \,0\nonumber \\
\text{subject to } & c_{1}\ge0,c_{2}\ge0,c_{3}\ge0,c_{4}\ge0\label{eq:SDP-3}\\
 & B\succeq0.\nonumber 
\end{align}
These numerical points $(\alpha,b_{13},\dots,c_{4})$ are obtained
when running the SDP \eqref{eq:SDP-3} in Mathematica for $\alpha$
taking values of $0.92,1.1,1.28,1.46,1.64$ and $1.81$, respectively.
Then we fit these points using polynomial functions of degree $6$.

For example, upon rounding off the numerical values to an appropriate
decimal precision, the points $(\alpha,b_{13})$ obtained by the optimization
process are 
\begin{align*}
L_{2} & :=\{(0.92,-4.430679),(1.1,-3.530372),(1.28,-2.677838),(1.46,-1.853103),\\
 & \qquad\qquad(1.64,-1.036544),(1.81,-0.273122)\}.
\end{align*}
Then we use a sixth-order polynomial function to fit $L_{2}$ and
obtain the curve-fitting function 
\[
\hat{b}_{13}(\alpha)=\left(-91252+35904\alpha+35890\alpha^{2}-19791\alpha^{3}-9164\alpha^{4}+10537\alpha^{5}-2361\alpha^{6}\right)/10000.
\]
Similarly, for other points $(\alpha,b_{14}),\dots,(\alpha,c_{4})$,
indeed, all curves could be fitted simultaneously by Mathematica,
and the resultant functions are
\begin{align*}
\hat{b}_{14}(\alpha) & =(189476+23039\alpha-175141\alpha^{2}+41344\alpha^{3}+44217\alpha^{4}-27814\alpha^{5}+5007\alpha^{6})/10000,\\
\hat{b}_{15}(\alpha) & =(-180930+305354\alpha-217895\alpha^{2}+36990\alpha^{3}+58957\alpha^{4}-39666\alpha^{5}+7379\alpha^{6})/10000,\\
\hat{b}_{16}(\alpha) & =(250345-840213\alpha+1015652\alpha^{2}-308670\alpha^{3}-284395\alpha^{4}+229172\alpha^{5}-46186\alpha^{6})/10000,\\
\hat{b}_{18}(\alpha) & =(-3101+132016\alpha-326706\alpha^{2}+134523\alpha^{3}+93089\alpha^{4}-87288\alpha^{5}+18566\alpha^{6})/10000,\\
\hat{b}_{19}(\alpha) & =(-905294+3493036\alpha-4491592\alpha^{2}+1464337\alpha^{3}+1256145\alpha^{4}-1066012\alpha^{5}+224800\alpha^{6})/10000,\\
\hat{b}_{110}(\alpha) & =(-167116+229222\alpha+162008\alpha^{2}-239860\alpha^{3}-53641\alpha^{4}+132827\alpha^{5}-38420\alpha^{6})/10000,\\
\hat{b}_{23}(\alpha) & =(105908+97501\alpha-265476\alpha^{2}+87805\alpha^{3}+72522\alpha^{4}-58267\alpha^{5}+11928\alpha^{6})/10000,\\
\hat{b}_{24}(\alpha) & =(-346450+119563\alpha+121389\alpha^{2}-5240\alpha^{3}-27137\alpha^{4}-2197\alpha^{5}+2547\alpha^{6})/10000,\\
\hat{b}_{25}(\alpha) & =(473201-1035341\alpha+872198\alpha^{2}-167048\alpha^{3}-235129\alpha^{4}+159936\alpha^{5}-29918\alpha^{6})/10000,\\
\hat{b}_{26}(\alpha) & =(-218012+774319\alpha-1053192\alpha^{2}+383600\alpha^{3}+300441\alpha^{4}-268841\alpha^{5}+56616\alpha^{6})/10000,\\
\hat{b}_{27}(\alpha) & =(587831-1997272\alpha+2510957\alpha^{2}-863328\alpha^{3}-699623\alpha^{4}+631534\alpha^{5}-137285\alpha^{6})/10000,\\
\hat{b}_{28}(\alpha) & =(-157085+154690\alpha+317156\alpha^{2}-243313\alpha^{3}-104416\alpha^{4}+128911\alpha^{5}-29406\alpha^{6})/10000,\\
\hat{b}_{29}(\alpha) & =(1225754-5024686\alpha+6827844\alpha^{2}-2392199\alpha^{3}-1919021\alpha^{4}+1705479\alpha^{5}-364792\alpha^{6})/10000,\\
\hat{b}_{210}(\alpha) & =(531222-1137657\alpha+411707\alpha^{2}+283843\alpha^{3}-85384\alpha^{4}-101700\alpha^{5}+39106\alpha^{6})/10000,\\
\hat{b}_{35}(\alpha) & =(164610-232680\alpha+71531\alpha^{2}+35584\alpha^{3}-14442\alpha^{4}-5843\alpha^{5}+2282\alpha^{6})/10000,\\
\hat{b}_{36}(\alpha) & =(-72349+183502\alpha-209314\alpha^{2}+69998\alpha^{3}+58837\alpha^{4}-51981\alpha^{5}+11058\alpha^{6})/10000,\\
\hat{b}_{38}(\alpha) & =(-65485+191390\alpha-102360\alpha^{2}-16302\alpha^{3}+21592\alpha^{4}-4711\alpha^{5}+334\alpha^{6})/10000,\\
\hat{b}_{39}(\alpha) & =(-237774+646429\alpha-500069\alpha^{2}+42196\alpha^{3}+118614\alpha^{4}-73719\alpha^{5}+17191\alpha^{6})/10000,\\
\hat{b}_{310}(\alpha) & =(130292-402176\alpha+344316\alpha^{2}-35571\alpha^{3}-88379\alpha^{4}+46327\alpha^{5}-7687\alpha^{6})/10000,\\
\hat{b}_{45}(\alpha) & =(-555060+1103729\alpha-782340\alpha^{2}+63500\alpha^{3}+203177\alpha^{4}-104028\alpha^{5}+16589\alpha^{6})/10000,\\
\hat{b}_{46}(\alpha) & =(256607-750153\alpha+903638\alpha^{2}-303021\alpha^{3}-253643\alpha^{4}+222514\alpha^{5}-47270\alpha^{6})/10000,\\
\hat{b}_{48}(\alpha) & =(-47232+359306\alpha-852339\alpha^{2}+410441\alpha^{3}+259946\alpha^{4}-252015\alpha^{5}+52435\alpha^{6})/10000,\\
\hat{b}_{49}(\alpha) & =(-776042+3211664\alpha-4564779\alpha^{2}+1705665\alpha^{3}+1309323\alpha^{4}-1178244\alpha^{5}+246089\alpha^{6})/10000,\\
\hat{b}_{410}(\alpha) & =(146608-674489\alpha+1225886\alpha^{2}-591053\alpha^{3}-358918\alpha^{4}+386564\alpha^{5}-87940\alpha^{6})/10000,\\
\hat{b}_{56}(\alpha) & =(-212195+621958\alpha-650747\alpha^{2}+160209\alpha^{3}+177993\alpha^{4}-130945\alpha^{5}+25633\alpha^{6})/10000,\\
\hat{b}_{57}(\alpha) & =(1636258-4762460\alpha+4731637\alpha^{2}-1069804\alpha^{3}-1286924\alpha^{4}+910874\alpha^{5}-172598\alpha^{6})/10000,\\
\hat{b}_{58}(\alpha) & =(373592-1191822\alpha+1381783\alpha^{2}-404760\alpha^{3}-386732\alpha^{4}+304993\alpha^{5}-60278\alpha^{6})/10000,\\
\hat{b}_{59}(\alpha) & =(849263-2834446\alpha+3316064\alpha^{2}-996486\alpha^{3}-926530\alpha^{4}+748482\alpha^{5}-150180\alpha^{6})/10000,\\
\hat{b}_{67}(\alpha) & =(247305-846256\alpha+853751\alpha^{2}-149831\alpha^{3}-228476\alpha^{4}+137913\alpha^{5}-23938\alpha^{6})/10000,\\
\hat{b}_{69}(\alpha) & =(-490805+1729089\alpha-2070698\alpha^{2}+656732\alpha^{3}+569796\alpha^{4}-495499\alpha^{5}+107130\alpha^{6})/10000,\\
\hat{b}_{610}(\alpha) & =(-34696-24941\alpha+265660\alpha^{2}-197466\alpha^{3}-82675\alpha^{4}+117867\alpha^{5}-29069\alpha^{6})/10000,\\
\hat{b}_{77}(\alpha) & =(1963506-8820363\alpha+12481179\alpha^{2}-4438798\alpha^{3}-3447795\alpha^{4}+3192710\alpha^{5}-717188\alpha^{6})/10000,\\
\hat{b}_{78}(\alpha) & =(-798409+2258254\alpha-1983151\alpha^{2}+283227\alpha^{3}+533372\alpha^{4}-294422\alpha^{5}+41938\alpha^{6})/10000,\\
\hat{b}_{79}(\alpha) & =(3003706-11639204\alpha+14866060\alpha^{2}-4806471\alpha^{3}-4067396\alpha^{4}+3590054\alpha^{5}-793636\alpha^{6})/10000,\\
\hat{b}_{710}(\alpha) & =(2167103-5896234\alpha+5186510\alpha^{2}-1007159\alpha^{3}-1383214\alpha^{4}+957062\alpha^{5}-178679\alpha^{6})/10000,\\
\hat{b}_{89}(\alpha) & =(1520755-5407036\alpha+6563579\alpha^{2}-2113353\alpha^{3}-1818437\alpha^{4}+1576022\alpha^{5}-336478\alpha^{6})/10000,\\
\hat{b}_{99}(\alpha) & =(7741109-28812524\alpha+35525537\alpha^{2}-11166017\alpha^{3}-9616108\alpha^{4}+8519970\alpha^{5}-1922915\alpha^{6})/10000,\\
\hat{c}_{1}(\alpha) & =(8489-61368\alpha+103126\alpha^{2}-43353\alpha^{3}-28584\alpha^{4}+29984\alpha^{5}-7198\alpha^{6})/10000,\\
\hat{c}_{2}(\alpha) & =(33383-160476\alpha+232596\alpha^{2}-88736\alpha^{3}-63744\alpha^{4}+63678\alpha^{5}-15070\alpha^{6})/10000,\\
\hat{c}_{3}(\alpha) & =(-217390+740690\alpha-859895\alpha^{2}+251505\alpha^{3}+238219\alpha^{4}-191600\alpha^{5}+39123\alpha^{6})/10000,\\
\hat{c}_{4}(\alpha) & =(-194197+567924\alpha-575575\alpha^{2}+141205\alpha^{3}+159513\alpha^{4}-114845\alpha^{5}+20978\alpha^{6})/10000.
\end{align*}
Replacing $b_{13},\dots,c_{4}$ with $\hat{b}_{13},\dots,\hat{c}_{4}$
respectively, the matrix $B$ then turns into a matrix $\hat{B}$
with only one parameter $\alpha$. To show that $\hat{B}$ is a desired
choice, it is sufficient to show that $\hat{B}$ is positive definite,
or equivalently, its all sequential principal minors are positive.
Let the sequential principal minors of matrix $\hat{B}$ of orders
one to ten be $\hat{\mathsf{B}}_{1},\dots,\hat{\mathsf{B}}_{10}$
respectively. Obviously, $\hat{\mathsf{B}}_{1}=6>0$. As for $\hat{\mathsf{B}}_{2}$,
we have
\begin{align*}
\hat{\mathsf{B}}_{2} & =\begin{vmatrix}6 & \begin{array}{cc}
-4(-\frac{22813}{2500}+\frac{2244\alpha}{625}+\frac{3589\alpha^{2}}{1000}-\frac{19791\alpha^{3}}{10000}-\frac{2291\alpha^{4}}{2500}\\
+\frac{10537\alpha^{5}}{10000}-\frac{2361\alpha^{6}}{10000})+24(-2+\alpha)
\end{array}\\
\\
\begin{array}{cc}
-4(-\frac{22813}{2500}+\frac{2244\alpha}{625}+\frac{3589\alpha^{2}}{1000}-\frac{19791\alpha^{3}}{10000}-\frac{2291\alpha^{4}}{2500}\\
+\frac{10537\alpha^{5}}{10000}-\frac{2361\alpha^{6}}{10000})+24(-2+\alpha)
\end{array} & \begin{array}{cc}
12(6-5\alpha+\alpha^{2})-4(-3+\alpha)(-\frac{22813}{2500}+\frac{2244\alpha}{625}\\
+\frac{3589\alpha^{2}}{1000}-\frac{19791\alpha^{3}}{10000}-\frac{2291\alpha^{4}}{2500}+\frac{10537\alpha^{5}}{10000}-\frac{2361\alpha^{6}}{10000})\\
+2(\frac{47369}{2500}+\frac{23039\alpha}{10000}-\frac{175141\alpha^{2}}{10000}+\frac{2584\alpha^{3}}{625}+\frac{44217\alpha^{4}}{10000}\\
-\frac{13907\alpha^{5}}{5000}+\frac{5007\alpha^{6}}{10000})
\end{array}
\end{vmatrix}\\
 & =\frac{1}{6250000}(-811717504+2292676116\alpha-2431216156\alpha^{2}+1748649216\alpha^{3}-1498854128\alpha^{4}\\
 & +776151140\alpha^{5}+682907899\alpha^{6}-1197441620\alpha^{7}+502569218\alpha^{8}+99669034\alpha^{9}\\
 & -154300777\alpha^{10}+49755714\alpha^{11}-5574321\alpha^{12}).
\end{align*}
This is a polynomial function of $\alpha$. It is straightforward
to verify that $\hat{\mathsf{B}}_{2}$ is positive whenever $\alpha\in(0.75,2.38)$.
Indeed, using Mathematica, it is easy to solve the polynomial equation
$6250000\hat{\mathsf{B}}_{2}=0$ with integral coefficients, and we
find that there are only two real roots  $\alpha_{21}\in(0.74,0.75)$
and  $\alpha_{22}\in(2.38,2.39)$, thus, $\hat{\mathsf{B}}_{2}>0$
whenever $\alpha\in(0.75,2.38)$. In fact, we use the function ``CountRoots''
(which is symbolic computation) in Mathematica to count the number
of real roots for a polynomial with integral coefficients, and then
locate each real root in arbitrary precision by observing the sign
change of the polynomial around the root. 

   Similarly, for $\hat{\mathsf{B}}_{3},\dots,\hat{\mathsf{B}}_{10}$,
we have
\[
\hat{\mathsf{B}}_{3}>0,\text{whenever }\alpha\in(0.92,1.82),\quad\hat{\mathsf{B}}_{4}>0,\text{whenever }\alpha\in(0.92,1.81),
\]
\[
\hat{\mathsf{B}}_{5}>0,\text{whenever }\alpha\in(0.93,1.76),\quad\hat{\mathsf{B}}_{6}>0,\text{whenever }\alpha\in(0.93,1.76),
\]
\[
\hat{\mathsf{B}}_{7}>0,\text{whenever }\alpha\in(0.93,1.76),\quad\hat{\mathsf{B}}_{8}>0,\text{whenever }\alpha\in(0.93,1.76),
\]
\[
\hat{\mathsf{B}}_{9}>0,\text{whenever }\alpha\in(0.93,1.76),\quad\hat{\mathsf{B}}_{10}>0,\text{whenever }\alpha\in(0.93,1.76).
\]
Therefore, for $\alpha\in(0.93,1.76)$, $\hat{B}$ is positive definite.

\section{\label{sec:Proof of 4th Tsallis entropy}Proof of Theorem \ref{thm:4th Tsallis entropy}}

\subsection{Case of $0.93<\alpha<1.98$}

By \eqref{eq:-13} and \eqref{eq:4th derivative},
\begin{align}
\frac{1}{\int p^{\alpha}dx}\frac{\partial^{4}\hat{h}_{\alpha}(p)}{\partial t^{4}} & =\frac{\alpha}{12}\left(\frac{3}{28}(\alpha-2)(\alpha-3)(\alpha-4)(\alpha-5)(\alpha-6)(\alpha-7)\mathbb{E}_{\alpha}[\bar{p}_{1}^{8}]\right.\nonumber \\
 & \qquad-9(\alpha-2)(\alpha-3)(\alpha-4)(\alpha-5)\mathbb{E}_{\alpha}[\bar{p}_{1}^{4}\bar{p}_{2}^{2}]-24(\alpha-2)(\alpha-3)(\alpha-4)\mathbb{E}_{\alpha}[\bar{p}_{1}^{2}\bar{p}_{2}^{3}]\\
 & \qquad-9(\alpha-2)(\alpha-3)\mathbb{E}_{\alpha}[\bar{p}_{2}^{4}]+12(\alpha-2)(\alpha-3)\mathbb{E}_{\alpha}[\bar{p}_{1}^{2}\bar{p}_{3}^{2}]\nonumber \\
 & \qquad+24(\alpha-2)\mathbb{E}_{\alpha}[\bar{p}_{2}\bar{p}_{3}^{2}]-6\mathbb{E}_{\alpha}[\bar{p}_{4}^{2}]\bigg).\label{eq:4th derivative of Tsallis entropy}
\end{align}
Following the steps in the proof of Theorem \ref{thm:sign of 4th derivative},
we would like to write \eqref{eq:4th derivative of Tsallis entropy}
in the following more general form: 
\[
P:=-\frac{\alpha}{12}\mathbb{E}_{\alpha}[Q],
\]
where $Q=\overrightarrow{y}B\overrightarrow{y}^{\top}$ with a positive
semidefinite matrix $B$ and the vector $\overrightarrow{y}:=\left(\bar{p}_{4},\bar{p}_{1}\bar{p}_{3},\bar{p}_{2}^{2},\bar{p}_{1}^{2}\bar{p}_{2},\bar{p}_{1}^{4}\right)^{\top}$.
We choose symmetric $B$ as $\left(b_{i,j}\right)_{1\le i,j\le5}$,
such that
\[
b_{11}=6,\;b_{12}=24(-2+\alpha)-4b_{13},\;b_{22}=12(6-5\alpha+\alpha^{2})-4(\alpha-3)b_{13}+2b_{14},
\]
\begin{align*}
b_{33}=9(6-5\alpha+\alpha^{2})-\frac{2}{3}(\alpha-3)b_{13}-\frac{4}{3}b_{14}+\frac{2}{3}b_{23},\;b_{34} & =12(-24+26\alpha-9\alpha^{2}+\alpha^{3})-\frac{1}{3}(12-7\alpha+\alpha^{2})b_{13}\\
 & \qquad-\frac{13}{6}(\alpha-4)b_{14}-6b_{15}-\frac{1}{3}(4-\alpha)b_{23}+\frac{3}{2}b_{24},
\end{align*}
\[
b_{44}=9(\alpha-5)(\alpha-4)(\alpha-3)(\alpha-2)-(20-9\alpha+\alpha^{2})b_{14}-14(\alpha-5)b_{15}-(5-\alpha)b_{24}+10b_{25}-2b_{35},
\]
\[
b_{55}=-\frac{3}{28}(\alpha-7)(\alpha-6)(\alpha-5)(\alpha-4)(\alpha-3)(\alpha-2)+\frac{2}{7}(\alpha-7)(\alpha-6)(\alpha-5)b_{15}-\frac{2}{7}(\alpha-7)(\alpha-6)b_{25}+\frac{2}{7}(\alpha-7)b_{45}.
\]
Using Lemma \ref{lem:3rd derivative-1} and Lemma \ref{lem:4th derivative-1},
it is straightforward to check that $P$ is equal to \eqref{eq:4th derivative of Tsallis entropy},
when $B$ is taken as the above expression. Hence, to establish the
non-positivity of $P$, it is suffice to ensure the positive semidefiniteness
of $B$ by choosing suitable values of $(b_{13}(\alpha),b_{14}(\alpha),b_{15}(\alpha),b_{23}(\alpha),b_{24}(\alpha),b_{25}(\alpha),b_{35}(\alpha),b_{45}(\alpha))$,
depending on the value of $\alpha$.

Observe that \eqref{eq:4th derivative of Tsallis entropy} is \eqref{eq:4th derivative}
minus all the product terms of integrals. And matrix $B$ here is
actually a submatrix consisting of the first five rows and the first
five columns of $B$ in Section \ref{sec:Proof of sign of 4th derivative}.
Thus, the construction we considered here is just a reduction of Section
\ref{sec:Proof of sign of 4th derivative}. It follows that the region
of $\alpha$ for $\frac{\partial^{4}\hat{h}_{\alpha}(p)}{\partial t^{4}}\le0$
is wider than that for $\frac{\partial^{4}h_{\alpha}(p)}{\partial t^{4}}\le0$.
As a result, $\frac{\partial^{4}\hat{h}_{\alpha}(p)}{\partial t^{4}}\le0$
whenever $\alpha\in(0.93,1.76).$ Then, for $\alpha\in(1.65,1.98)$,
like the proof in the case of Rényi entropy, we use the curve-fitting
method to find the desired choice of $(b_{13}(\alpha),\dotsm,b_{45}(\alpha))$
to ensure the positive semidefiniteness of $B$. To this end, we use
optimization function ``SemidefiniteOptimization'' in Mathematica
to obtain numerical points $(b_{13},\dots,b_{45})$ that meet the
requirements that $B$ is positive semidefinite. Indeed, recall \eqref{eq:SDP-1},
the problem of semidefinite optimization we concerned is 
\begin{align}
\min_{B\in\mathbb{S}^{5}} & \,0\nonumber \\
\text{subject to } & B\succeq0.\label{eq:SDP-4}
\end{align}
These numerical points $(\alpha,b_{13},\dots,b_{45})$ are obtained
when running the SDP \eqref{eq:SDP-4} in Mathematica for $\alpha$
taking value of $1.75,1.85,1.95$ and $1.99$, respectively. Then
we fit these points together with the point $(\alpha,b_{13},\dots,b_{45})=(2,0,0,0,0,0,0,0)$,
which is corresponding to taking $\alpha=2$ into \eqref{eq:4th derivative of Tsallis entropy},
using polynomial functions of degree $2$.

For example, upon rounding off the numerical values to an appropriate
decimal precision, the points $(\alpha,b_{13})$ obtained by the optimization
process and the point $(\alpha,b_{13})=(2,0)$,which corresponds to
the case of $\alpha=2$, are 
\begin{align*}
L_{3} & :=\{(1.75,-0.961764),(1.85,-0.588612),(1.95,-0.206353),(1.99,-0.038556),\\
 & \qquad\qquad(2,0)\}.
\end{align*}
Then we use a second-order polynomial function to fit $L_{3}$ and
obtain the curve-fitting function 
\[
\hat{b}_{13}(\alpha)=\left(-43159+2316\alpha+9631\alpha^{2}\right)/10000.
\]
Similarly, for other points of $(\alpha,b_{14}),\dots,(\alpha,b_{45})$,
indeed, all curves could be fitted simultaneously by Mathematica,
the other curve-fitting functions found are
\begin{align*}
\hat{b}_{14}(\alpha)=(389887-350246\alpha+77641\alpha^{2})/10000, & \;\hat{b}_{15}(\alpha)=(-105282+101670\alpha-24505\alpha^{2})/10000,\\
\hat{b}_{23}(\alpha)=(281922-283833\alpha+71413\alpha^{2})/10000, & \;\hat{b}_{24}(\alpha)=(-643827+644561\alpha-161257\alpha^{2})/10000,\\
\hat{b}_{25}(\alpha)=(129980-134523\alpha+34731\alpha^{2})/10000, & \;\hat{b}_{35}(\alpha)=(39405-30388\alpha+5338\alpha^{2})/10000,\\
\hat{b}_{45}(\alpha)=(-63338+ & 10256\alpha+10729\alpha^{2})/10000.
\end{align*}
Replacing $b_{13},\dots,b_{45}$ with $\hat{b}_{13},\dots,\hat{b}_{45}$
respectively, the matrix $B$ then turns into a matrix $\hat{B}$
with only one parameter $\alpha$. To show that $\hat{B}$ is a desired
choice, it is sufficient to show that $\hat{B}$ is positive definite,
or equivalently, its all sequential principal minors are positive.
Let the sequential principal minors of matrix $\hat{B}$ of orders
one to ten be respectively $\hat{\mathsf{B}}_{1},\dots,\hat{\mathsf{B}}_{10}$.
Obviously, $\hat{\mathsf{B}}_{1}=6>0$. As for $\hat{\mathsf{B}}_{2}$,
we have
\begin{align*}
\hat{\mathsf{B}}_{2} & =\begin{vmatrix}6 & \begin{array}{cc}
-4(-\frac{43159}{10000}+\frac{579\alpha}{2500}+\frac{9631\alpha^{2}}{10000})\\
+24(\alpha-2)
\end{array}\\
\\
\begin{array}{cc}
-4(-\frac{43159}{10000}+\frac{579\alpha}{2500}+\frac{9631\alpha^{2}}{10000})\\
+24(\alpha-2)
\end{array} & \begin{array}{cc}
-4(\alpha-3)(-\frac{43159}{10000}+\frac{579\alpha}{2500}+\frac{9631\alpha^{2}}{10000})\\
+2(\frac{389887}{10000}-\frac{175123\alpha}{5000}+\frac{77641\alpha^{2}}{10000})\\
+12(6-5\alpha+\alpha^{2})
\end{array}
\end{vmatrix}\\
 & =\frac{1}{12500000}(-4445083562+9479504976\alpha-6753185396\alpha^{2}+1933288416\alpha^{3}-185512322\alpha^{4}).
\end{align*}
This is a polynomial function of $\alpha$. Using Mathematica, it
is easy to verify that $\hat{\mathsf{B}}_{2}$ is positive whenever
$\alpha\in(0.98,1.99)$.

Similarly, 
\begin{align*}
\hat{\mathsf{B}}_{3} & =-\frac{5172957560007319}{1500000000000}+\frac{4677340915427557}{375000000000}\alpha-\frac{9398568577694853}{500000000000}\alpha^{2}+\frac{11460279945639217}{750000000000}\alpha^{3}\\
 & \qquad-\frac{718307640500547}{100000000000}\alpha^{4}+\frac{90122064461137}{46875000000}\alpha^{5}-\frac{78856928542633}{300000000000}\alpha^{6}+\frac{9826680452501}{750000000000}\alpha^{7}\\
 & >0,
\end{align*}
whenever $\alpha\in(1.03,1.99)$, and
\[
\hat{\mathsf{B}}_{4}>0,\text{whenever }\alpha\in(1.31,1.99),\quad\hat{\mathsf{B}}_{5}>0,\text{whenever }\alpha\in(1.65,1.98).
\]
Therefore, for $\alpha\in(1.65,1.98)$, $\hat{B}$ is positive definite,
completing the proof for the case of $\alpha\in(0.93,1.98)$.

\subsection{Case of $1.97<\alpha\le2$}

For $\alpha\in(1.97,2]$, we choose 
\begin{align*}
\tilde{b}_{13}(\alpha)=\frac{42163}{10000}(\alpha-2), & \;\tilde{b}_{14}(\alpha)=\frac{36667}{10000}(2-\alpha),\\
\tilde{b}_{15}(\alpha)=\frac{3155}{10000}(\alpha-2), & \;\tilde{b}_{23}(\alpha)=\frac{7699}{10000}(\alpha-2),\\
\tilde{b}_{24}(\alpha)=\frac{9191}{10000}(2-\alpha), & \;\tilde{b}_{25}(\alpha)=\frac{6194}{10000}(\alpha-2),\\
\tilde{b}_{35}(\alpha)=\frac{10374}{10000}(2-\alpha), & \;\tilde{b}_{45}(\alpha)=\frac{56793}{10000}(\alpha-2).
\end{align*}
This choice of $(b_{13},\dots,b_{45})$ is obtained by the same method
used for Case 2.

Substituting $\tilde{a}(\alpha),\tilde{b}(\alpha),\tilde{c}(\alpha),\tilde{d}(\alpha),\tilde{e}(\alpha)$
into the matrix $B$ yields a new matrix $\tilde{B}$. Similarly,
it is sufficient to show that for $\alpha\in(1.97,2]$, $\tilde{B}$
is positive semidefinite. To this end, we consider all sequential
principal minors of matrix $\tilde{B}$. Let $\tilde{\mathsf{B}}_{1},\tilde{\mathsf{B}}_{2},\tilde{\mathsf{B}}_{3},\tilde{\mathsf{B}}_{4}$
and $\tilde{\mathsf{B}}_{5}$ denote the sequential principal minors
of matrix $\tilde{B}$ of orders one to five. Obviously, $\tilde{\mathsf{B}}_{1}=6>0$.
As for $\tilde{\mathsf{B}}_{2}$, we have 
\begin{align*}
\tilde{\mathsf{B}}_{2} & =\begin{vmatrix}6 & \begin{array}{cc}
-4(-\frac{42163}{5000}+\frac{42163\alpha}{10000})+24(\alpha-2)\end{array}\\
\\
\begin{array}{cc}
-4(-\frac{42163}{5000}+\frac{42163\alpha}{10000})+24(\alpha-2)\end{array} & \begin{array}{cc}
-4(\alpha-3)(-\frac{42163}{5000}+\frac{42163\alpha}{10000})+2(\frac{36667}{5000}-\frac{36667\alpha}{10000})\\
+12(6-5\alpha+\alpha^{2})
\end{array}
\end{vmatrix}\\
 & =\frac{1}{12500000}(-3634598552+3819713552\alpha-1001207138\alpha^{2}).
\end{align*}
which is positive whenever $\alpha\in(1.82,2)$. Similarly, 
\begin{align*}
\tilde{\mathsf{B}}_{3} & =-\frac{438001032453021}{62500000000}+\frac{5175798706543321}{375000000000}\alpha-\frac{509153633512039}{50000000000}\alpha^{2}\\
 & \qquad+\frac{1667741758389153}{500000000000}\alpha^{3}-\frac{306929095417783}{750000000000}\alpha^{4}\\
 & >0,
\end{align*}
whenever $\alpha\in(1.92,2)$, 
\begin{align*}
\tilde{\mathsf{B}}_{4} & =\frac{106907812000023320531}{31250000000000}-\frac{11941261263467145196009}{937500000000000}\alpha+\frac{57745805445309793475201}{2812500000000000}\alpha^{2}\\
 & \qquad-\frac{52599610348687644579377}{2812500000000000}\alpha^{3}+\frac{236684569508655171493181}{22500000000000000}\alpha^{4}\\
 & \qquad-\frac{168160914586096868096609}{45000000000000000}\alpha^{5}+\frac{36790596886486158864649}{45000000000000000}\alpha^{6}\\
 & \qquad-\frac{2261963247429711691417}{22500000000000000}\alpha^{7}+\frac{29853438100600780261}{5625000000000000}\alpha^{8}\\
 & >0,
\end{align*}
whenever $\alpha\in(1.95,2)$, and
\begin{align*}
\tilde{\mathsf{B}}_{5} & =-\frac{(-2+\alpha)^{4}}{12600000000000000000000}(1251740234822477224840391839554\\
 & \qquad-4299383882809915655476099845165\alpha+6439628610978696497648403899508\alpha^{2}\\
 & \qquad-5534948519611909347505324758619\alpha^{3}+3022633030214774759148122411046\alpha^{4}\\
 & \qquad-1096367306122439673667192658340\alpha^{5}+267818100227304439977449692688\alpha^{6}\\
 & \qquad-43579616547056980886865291824\alpha^{7}+4528942194596341068858708880\alpha^{8}\\
 & \qquad-271849472584522259475180000\alpha^{9}+7164825144144187262640000\alpha^{10})\\
 & >0,
\end{align*}
whenever $\alpha\in(1.97,2)$. 

Therefore, for $\alpha\in(1.97,2)$, $\tilde{B}$ is positive definite.
Moreover, for $\alpha=2$, $\tilde{B}$ turns to be 
\[
\begin{pmatrix}6 & 0 & 0 & 0 & 0\\
0 & 0 & 0 & 0 & 0\\
0 & 0 & 0 & 0 & 0\\
0 & 0 & 0 & 0 & 0\\
0 & 0 & 0 & 0 & 0
\end{pmatrix},
\]
which is obviously positive semidefinite. We complete the proof for
the case of $\alpha\in(1.97,2]$. 

\section{\label{sec:Proof-of-Proposition}Proof of Proposition \ref{prop:sign}}

To prove the desired results, we first estimate $h_{\alpha}(X_{t})$
in terms of $(\alpha,t)$. Let $\sigma^{2}$ be the variance of
$X$. Denote $G_{t}$ as a Gaussian random variable with variance
$t$ and denote $\varphi_{t}$ as its density. Then, for $\alpha>1$,
\begin{align*}
e^{(1-\alpha)h_{\alpha}(G_{t})} & =\int\varphi_{t}(z)^{\alpha}dz\\
 & =\int f(x)\int\varphi_{t}(y-x)^{\alpha}dydx\\
 & \ge\int\left(\int f(x)\varphi_{t}(y-x)dx\right)^{\alpha}dy\\
 & =\int p(y)^{\alpha}dy\\
 & =e^{(1-\alpha)h_{\alpha}(X_{t})},
\end{align*}
where the inequality follows by Jensen's inequality. So, for $\alpha>1$,
\begin{equation}
h_{\alpha}(G_{t})\le h_{\alpha}(X_{t}).\label{eq:-1-1}
\end{equation}
Similarly, it can be shown that this inequality also holds for $0<\alpha<1$.
By taking $\alpha\to1$, this inequality still holds for $\alpha=1$.
So, for all for $\alpha>0$, 
\begin{align}
h_{\alpha}(X_{t}) & \ge\frac{1}{2}\log(2\pi t)+\frac{1}{2(\alpha-1)}\log\alpha.\label{eq:-6}
\end{align}

By the fact that Gaussian distributions maximize the entropy over
all distributions with the same variance and the fact that Rényi entropy
is nonincreasing in its order, it holds that for $\alpha>1$, 
\begin{align}
h_{\alpha}(X_{t}) & \le h(X_{t})\nonumber \\
 & \le h(G_{t+\sigma^{2}})\nonumber \\
 & =\frac{1}{2}\log(2\pi(t+\sigma^{2}))+\frac{1}{2(\alpha-1)}\log\alpha.\label{eq:-17}
\end{align}

For $\frac{1}{3}<\alpha<1$, it is known that the probability distributions
that maximizes the Rényi entropy $h_{\alpha}$ over all distributions
with the same variance are the Pearson type VII laws (also called
Student-s laws); see more details in \cite{costa2003solutions}. Denote
$\tilde{G}_{t}$ as a Pearson type VII random variable with variance
$t$ and its density is 
\[
f_{\alpha,t}(x):=\frac{\Gamma\left(\frac{1}{1-\alpha}\right)}{\Gamma\left(\frac{1+\alpha}{2(1-\alpha)}\right)}\frac{1}{\sqrt{\pi b(t)}}\left(1+\frac{x^{2}}{b(t)}\right)^{\frac{1}{\alpha-1}},
\]
where $b(t)=\frac{(3\alpha-1)t}{1-\alpha}$ and  $\Gamma$ is the
Gamma function. So, for $\frac{1}{3}<\alpha<1$,
\begin{align}
h_{\alpha}(X_{t}) & \le h_{\alpha}(\tilde{G}_{t+\sigma^{2}})\nonumber \\
 & =\frac{1}{1-\alpha}\log\int f_{\alpha,t+\sigma^{2}}^{\alpha}(x)dx\nonumber \\
 & =\frac{1}{1-\alpha}\log\frac{2\alpha}{3\alpha-1}+\log\frac{\Gamma\left(\frac{1+\alpha}{2(1-\alpha)}\right)}{\Gamma\left(\frac{1}{1-\alpha}\right)}+\frac{1}{2}\log\frac{\pi(3\alpha-1)(t+\sigma^{2})}{1-\alpha},\label{eq:-18}
\end{align}
where the expression in the last line is obtained by using Mathematica.

We now turn back to proving Proposition \ref{prop:sign}. Since $\frac{\partial^{k+1}}{\partial t^{k+1}}h_{\alpha}(X_{t})\le0$,
it holds that $\frac{\partial^{k}}{\partial t^{k}}h_{\alpha}(X_{t})$
is nonincreasing in $t$. So, $\lim_{t\to+\infty}\frac{\partial^{k}}{\partial t^{k}}h_{\alpha}(X_{t})$
exists (and might be equal to $-\infty$). By L'Hospital's rule,
\[
\lim_{t\to+\infty}\frac{\frac{\partial^{k-1}}{\partial t^{k-1}}h_{\alpha}(X_{t})}{t}=\lim_{t\to+\infty}\frac{\partial^{k}}{\partial t^{k}}h_{\alpha}(X_{t}).
\]
Repeating applying L'Hospital's rule, we obtain that 
\begin{align*}
\lim_{t\to+\infty}\frac{2!\frac{\partial^{k-2}}{\partial t^{k-2}}h_{\alpha}(X_{t})}{t^{2}} & =\lim_{t\to+\infty}\frac{\frac{\partial^{k-1}}{\partial t^{k-1}}h_{\alpha}(X_{t})}{t},\\
\lim_{t\to+\infty}\frac{3!\frac{\partial^{k-3}}{\partial t^{k-3}}h_{\alpha}(X_{t})}{t^{3}} & =\lim_{t\to+\infty}\frac{2!\frac{\partial^{k-2}}{\partial t^{k-2}}h_{\alpha}(X_{t})}{t^{2}},\\
 & \vdots\\
\lim_{t\to+\infty}\frac{k!h_{\alpha}(X_{t})}{t^{k}} & =\lim_{t\to+\infty}\frac{(k-1)!\frac{\partial}{\partial t}h_{\alpha}(X_{t})}{t^{k-1}}.
\end{align*}
Hence,
\[
\lim_{t\to+\infty}\frac{k!h_{\alpha}(X_{t})}{t^{k}}=\lim_{t\to+\infty}\frac{\partial^{k}}{\partial t^{k}}h_{\alpha}(X_{t}).
\]

On the other hand, from \eqref{eq:-6}, \eqref{eq:-17}, and \eqref{eq:-18},
we see that for $\alpha>\frac{1}{3}$, 
\[
\lim_{t\to+\infty}\frac{k!h_{\alpha}(X_{t})}{t^{k}}=0.
\]
So, 
\[
\lim_{t\to+\infty}\frac{\partial^{k}}{\partial t^{k}}h_{\alpha}(X_{t})=0.
\]
Combined with that $\frac{\partial^{k}}{\partial t^{k}}h_{\alpha}(X_{t})$
is nonincreasing in $t$, this implies $\frac{\partial^{k}}{\partial t^{k}}h_{\alpha}(X_{t})\ge0$
for all $t>0$. 

\section{\label{sec:Proof of concavity of REP-1}Proof of Theorem \ref{thm:REP-1}}

The proofs of Statements 1 and 2 are almost identical. So, we only
provide the former.

By definition, 
\[
\frac{\partial N_{\alpha}^{\frac{1}{2}}(X_{t})}{\partial t}=\frac{\partial\exp[h_{\alpha}(X_{t})]}{\partial t}=e^{h_{\alpha}(p)}\frac{\partial h_{\alpha}(p)}{\partial t}.
\]
Taking the derivative w.r.t. $t$ again, we have 
\[
\frac{\partial^{2}N_{\alpha}^{\frac{1}{2}}(X_{t})}{\partial t^{2}}=e^{h_{\alpha}(p)}\left(\frac{\partial h_{\alpha}(p)}{\partial t}\right)^{2}+e^{h_{\alpha}(p)}\frac{\partial^{2}h_{\alpha}(p)}{\partial t^{2}}.
\]
Thus, the concavity of $N_{\alpha}^{\frac{1}{2}}(X_{t})$ is equivalent
to 
\[
e^{h_{\alpha}(p)}\left(\frac{\partial h_{\alpha}(p)}{\partial t}\right)^{2}+e^{h_{\alpha}(p)}\frac{\partial^{2}h_{\alpha}(p)}{\partial t^{2}}\le0,
\]
that is, 
\begin{equation}
\left(\frac{\partial h_{\alpha}(p)}{\partial t}\right)^{2}+\frac{\partial^{2}h_{\alpha}(p)}{\partial t^{2}}\le0.\label{eq:REP-1-1}
\end{equation}

By Theorem \ref{thm:1st derivative} and Theorem \ref{thm:2nd derivative},

\begin{align*}
\left(\frac{\partial h_{\alpha}(p)}{\partial t}\right)^{2}+\frac{\partial^{2}h_{\alpha}(p)}{\partial t^{2}} & =\left(\frac{\alpha}{2}\mathbb{E}_{\alpha}[\bar{p}_{1}^{2}]\right)^{2}+\frac{\alpha}{12}\left((\alpha-2)(\alpha-3)\mathbb{E}_{\alpha}[\bar{p}_{1}^{4}]-6\mathbb{E}_{\alpha}[\bar{p}_{2}^{2}]+3\alpha(\alpha-1)(\mathbb{E}_{\alpha}[\bar{p}_{1}^{2}])^{2}\right)\\
 & =\frac{\alpha(\alpha-2)(\alpha-3)}{12}\mathbb{E}_{\alpha}[\bar{p}_{1}^{4}]-\frac{\alpha}{2}\mathbb{E}_{\alpha}[\bar{p}_{2}^{2}]+\frac{\alpha^{3}}{4}\left(\mathbb{E}_{\alpha}[\bar{p}_{1}^{2}]\right)^{2}.
\end{align*}
Thus \eqref{eq:REP-1-1} is equivalent to 
\begin{equation}
\frac{(\alpha-2)(\alpha-3)}{12}\mathbb{E}_{\alpha}[\bar{p}_{1}^{4}]-\frac{1}{2}\mathbb{E}_{\alpha}[\bar{p}_{2}^{2}]+\frac{\alpha^{2}}{4}\left(\mathbb{E}_{\alpha}[\bar{p}_{1}^{2}]\right)^{2}\le0.\label{eq:REP-1-2}
\end{equation}

We will express 
\begin{equation}
\frac{(\alpha-2)(\alpha-3)}{12}\mathbb{E}_{\alpha}[\bar{p}_{1}^{4}]-\frac{1}{2}\mathbb{E}_{\alpha}[\bar{p}_{2}^{2}]+\frac{\alpha^{2}}{4}\left(\mathbb{E}_{\alpha}[\bar{p}_{1}^{2}]\right)^{2}\label{eq:REP-1-3}
\end{equation}
as 
\begin{equation}
H:=-\mathbb{E}_{\alpha}[(a\bar{p}_{2}+b\bar{p}_{1}^{2}+c\mathbb{E}_{\alpha}[\bar{p}_{1}^{2}])^{2}].\label{eq:-9}
\end{equation}

Expanding this expression yields that 
\begin{align}
H & =-\mathbb{E}_{\alpha}[(a\bar{p}_{2}+b\bar{p}_{1}^{2})^{2}]-c^{2}\left(\mathbb{E}_{\alpha}[\bar{p}_{1}^{2}]\right)^{2}-2c\mathbb{E}_{\alpha}[\bar{p}_{1}^{2}]\mathbb{E}_{\alpha}[(a\bar{p}_{2}+b\bar{p}_{1}^{2})]\nonumber \\
 & =-a^{2}\mathbb{E}_{\alpha}[\bar{p}_{2}^{2}]-(b^{2}-\frac{2}{3}(\alpha-3)ab)\mathbb{E}_{\alpha}[\bar{p}_{1}^{4}]-(c^{2}+2c(b-(\alpha-1)a))\left(\mathbb{E}_{\alpha}[\bar{p}_{1}^{2}]\right)^{2},\label{eq:REP-1-4}
\end{align}
where \eqref{eq:REP-1-4} follows from \eqref{eq:p^alpha-3p_1^2p_2}
and $\mathbb{E}_{\alpha}\left[\bar{p}_{2}\right]=-(\alpha-1)\mathbb{E}_{\alpha}\left[\bar{p}_{1}^{2}\right]$.

Comparing the coefficients in \eqref{eq:REP-1-4} and the ones in
\eqref{eq:REP-1-3}, we choose $a,b,c$ such that 
\begin{equation}
\begin{cases}
-a^{2}=-\frac{1}{2}\\
-(b^{2}-\frac{2}{3}(\alpha-3)ab)=\frac{(\alpha-2)(\alpha-3)}{12}\\
-(c^{2}+2c(b-(\alpha-1)a))=\frac{\alpha^{2}}{4}
\end{cases}.\label{eq:REP-1-5}
\end{equation}
A solution to these equations is 
\begin{equation}
\begin{cases}
a=\frac{1}{\sqrt{2}}\\
b=\frac{1}{6}(-3\sqrt{2}+\sqrt{2}\alpha-\sqrt{3\alpha-\alpha^{2}})\\
c=\frac{1}{6}(2\sqrt{2}\alpha+\sqrt{3\alpha-\alpha^{2}}+\sqrt{3\alpha-2\alpha^{2}+4\sqrt{2}\alpha\sqrt{3\alpha-\alpha^{2}}})
\end{cases}.\label{eq:REP-1-6}
\end{equation}
It is easy to check that when $0\le\alpha\le\frac{3}{2}+\sqrt{2}$,
\[
\begin{cases}
3\alpha-\alpha^{2}\ge0\\
3\alpha-2\alpha^{2}+4\sqrt{2}\alpha\sqrt{3\alpha-\alpha^{2}}\ge0
\end{cases}.
\]
Thus, when $\alpha\in(0,\frac{3}{2}+\sqrt{2}]$, \eqref{eq:REP-1-6}
is a real root of equations in \eqref{eq:REP-1-5}. Hence, the expression
in \eqref{eq:REP-1-3} can be expressed the one in \eqref{eq:-9},
which implies \eqref{eq:REP-1-2}, i.e.,$\frac{\partial^{2}N_{\alpha}^{\frac{1}{2}}(Y_{t})}{\partial t^{2}}\le0,$
whenever $\alpha\in(0,\frac{3}{2}+\sqrt{2}]$. 
\begin{rem}
When we consider the concavity of $N_{\alpha}(X_{t})$, the proof
above is only valid for $\alpha=1$, yielding no new results. 
\end{rem}

\section{\label{sec:conjecture-1-1} Proof of Proposition \ref{prop:For-any-}}

We follow proof steps similar to the ones in \cite{wang2024entropy}.
Denote $y(t)=2h_{\alpha}(p)$. Denote $B_{m}(x_{1},\dots,x_{m})$
or $B_{m}(x_{1},x_{2},\dots)$ the complete exponential Bell polynomials.
By Faa di Bruno's formula, for any $k\ge1$,
\[
\frac{d^{k}}{dt^{k}}N_{\alpha}(p)=\frac{d^{k}}{dt^{k}}e^{y(t)}=e^{y(t)}B_{k}(\dot{y},y^{(2)},y^{(3)},\dots)
\]
where $y^{(2)}=\frac{d}{dt}\dot{y},y^{(3)}=\frac{d}{dt}y^{(2)}$,
etc. By property of the Bell polynomials,
\[
\forall m,\forall\beta\in\mathbb{R},\;B_{m}(\beta x_{1},\beta^{2}x_{2},\beta^{3}x_{3},\dots)=\beta^{m}B_{m}(x_{1},x_{2},\dots).
\]
Letting $\beta=-1$ and $Y_{k}=(-1)^{k}y^{(k)}$ for all $k\ge1$
yields that 
\begin{align*}
(-1)^{k-1}\frac{d^{k}}{dt^{k}}N_{\alpha}(p) & =(-1)^{k-1}e^{y(t)}B_{k}(\dot{y},y^{(2)},y^{(3)},\dots)\\
 & =-e^{y(t)}B_{k}(-\dot{y},y^{(2)},-y^{(3)},\dots)\\
 & =-e^{y(t)}B_{k}(Y_{1},Y_{2},Y_{3},\dots).
\end{align*}

Fix $K\in\mathbb{N}^{+}$ and suppose that $(-1)^{k-1}\frac{d^{k}}{dt^{k}}N_{\alpha}(p)\ge0$
for all $k\le K$, i.e., $B_{k}(Y_{1},Y_{2},Y_{3},\dots)\le0$ for
all $k\le K$. Then by Lemma 2 in \cite{wang2024entropy},
\[
Y_{k}\le-(k-1)!(-Y_{1})^{k}\quad\text{for all\;}1\le k\le K,
\]
i.e.,
\[
(-1)^{k-1}y^{(k)}\ge(k-1)!\dot{y}^{k}.
\]
Note that $\dot{y}\ge0$, which completes the proof.

\section{\label{sec:conjecture-1} Proof of Fact \ref{thm:Tsallis2}}

It is easy to verify that $\frac{\partial\hat{h}_{2}(p)}{\partial t}=\int p_{1}^{2}dx$.
By mathematical induction, suppose $\frac{\partial^{m}\hat{h}_{2}(p)}{\partial t^{m}}=(-1)^{m-1}\int p_{m}^{2}dx$,
for some $m\in\mathbb{Z}_{+}$. Then taking the derivative on both
sides, we have 
\begin{align*}
\frac{\partial^{m+1}\hat{h}_{2}(p)}{\partial t^{m+1}} & =(-1)^{m-1}\int2p_{m}(p_{m})_{t}dx\\
 & =(-1)^{m-1}\int p_{m}p_{m+2}dx\\
 & =(-1)^{m-1}\int p_{m}dp_{m+1}\\
 & =(-1)^{m-1}p_{m}p_{m+1}\bigg|_{-\infty}^{+\infty}-(-1)^{m-1}\int p_{m+1}^{2}dx\\
 & =(-1)^{m}\int p_{m+1}^{2}dx,
\end{align*}
where the second equality follows from \eqref{eq:heat equation}.

\appendices{}

\section{\label{sec:-Proof-of} Proof of Lemma \ref{lem:k>=00003D00003D00003D1}}

From Appendix A of \cite{Cheng2015Higher}, 
\[
p^{(n)}(x,t)=\int f(y)\frac{1}{\sqrt{2\pi t}}e^{-\frac{(x-y)^{2}}{2t}}H_{n}(x-y)dy.
\]
where $H_{0}=1$ and $H_{n}$ satisfies the recursion formula 
\[
H_{n}(x-y)=-\frac{x-y}{t}H_{n-1}(x-y)+\frac{\partial}{\partial x}H_{n-1}(x-y).
\]
In general, 
\[
H_{n}(x-y)=\sum_{j=0}^{n}\beta_{j}(t)(x-y)^{j},
\]
where $\beta_{j}(t),0\le j\le n$ are continuous functions of $t$.
Hence, 
\[
\frac{p^{(n)}}{p}=\sum_{j=0}^{n}\beta_{j}(t)\mathbb{E}\left[(X_{t}-X)^{j}|X_{t}=x\right],
\]
Denote $\beta(t):=\sum_{j=0}^{n}|\beta_{j}(t)|$. By Jensen’s inequality,
it holds that 
\begin{align}
\left|\frac{p^{(n)}}{p}\right|^{k} & \le\left(\sum_{j=0}^{n}|\beta_{j}(t)|\mathbb{E}\left[|X_{t}-X|^{j}|X_{t}=x\right]\right)^{k}\nonumber \\
 & \le\beta^{k}(t)\sum_{j=0}^{n}\frac{|\beta_{j}(t)|}{\beta(t)}\left(\mathbb{E}\left[|X_{t}-X|^{j}|X_{t}=x\right]\right)^{k}\nonumber \\
 & \le\beta^{k-1}(t)\sum_{j=0}^{n}|\beta_{j}(t)|\mathbb{E}\left[|X_{t}-X|^{jk}|X_{t}=x\right].\label{eq:-4}
\end{align}

We now divide the remaining proof into the case $\alpha\ge1$ and
the case $0<\alpha<1$.
\begin{itemize}
\item Case $\alpha\ge1$: 
\end{itemize}
Denote 
\[
\varphi(x,t)=\frac{1}{\sqrt{2\pi t}}e^{-\frac{x^{2}}{2t}}.
\]

Then, \eqref{eq:-4} implies 
\begin{align}
p^{\alpha}\left|\frac{p^{(n)}}{p}\right|^{k} & \le\beta^{k-1}(t)\sum_{j=0}^{n}|\beta_{j}(t)|M_{jk}(t),\label{eq:-21}
\end{align}
where 
\[
M_{jk}(t):=p^{\alpha}\mathbb{E}\left[|X_{t}-X|^{jk}|X_{t}=x\right]=p^{\alpha-1}(x,t)\int|x-y|^{jk}f(y)\varphi(x-y,t)dy.
\]

Let $0<a<b<+\infty$ and $\epsilon>0$. Then, there always exists
a positive and continuous function $A_{jk}$ such that 
\begin{equation}
\left|x^{jk}\varphi(x,t)\right|\le\frac{A_{jk}(t)}{\sqrt{2\pi t}}e^{-\frac{x^{2}}{2(t+\epsilon)}},\quad\forall x.\label{eq:-7-1}
\end{equation}
This is because, $\frac{x^{jk}\varphi(x,t)}{\frac{1}{\sqrt{2\pi t}}e^{-\frac{x^{2}}{2(t+\epsilon)}}}\to0$
as $|x|\to\infty$. The inequality in \eqref{eq:-7-1} yields that
for any $t\in[a,b]$, 
\[
|x|^{jk}\varphi(x,t)\le\frac{B_{jk}(a,b,\epsilon)}{\sqrt{2\pi(b+\epsilon)}}e^{-\frac{x^{2}}{2(b+\epsilon)}}=B_{jk}(a,b,\epsilon)\varphi(x,b+\epsilon),\quad\forall x,
\]
where $B_{jk}(a,b,\epsilon):=\max_{t\in[a,b]}A_{jk}(t)\sqrt{\frac{b+\epsilon}{t}}$
is finite. That is, $x\mapsto|x|^{jk}\varphi(x,t)$ is uniformly bounded
on $[a,b]$. This further implies that $M_{jk}(t)$ is also uniformly
bounded: 
\begin{align*}
M_{jk}(t) & \le B_{jk}(a,b,\epsilon)p^{\alpha-1}(x,t)p(x,b+\epsilon).
\end{align*}
In addition, 
\begin{align*}
p(x,t) & =\int f(y)\varphi(x-y,t)dy\\
 & \le\sqrt{\frac{b+\epsilon}{a}}\int f(y)\varphi(x-y,b+\epsilon)dy\\
 & =\sqrt{\frac{b+\epsilon}{a}}p(x,b+\epsilon).
\end{align*}
Hence, 
\begin{align}
M_{jk}(t) & \le C_{jk}(a,b,\epsilon)p^{\alpha}(x,b+\epsilon),\label{eq:-20}
\end{align}
where $C_{jk}(a,b,\epsilon)=B_{jk}(a,b,\epsilon)\left(\frac{b+\epsilon}{a}\right)^{\alpha-1}$. 

Substituting \eqref{eq:-20} into \eqref{eq:-21} yields that for
any $t\in[a,b]$, 
\begin{align*}
p^{\alpha}\left|\frac{p^{(n)}}{p}\right|^{k} & \le\left(\beta^{k-1}(t)\sum_{j=0}^{n}|\beta_{j}(t)|C_{jk}(a,b,\epsilon)\right)p^{\alpha}(x,b+\epsilon),\quad\forall x.
\end{align*}
Since $\beta^{k-1}(t)\sum_{j=0}^{n}|\beta_{j}(t)|C_{jk}(a,b,\epsilon)$
is continuous in $t\in[a,b]$, it is also bounded on this interval. 

By the inequality of arithmetic and geometric means, 
\begin{align*}
\prod_{i=1}^{r}\left|\frac{p^{(n_{i})}}{p}\right|^{k_{i}} & \le\frac{1}{r}\sum_{i=1}^{r}\left|\frac{p^{(n_{i})}}{p}\right|^{rk_{i}}.
\end{align*}
Applying the inequality in for each case term $\left|\frac{p^{(n_{i})}}{p}\right|^{rk_{i}}$,
we obtain 
\begin{align*}
p^{\alpha}\prod_{i=1}^{r}\left|\frac{p^{(n_{i})}}{p}\right|^{k_{i}} & \le Cp^{\alpha}(x,b+\epsilon),
\end{align*}
for some constant $C$ independent of $(x,t)$, completing the proof
for the case $\alpha\ge1$.
\begin{itemize}
\item Case $0<\alpha<1$: 
\end{itemize}
For the case $0<\alpha<1$, we rewrite 
\begin{align*}
M_{jk}(t) & =\int|x-y|^{jk}\frac{f(y)\varphi(x-y,t)}{\int f(y')\varphi(x-y',t)dy'}dy\left(\int f(y')\varphi(x-y',t)dy'\right)^{\alpha}\\
 & =\int|x-y|^{jk}\left(f(y)\varphi(x-y,t)\right)^{\alpha}\left(\frac{f(y)\varphi(x-y,t)}{\int f(y')\varphi(x-y',t)dy'}\right)^{1-\alpha}dy.
\end{align*}

Let $0<a<b<+\infty$ and $\epsilon>0$. Then, there always exists
a positive and continuous function $A_{jk}$ such that 
\begin{equation}
\left|x^{jk}\varphi^{\alpha}(x,t)\right|\le A_{jk}(t)\left(\frac{1}{\sqrt{2\pi t}}e^{-\frac{x^{2}}{2(t+\epsilon)}}\right)^{\alpha},\quad\forall x.\label{eq:-7}
\end{equation}
This is because, $\frac{x^{jk}\varphi^{\alpha}(x,t)}{\left(\frac{1}{\sqrt{2\pi t}}e^{-\frac{x^{2}}{2(t+\epsilon)}}\right)^{\alpha}}\to0$
as $|x|\to\infty$. The inequality in \eqref{eq:-7} yields that for
any $t\in[a,b]$, 
\[
|x|^{jk}\varphi^{\alpha}(x,t)\le B_{jk}(a,b,\epsilon)\left(\frac{1}{\sqrt{2\pi(b+\epsilon)}}e^{-\frac{x^{2}}{2(b+\epsilon)}}\right)^{\alpha}=B_{jk}(a,b,\epsilon)\varphi^{\alpha}(x,b+\epsilon),\quad\forall x,
\]
where $B_{jk}(a,b,\epsilon):=\max_{t\in[a,b]}A_{jk}(t)\left(\frac{b+\epsilon}{t}\right)^{\alpha/2}$
is finite. That is, $x\mapsto|x|^{jk}\varphi^{\alpha}(x,t)$ is uniformly
bounded on $[a,b]$. This further implies that $M_{jk}(t)$ is also
uniformly bounded: 
\begin{align*}
M_{jk}(t) & \le B_{jk}(a,b,\epsilon)\int\left(f(y)\varphi(x-y,(1+\epsilon)b)\right)^{\alpha}\left(\frac{f(y)\varphi(x-y,t)}{\int f(y')\varphi(x-y',t)dy'}\right)^{1-\alpha}dy\\
 & \le B_{jk}(a,b,\epsilon)\left(\int f(y)\varphi(x-y,(1+\epsilon)b)dy\right)^{\alpha}\left(\int\frac{f(y)\varphi(x-y,t)}{\int f(y')\varphi(x-y',t)dy'}dy\right)^{1-\alpha}\\
 & =B_{jk}(a,b,\epsilon)p^{\alpha}(x,b+\epsilon),
\end{align*}
where the second inequality follows by Hölder's inequality for $0<\alpha<1$.
Therefore, for any $t\in[a,b]$, 
\begin{align*}
p^{\alpha}\left|\frac{p^{(n)}}{p}\right|^{k} & \le\left(\beta^{k-1}(t)\sum_{j=0}^{n}|\beta_{j}(t)|B_{jk}(a,b,\epsilon)\right)p^{\alpha}(x,b+\epsilon),\quad\forall x.
\end{align*}
Similarly to the case $\alpha\ge1$, the factor $\beta^{k-1}(t)\sum_{j=0}^{n}|\beta_{j}(t)|B_{jk}(a,b,\epsilon)$
is also uniformly bounded on $t\in[a,b]$. Then, still by using the
inequality of arithmetic and geometric means, we obtained the desired
inequality for $0<\alpha<1$.

\section{\label{sec:Appendix-(p^alpha)_t} Proof of Lemma \ref{lem:(p^alpha)_t}}

By the chain rule of derivatives, 
\begin{align*}
\frac{\partial(\int p^{\alpha-2}p_{1}^{2}dx)}{\partial t} & =\int(\alpha-2)p^{\alpha-3}p_{t}p_{1}^{2}+p^{\alpha-2}\cdot2p_{1}p_{xt}dx.
\end{align*}

We evaluate the integrals at the RHS above as follows. The proof of
Lemma \ref{lem:2nd derivative-1} will be given later. 
\begin{lem}
\label{lem:2nd derivative-1}For $t>0$, 
\begin{align}
\int p^{\alpha-3}p_{t}p_{1}^{2}dx & =-\frac{\alpha-3}{6}\int p^{\alpha-4}p_{1}^{4}dx,\label{eq:p^alpha-3p_tp_1^2}\\
\int p^{\alpha-2}p_{1}p_{xt}dx & =\frac{(\alpha-2)(\alpha-3)}{6}\int p^{\alpha-4}p_{1}^{4}dx-\frac{1}{2}\int p^{\alpha-2}p_{2}^{2}dx.\label{eq:p^alpha-2p_1p_xt}
\end{align}
\end{lem}
By Lemma \ref{lem:2nd derivative-1}, 
\begin{align*}
\frac{\partial(\int p^{\alpha-2}p_{1}^{2}dx)}{\partial t} & =-\frac{(\alpha-2)(\alpha-3)}{6}\int p^{\alpha-4}p_{1}^{4}dx+\frac{(\alpha-2)(\alpha-3)}{3}\int p^{\alpha-4}p_{1}^{4}dx-\int p^{\alpha-2}p_{2}^{2}dx\\
 & =\frac{(\alpha-2)(\alpha-3)}{6}\int p^{\alpha-4}p_{1}^{4}dx-\int p^{\alpha-2}p_{2}^{2}dx.
\end{align*}

As for $\frac{\partial(\int p^{\alpha}dx)}{dt}$, we have 
\begin{align*}
\frac{\partial(\int p^{\alpha}dx)}{dt} & =\alpha\int p^{\alpha-1}p_{t}dx\\
 & =\frac{\alpha}{2}\int p^{\alpha-1}p_{2}dx\\
 & =\frac{\alpha}{2}\int p^{\alpha-1}dp_{1}\\
 & =\frac{\alpha}{2}p^{\alpha-1}p_{1}\bigg|_{-\infty}^{+\infty}-\frac{\alpha(\alpha-1)}{2}\int p^{\alpha-2}p_{1}^{2}dx\\
 & =-\frac{\alpha(\alpha-1)}{2}\int p^{\alpha-2}p_{1}^{2}dx,
\end{align*}
where the second equality follows from heat equation \eqref{eq:heat equation}. 
\begin{IEEEproof}[Proof of Lemma \ref{lem:2nd derivative-1}]
Using integration by parts,

\begin{align}
\int p^{\alpha-3}p_{t}p_{1}^{2}dx & =\frac{1}{2}\int p^{\alpha-3}p_{2}p_{1}^{2}dx\nonumber \\
 & =\frac{1}{2}\int p^{\alpha-3}p_{1}^{2}dp_{1}\nonumber \\
 & =\frac{1}{6}\int p^{\alpha-3}dp_{1}^{3}\\
 & =\frac{1}{6}p^{\alpha-3}p_{1}^{3}\bigg|_{-\infty}^{+\infty}-\frac{1}{6}\int(\alpha-3)p^{\alpha-4}p_{1}^{4}dx\\
 & =-\frac{\alpha-3}{6}\int p^{\alpha-4}p_{1}^{4}dx
\end{align}

Similarly, 
\begin{align}
\int p^{\alpha-2}p_{1}p_{xt}dx & =\frac{1}{2}\int p^{\alpha-2}p_{1}p_{3}dx\nonumber \\
 & =\frac{1}{2}\int p^{\alpha-2}p_{1}dp_{2}\nonumber \\
 & =\frac{1}{2}p^{\alpha-2}p_{1}p_{2}\bigg|_{-\infty}^{+\infty}-\frac{1}{2}\int((\alpha-2)p^{\alpha-3}p_{1}^{2}+p^{\alpha-2}p_{2})p_{2}dx\nonumber \\
 & =-\frac{\alpha-2}{2}\int p^{\alpha-3}p_{1}^{2}p_{2}dx-\frac{1}{2}\int p^{\alpha-2}p_{2}^{2}dx,\label{eq:p^alpha-2p_2^2}
\end{align}
As for $\int p^{\alpha-3}p_{1}^{2}p_{2}dx$, we have 
\begin{align*}
\int p^{\alpha-3}p_{1}^{2}p_{2}dx & =\int p^{\alpha-3}p_{1}^{2}dp_{1}\\
 & =p^{\alpha-3}p_{1}^{3}\bigg|_{-\infty}^{+\infty}-\int((\alpha-3)p^{\alpha-4}p_{1}^{3}+p^{\alpha-3}\cdot2p_{1}p_{2})p_{1}dx\\
 & =-(\alpha-3)\int p^{\alpha-4}p_{1}^{4}dx-2\int p^{\alpha-3}p_{1}^{2}p_{2}dx,
\end{align*}
thus, 
\begin{equation}
\int p^{\alpha-3}p_{1}^{2}p_{2}dx=-\frac{\alpha-3}{3}\int p^{\alpha-4}p_{1}^{4}dx,\label{eq:p_1^2p_2p^alpha-3}
\end{equation}
then substitute \eqref{eq:p_1^2p_2p^alpha-3} into \eqref{eq:p^alpha-2p_2^2},
we obtain \eqref{eq:p^alpha-2p_1p_xt}. 
\end{IEEEproof}

\section{\label{sec:Appendix-E_alphap_1^4} Proof of Lemma \ref{lem:E_alphap_1^4}}

We need the following two lemmas, whose proofs will be given later. 
\begin{lem}
\label{lem:3rd derivative-1}For $t>0$, 
\begin{align}
\int p^{\alpha-5}p_{1}^{4}p_{2}dx & =-\frac{\alpha-5}{5}\int p^{\alpha-6}p_{1}^{6}dx,\label{eq:p^alpha-5p_1^4p_2}\\
\int p^{\alpha-4}p_{1}^{3}p_{3}dx & =\frac{(\alpha-4)(\alpha-5)}{5}\int p^{\alpha-6}p_{1}^{6}dx-3\int p^{\alpha-4}p_{1}^{2}p_{2}^{2}dx,\label{eq:p^alpha-4p_1^3p_3}\\
\int p^{\alpha-3}p_{1}p_{2}p_{3}dx & =-\frac{\alpha-3}{2}\int p^{\alpha-4}p_{1}^{2}p_{2}^{2}dx-\frac{1}{2}\int p^{\alpha-3}p_{2}^{3}dx,\label{eq:p^alpha-3p_1p_2p_3}\\
\int p^{\alpha-2}p_{2}p_{4}dx & =\frac{(\alpha-2)(\alpha-3)}{2}\int p^{\alpha-4}p_{1}^{2}p_{2}^{2}dx+\frac{\alpha-2}{2}\int p^{\alpha-3}p_{2}^{3}dx-\int p^{\alpha-2}p_{3}^{2}dx,\label{eq:p^alpha-2p_2p_4}\\
\int p^{\alpha-3}p_{1}^{2}p_{2}dx & =-\frac{\alpha-3}{3}\int p^{\alpha-4}p_{1}^{4}dx,\label{eq:p^alpha-3p_1^2p_2}\\
\int p^{\alpha-2}p_{1}p_{3}dx & =\frac{(\alpha-2)(\alpha-3)}{3}\int p^{\alpha-4}p_{1}^{4}dx-\int p^{\alpha-2}p_{2}^{2}dx,\label{eq:p^alpha-2p_1p_3}\\
\int p^{\alpha-1}p_{4}dx & =-\frac{(\alpha-1)(\alpha-2)(\alpha-3)}{3}\int p^{\alpha-4}p_{1}^{4}dx+(\alpha-1)\int p^{\alpha-2}p_{2}^{2}dx,\label{eq:p^alpha-1p_4}\\
\int p^{\alpha-3}p_{1}^{2}p_{4}dx & =-\frac{(\alpha-3)(\alpha-4)(\alpha-5)}{5}\int p^{\alpha-6}p_{1}^{6}dx+4(\alpha-3)\int p^{\alpha-4}p_{1}^{2}p_{2}^{2}dx+\int p^{\alpha-3}p_{2}^{3}dx.\label{eq:p^alpha-3p_1^2p_4}
\end{align}
\end{lem}
\begin{lem}
\label{lem:3rd derivative-2}For $t>0$, 
\begin{align}
\frac{\partial\int p^{\alpha-4}p_{1}^{4}dx}{\partial t} & =\frac{3(\alpha-4)(\alpha-5)}{10}\int p^{\alpha-6}p_{1}^{6}dx-6\int p^{\alpha-4}p_{1}^{2}p_{2}^{2}dx,\label{eq:p^alpha-4p_1^4-1}\\
\frac{\partial\int p^{\alpha-2}p_{2}^{2}dx}{\partial t} & =\frac{(\alpha-2)(\alpha-3)}{2}\int p^{\alpha-4}p_{1}^{2}p_{2}^{2}dx+(\alpha-2)\int p^{\alpha-3}p_{2}^{3}dx-\int p^{\alpha-2}p_{3}^{2}dx.\label{eq:p^alpha-2p_2^2-1}
\end{align}
\end{lem}
Note that $\mathbb{E}_{\alpha}[\bar{p}_{1}^{4}]=\frac{\int p^{\alpha-4}p_{1}^{4}dx}{\int p^{\alpha}dx}$,
and thus 
\begin{align*}
\frac{\partial\mathbb{E}_{\alpha}[\bar{p}_{1}^{4}]}{\partial t} & =\frac{(\int p^{\alpha-4}p_{1}^{4}dx)_{t}\int p^{\alpha}dx-\int p^{\alpha-4}p_{1}^{4}dx\cdot(\int p^{\alpha}dx)_{t}}{(\int p^{\alpha}dx)^{2}}\\
 & =\frac{\frac{3(\alpha-4)(\alpha-5)}{10}\int p^{\alpha-6}p_{1}^{6}dx-6\int p^{\alpha-4}p_{1}^{2}p_{2}^{2}dx}{\int p^{\alpha}dx}+\frac{\frac{\alpha(\alpha-1)}{2}\int p^{\alpha-4}p_{1}^{4}dx\cdot\int p^{\alpha-2}p_{1}^{2}dx}{(\int p^{\alpha}dx)^{2}}\\
 & =\frac{3(\alpha-4)(\alpha-5)}{10}\mathbb{E}_{\alpha}[\bar{p}_{1}^{6}]-6\mathbb{E}_{\alpha}[\bar{p}_{1}^{2}\bar{p}_{2}^{2}]+\frac{\alpha(\alpha-1)}{2}\mathbb{E}_{\alpha}[\bar{p}_{1}^{4}]\mathbb{E}_{\alpha}[\bar{p}_{1}^{2}],
\end{align*}
where the second equality follows from \eqref{eq:p^alpha-4p_1^4-1}
and \eqref{eq:p^alpha}.

Note that $\mathbb{E}_{\alpha}[\bar{p}_{2}^{2}]=\frac{\int p^{\alpha-2}p_{2}^{2}dx}{\int p^{\alpha}dx}$,
and thus 
\begin{align*}
\frac{\partial\mathbb{E}_{\alpha}[\bar{p}_{2}^{2}]}{\partial t} & =\frac{(\int p^{\alpha-2}p_{2}^{2}dx)_{t}\int p^{\alpha}dx-\int p^{\alpha-2}p_{2}^{2}dx\cdot(\int p^{\alpha}dx)_{t}}{(\int p^{\alpha}dx)^{2}}\\
 & =\frac{\frac{(\alpha-2)(\alpha-3)}{2}\int p^{\alpha-4}p_{1}^{2}p_{2}^{2}dx+(\alpha-2)\int p^{\alpha-3}p_{2}^{3}dx-\int p^{\alpha-2}p_{3}^{2}dx}{\int p^{\alpha}dx}\\
 & +\frac{\frac{\alpha(\alpha-1)}{2}\int p^{\alpha-2}p_{1}^{2}dx\cdot\int p^{\alpha-2}p_{2}^{2}dx}{(\int p^{\alpha}dx)^{2}}\\
 & =\frac{(\alpha-2)(\alpha-3)}{2}\mathbb{E}_{\alpha}[\bar{p}_{1}^{2}\bar{p}_{2}^{2}]+(\alpha-2)\mathbb{E}_{\alpha}[\bar{p}_{2}^{3}]-\mathbb{E}_{\alpha}[\bar{p}_{3}^{2}]+\frac{\alpha(\alpha-1)}{2}\mathbb{E}_{\alpha}[\bar{p}_{1}^{2}]\mathbb{E}_{\alpha}[\bar{p}_{2}^{2}],
\end{align*}
where the second equality follows from \eqref{eq:p^alpha-2p_2^2-1}
and \eqref{eq:p^alpha}.

Note that $\mathbb{E}_{\alpha}[\bar{p}_{1}^{2}]=\frac{\int p^{\alpha-2}p_{1}^{2}dx}{\int p^{\alpha}dx}$,
and thus 
\begin{align*}
\frac{\partial\mathbb{E}_{\alpha}[\bar{p}_{1}^{2}]}{\partial t} & =\frac{(\int p^{\alpha-2}p_{1}^{2}dx)_{t}\cdot\int p^{\alpha}dx-\int p^{\alpha-2}p_{1}^{2}dx\cdot(\int p^{\alpha}dx)_{t}}{(\int p^{\alpha}dx)^{2}}\\
 & =\frac{\frac{(\alpha-2)(\alpha-3)}{6}\int p^{\alpha-4}p_{1}^{4}dx-\int p^{\alpha-2}p_{2}^{2}dx}{\int p^{\alpha}dx}+\frac{\frac{\alpha(\alpha-1)}{2}(\int p^{\alpha-2}p_{1}^{2}dx)^{2}}{(\int p^{\alpha}dx)^{2}}\\
 & =\frac{(\alpha-2)(\alpha-3)}{6}\mathbb{E}_{\alpha}[\bar{p}_{1}^{4}]-\mathbb{E}_{\alpha}[\bar{p}_{2}^{2}]+\frac{\alpha(\alpha-1)}{2}(\mathbb{E}_{\alpha}[\bar{p}_{1}^{2}])^{2},
\end{align*}
where the second equality follows from \eqref{eq:p^alpha-2p_1^2}
and \eqref{eq:p^alpha}. Therefore, we have 
\begin{align*}
\frac{\partial(\mathbb{E}_{\alpha}[\bar{p}_{1}^{2}])^{2}}{\partial t} & =2\mathbb{E}_{\alpha}[\bar{p}_{1}^{2}]\cdot\frac{\partial\mathbb{E}_{\alpha}[\bar{p}_{1}^{2}]}{\partial t}\\
 & =\mathbb{E}_{\alpha}[\bar{p}_{1}^{2}]\left(\frac{(\alpha-2)(\alpha-3)}{3}\mathbb{E}_{\alpha}[\bar{p}_{1}^{4}]-2\mathbb{E}_{\alpha}[\bar{p}_{2}^{2}]+\alpha(\alpha-1)(\mathbb{E}_{\alpha}[\bar{p}_{1}^{2}])^{2}\right).
\end{align*}

\begin{IEEEproof}[Proof of Lemma \ref{lem:3rd derivative-1}]
Similar to Lemma \ref{lem:(p^alpha)_t}, we have 
\begin{align*}
\int p^{\alpha-5}p_{1}^{4}p_{2}dx & =\int p^{\alpha-5}p_{1}^{4}dp_{1}\\
 & =p^{\alpha-5}p_{1}^{5}\bigg|_{-\infty}^{+\infty}-\int(\alpha-5)p^{\alpha-6}p_{1}^{6}+4p^{\alpha-5}p_{1}^{4}p_{2}dx\\
 & =-(\alpha-5)\int p^{\alpha-6}p_{1}^{6}dx-4\int p^{\alpha-5}p_{1}^{4}p_{2}dx,
\end{align*}
thus, 
\[
\int p^{\alpha-5}p_{1}^{4}p_{2}dx=-\frac{\alpha-5}{5}\int p^{\alpha-6}p_{1}^{6}dx.
\]
\begin{align*}
\int p^{\alpha-4}p_{1}^{3}p_{3}dx & =\int p^{\alpha-4}p_{1}^{3}dp_{2}\\
 & =p^{\alpha-4}p_{1}^{3}p_{2}\bigg|_{-\infty}^{+\infty}-\int(\alpha-4)p^{\alpha-5}p_{1}^{4}p_{2}+3p^{\alpha-4}p_{1}^{2}p_{2}^{2}dx\\
 & =\frac{(\alpha-4)(\alpha-5)}{5}\int p^{\alpha-6}p_{1}^{6}dx-3\int p^{\alpha-4}p_{1}^{2}p_{2}^{2}dx,
\end{align*}
where the last equality follows from \eqref{eq:p^alpha-5p_1^4p_2}.
\begin{align*}
\int p^{\alpha-3}p_{1}p_{2}p_{3}dx & =\frac{1}{2}\int p^{\alpha-3}p_{1}dp_{2}^{2}\\
 & =\frac{1}{2}p^{\alpha-3}p_{1}p_{2}^{2}\bigg|_{-\infty}^{+\infty}-\frac{1}{2}\int(\alpha-3)p^{\alpha-4}p_{1}^{2}p_{2}^{2}+p^{\alpha-3}p_{2}^{3}dx\\
 & =-\frac{\alpha-3}{2}\int p^{\alpha-4}p_{1}^{2}p_{2}^{2}dx-\frac{1}{2}\int p^{\alpha-3}p_{2}^{3}dx,
\end{align*}
\begin{align*}
\int p^{\alpha-2}p_{2}p_{4}dx & =\int p^{\alpha-2}p_{2}dp_{3}\\
 & =p^{\alpha-2}p_{2}p_{3}\bigg|_{-\infty}^{+\infty}-\int(\alpha-2)p^{\alpha-3}p_{1}p_{2}p_{3}+p^{\alpha-2}p_{3}^{2}dx\\
 & =\frac{(\alpha-2)(\alpha-3)}{2}\int p^{\alpha-4}p_{1}^{2}p_{2}^{2}dx+\frac{\alpha-2}{2}\int p^{\alpha-3}p_{2}^{3}dx-\int p^{\alpha-2}p_{3}^{2}dx,
\end{align*}
where the last equality follows from \eqref{eq:p^alpha-3p_1p_2p_3}.

Note that \eqref{eq:p^alpha-3p_1^2p_2} has been verified in \eqref{eq:p_1^2p_2p^alpha-3}.

\begin{align*}
\int p^{\alpha-2}p_{1}p_{3}dx & =\int p^{\alpha-2}p_{1}dp_{2}\\
 & =p^{\alpha-2}p_{1}p_{2}\bigg|_{-\infty}^{+\infty}-\int(\alpha-2)p^{\alpha-3}p_{1}^{2}p_{2}+p^{\alpha-2}p_{2}^{2}dx\\
 & =\frac{(\alpha-2)(\alpha-3)}{3}\int p^{\alpha-4}p_{1}^{4}dx-\int p^{\alpha-2}p_{2}^{2}dx,
\end{align*}
where the last equality follows from \eqref{eq:p^alpha-3p_1^2p_2}.
\begin{align*}
\int p^{\alpha-1}p_{4}dx & =\int p^{\alpha-1}dp_{3}\\
 & =p^{\alpha-1}p_{3}\bigg|_{-\infty}^{+\infty}-(\alpha-1)\int p^{\alpha-2}p_{1}p_{3}dx\\
 & =-\frac{(\alpha-1)(\alpha-2)(\alpha-3)}{3}\int p^{\alpha-4}p_{1}^{4}dx+(\alpha-1)\int p^{\alpha-2}p_{2}^{2}dx,
\end{align*}
where the last equality follows from \eqref{eq:p^alpha-2p_1p_3}.
\begin{align*}
\int p^{\alpha-3}p_{1}^{2}p_{4}dx & =\int p^{\alpha-3}p_{1}^{2}dp_{3}\\
 & =p^{\alpha-3}p_{1}^{2}p_{3}\bigg|_{-\infty}^{+\infty}-(\alpha-3)\int p^{\alpha-4}p_{1}^{3}p_{3}dx-2\int p^{\alpha-3}p_{1}p_{2}p_{3}dx\\
 & =-\frac{(\alpha-3)(\alpha-4)(\alpha-5)}{5}\int p^{\alpha-6}p_{1}^{6}dx+4(\alpha-3)\int p^{\alpha-4}p_{1}^{2}p_{2}^{2}dx+\int p^{\alpha-3}p_{2}^{3}dx,
\end{align*}
where the last equality follows from \eqref{eq:p^alpha-4p_1^3p_3}
and \eqref{eq:p^alpha-3p_1p_2p_3}.
\end{IEEEproof}
\begin{IEEEproof}[Proof of Lemma \ref{lem:3rd derivative-2}]
We have 
\begin{align*}
\frac{\partial\int p^{\alpha-4}p_{1}^{4}dx}{\partial t} & =\int(\alpha-4)p^{\alpha-5}\cdot\frac{1}{2}p_{2}p_{1}^{4}+p^{\alpha-4}\cdot4p_{1}^{3}p_{xt}dx\\
 & =\int\frac{\alpha-4}{2}p^{\alpha-5}p_{1}^{4}p_{2}+2p^{\alpha-4}p_{1}^{3}p_{3}dx\\
 & =-\frac{(\alpha-4)(\alpha-5)}{10}\int p^{\alpha-6}p_{1}^{6}dx+\frac{2(\alpha-4)(\alpha-5)}{5}\int p^{\alpha-6}p_{1}^{6}dx-6\int p^{\alpha-4}p_{1}^{2}p_{2}^{2}dx\\
 & =\frac{3(\alpha-4)(\alpha-5)}{10}\int p^{\alpha-6}p_{1}^{6}dx-6\int p^{\alpha-4}p_{1}^{2}p_{2}^{2}dx,
\end{align*}
where the first and the second equality come from heat equation \eqref{eq:heat equation},
the third equality follows from \eqref{eq:p^alpha-5p_1^4p_2} and
\eqref{eq:p^alpha-4p_1^3p_3}. 
\begin{align*}
\frac{\partial\int p^{\alpha-2}p_{2}^{2}dx}{\partial t} & =\int\frac{1}{2}(\alpha-2)p^{\alpha-3}p_{2}^{3}+p^{\alpha-2}p_{2}p_{4}dx\\
 & =\frac{\alpha-2}{2}\int p^{\alpha-3}p_{2}^{3}dx+\frac{(\alpha-2)(\alpha-3)}{2}\int p^{\alpha-4}p_{1}^{2}p_{2}^{2}dx\\
 & +\frac{\alpha-2}{2}\int p^{\alpha-3}p_{2}^{3}dx-\int p^{\alpha-2}p_{3}^{2}dx\\
 & =\frac{(\alpha-2)(\alpha-3)}{2}\int p^{\alpha-4}p_{1}^{2}p_{2}^{2}dx+(\alpha-2)\int p^{\alpha-3}p_{2}^{3}dx-\int p^{\alpha-2}p_{3}^{2}dx,
\end{align*}
where the first equality follows from heat equation \eqref{eq:heat equation},
and the second equality follows from \eqref{eq:p^alpha-2p_2p_4}. 
\end{IEEEproof}

\section{\label{sec:Appendix-F_alphap_1^6} Proof of Lemma \ref{lem:E_alphap_1^6}}

We need the following two lemmas, whose proofs will be given later. 
\begin{lem}
\label{lem:4th derivative-1}For $t>0$, 
\begin{align}
\int p^{\alpha-7}p_{2}p_{1}^{6}dx & =-\frac{\alpha-7}{7}\int p^{\alpha-8}p_{1}^{8}dx,\label{eq:p^alpha-7p_2p_1^6}\\
\int p^{\alpha-6}p_{1}^{5}p_{3}dx & =\frac{(\alpha-6)(\alpha-7)}{7}\int p^{\alpha-8}p_{1}^{8}dx-5\int p^{\alpha-6}p_{1}^{4}p_{2}^{2}dx,\label{eq:p^alpha-6p_1^5p_3}\\
\int p^{\alpha-5}p_{1}^{3}p_{2}p_{3}dx & =-\frac{\alpha-5}{2}\int p^{\alpha-6}p_{1}^{4}p_{2}^{2}dx-\frac{3}{2}\int p^{\alpha-5}p_{1}^{2}p_{2}^{3}dx,\label{eq:p^alpha-5p_1^3p_2p_3}\\
\int p^{\alpha-4}p_{1}p_{2}^{2}p_{3}dx & =-\frac{\alpha-4}{3}\int p^{\alpha-5}p_{1}^{2}p_{2}^{3}dx-\frac{1}{3}\int p^{\alpha-4}p_{2}^{4}dx,\label{eq:p^alpha-4p_1p_2^2p_3}\\
\int p^{\alpha-4}p_{1}^{2}p_{2}p_{4}dx & =\frac{(\alpha-4)(\alpha-5)}{2}\int p^{\alpha-6}p_{1}^{4}p_{2}^{2}dx+\frac{13}{6}(\alpha-4)\int p^{\alpha-5}p_{1}^{2}p_{2}^{3}dx+\frac{2}{3}\int p^{\alpha-4}p_{2}^{4}dx-\int p^{\alpha-4}p_{1}^{2}p_{3}^{2}dx,\label{eq:p^alpha-4p_1^2p_2p_4}\\
\int p^{\alpha-3}p_{2}^{2}p_{4}dx & =\frac{(\alpha-3)(\alpha-4)}{3}\int p^{\alpha-5}p_{1}^{2}p_{2}^{3}dx+\frac{\alpha-3}{3}\int p^{\alpha-4}p_{2}^{4}dx-2\int p^{\alpha-3}p_{2}p_{3}^{2}dx,\label{eq:p^alpha-3p_2^2p_4}\\
\int p^{\alpha-3}p_{1}p_{3}p_{4}dx & =-\frac{\alpha-3}{2}\int p^{\alpha-4}p_{1}^{2}p_{3}^{2}dx-\frac{1}{2}\int p^{\alpha-3}p_{2}p_{3}^{2}dx,\label{eq:p^alpha-3p_1p_3p_4}\\
\int p^{\alpha-2}p_{3}p_{5}dx & =\frac{(\alpha-2)(\alpha-3)}{2}\int p^{\alpha-4}p_{1}^{2}p_{3}^{2}dx+\frac{\alpha-2}{2}\int p^{\alpha-3}p_{2}p_{3}^{2}dx-\int p^{\alpha-2}p_{4}^{2}dx,\label{eq:p^alpha-2p_3p_5}\\
\int p^{\alpha-5}p_{1}^{4}p_{4}dx & =-\frac{(\alpha-5)(\alpha-6)(\alpha-7)}{7}\int p^{\alpha-8}p_{1}^{8}dx+7(\alpha-5)\int p^{\alpha-6}p_{1}^{4}p_{2}^{2}dx+6\int p^{\alpha-5}p_{1}^{2}p_{2}^{3}dx.\label{eq:p^=00005Calpha-5p_1^4p_4}
\end{align}
\end{lem}
\begin{lem}
\label{lem:4th derivative-2}For $t>0$, 
\begin{align}
\frac{\partial\int p^{\alpha-6}p_{1}^{6}dx}{\partial t} & =\frac{5}{14}(\alpha-6)(\alpha-7)\int p^{\alpha-8}p_{1}^{8}dx-15\int p^{\alpha-6}p_{1}^{4}p_{2}^{2}dx,\label{eq:p_1^6}\\
\frac{\partial\int p^{\alpha-4}p_{1}^{2}p_{2}^{2}dx}{\partial t} & =\frac{7}{3}(\alpha-4)\int p^{\alpha-5}p_{1}^{2}p_{2}^{3}dx+\frac{1}{3}\int p^{\alpha-4}p_{2}^{4}dx-\int p^{\alpha-4}p_{1}^{2}p_{3}^{2}dx+\frac{(\alpha-4)(\alpha-5)}{2}\int p^{\alpha-6}p_{1}^{4}p_{2}^{2}dx,\label{eq:p_1^2p_2^2}\\
\frac{\partial\int p^{\alpha-3}p_{2}^{3}dx}{\partial t} & =\frac{(\alpha-3)(\alpha-4)}{2}\int p^{\alpha-5}p_{1}^{2}p_{2}^{3}dx+(\alpha-3)\int p^{\alpha-4}p_{2}^{4}dx-3\int p^{\alpha-3}p_{2}p_{3}^{2}dx,\label{eq:p_2^3}\\
\frac{\partial\int p^{\alpha-2}p_{3}^{2}dx}{\partial t} & =\frac{(\alpha-2)(\alpha-3)}{2}\int p^{\alpha-4}p_{1}^{2}p_{3}^{2}dx+(\alpha-2)\int p^{\alpha-3}p_{2}p_{3}^{2}dx-\int p^{\alpha-2}p_{4}^{2}dx.\label{eq:p_3^2}
\end{align}
Note that $\mathbb{E}_{\alpha}[\bar{p}_{1}^{6}]=\frac{\int p^{\alpha-6}p_{1}^{6}dx}{\int p^{\alpha}dx}$,
thus
\begin{align*}
\frac{\partial\mathbb{E}_{\alpha}[\bar{p}_{1}^{6}]}{\partial t} & =\frac{(\int p^{\alpha-6}p_{1}^{6}dx)_{t}\int p^{\alpha}dx-\int p^{\alpha-6}p_{1}^{6}dx\cdot(\int p^{\alpha}dx)_{t}}{(\int p^{\alpha}dx)^{2}}\\
 & =\frac{\frac{5}{14}(\alpha-6)(\alpha-7)\int p^{\alpha-8}p_{1}^{8}dx-15\int p^{\alpha-6}p_{1}^{4}p_{2}^{2}dx}{\int p^{\alpha}dx}-\frac{\int p^{\alpha-6}p_{1}^{6}dx(-\frac{\alpha(\alpha-1)}{2})\int p^{\alpha-2}p_{1}^{2}dx}{(\int p^{\alpha}dx)^{2}}\\
 & =\frac{5}{14}(\alpha-6)(\alpha-7)\mathbb{E}_{\alpha}[\bar{p}_{1}^{8}]-15\mathbb{E}_{\alpha}[\bar{p}_{1}^{4}\bar{p}_{2}^{2}]+\frac{\alpha(\alpha-1)}{2}\mathbb{E}_{\alpha}[\bar{p}_{1}^{6}]\mathbb{E}_{\alpha}[\bar{p}_{1}^{2}],
\end{align*}
where the second equality follows from \eqref{eq:p_1^6} and \eqref{eq:p^alpha}.
\end{lem}
Note that $\mathbb{E}_{\alpha}[\bar{p}_{1}^{2}\bar{p}_{2}^{2}]=\frac{\int p^{\alpha-4}p_{1}^{2}p_{2}^{2}dx}{\int p^{\alpha}dx}$,
thus
\begin{align*}
\frac{\partial\mathbb{E}_{\alpha}[\bar{p}_{1}^{2}\bar{p}_{2}^{2}]}{\partial t} & =\frac{(\int p^{\alpha-4}p_{1}^{2}p_{2}^{2}dx)_{t}\int p^{\alpha}dx-\int p^{\alpha-4}p_{1}^{2}p_{2}^{2}dx(\int p^{\alpha}dx)_{t}}{(\int p^{\alpha}dx)^{2}}\\
 & =\frac{\frac{(\alpha-4)(\alpha-5)}{2}\int p^{\alpha-6}p_{1}^{4}p_{2}^{2}dx+\frac{7}{3}(\alpha-4)\int p^{\alpha-5}p_{1}^{2}p_{2}^{3}dx+\frac{1}{3}\int p^{\alpha-4}p_{2}^{4}dx-\int p^{\alpha-4}p_{1}^{2}p_{3}^{2}dx}{\int p^{\alpha}dx}\\
 & +\frac{\frac{\alpha(\alpha-1)}{2}\int p^{\alpha-2}p_{1}^{2}dx\int p^{\alpha-4}p_{1}^{2}p_{2}^{2}dx}{(\int p^{\alpha}dx)^{2}}\\
 & =\frac{(\alpha-4)(\alpha-5)}{2}\mathbb{E}_{\alpha}[\bar{p}_{1}^{4}\bar{p}_{2}^{2}]+\frac{7}{3}(\alpha-4)\mathbb{E}_{\alpha}[\bar{p}_{1}^{2}\bar{p}_{2}^{3}]+\frac{1}{3}\mathbb{E}_{\alpha}[\bar{p}_{2}^{4}]-\mathbb{E}_{\alpha}[\bar{p}_{1}^{2}\bar{p}_{3}^{2}]+\frac{\alpha(\alpha-1)}{2}\mathbb{E}_{\alpha}[\bar{p}_{1}^{2}]\mathbb{E}_{\alpha}[\bar{p}_{1}^{2}\bar{p}_{2}^{2}],
\end{align*}
where the second equality follows from \eqref{eq:p_1^2p_2^2} and
\eqref{eq:p^alpha}.

Note that $\mathbb{E}_{\alpha}[\bar{p}_{2}^{3}]=\frac{\int p^{\alpha-3}p_{2}^{3}dx}{\int p^{\alpha}dx}$,
thus
\begin{align*}
\frac{\partial\mathbb{E}_{\alpha}[\bar{p}_{2}^{3}]}{\partial t} & =\frac{(\int p^{\alpha-3}p_{2}^{3}dx)_{t}\int p^{\alpha}dx-\int p^{\alpha-3}p_{2}^{3}dx(\int p^{\alpha}dx)_{t}}{(\int p^{\alpha}dx)^{2}}\\
 & =\frac{\frac{(\alpha-3)(\alpha-4)}{2}\int p^{\alpha-5}p_{1}^{2}p_{2}^{3}dx+(\alpha-3)\int p^{\alpha-4}p_{2}^{4}dx-3\int p^{\alpha-3}p_{2}p_{3}^{2}dx}{\int p^{\alpha}dx}+\frac{\frac{\alpha(\alpha-1)}{2}\int p^{\alpha-3}p_{2}^{3}dx\int p^{\alpha-2}p_{1}^{2}dx}{(\int p^{\alpha}dx)^{2}}\\
 & =\frac{(\alpha-3)(\alpha-4)}{2}\mathbb{E}_{\alpha}[\bar{p}_{1}^{2}\bar{p}_{2}^{3}]+(\alpha-3)\mathbb{E}_{\alpha}[\bar{p}_{2}^{4}]-3\mathbb{E}_{\alpha}[\bar{p}_{2}\bar{p}_{3}^{2}]+\frac{\alpha(\alpha-1)}{2}\mathbb{E}_{\alpha}[\bar{p}_{2}^{3}]\mathbb{E}_{\alpha}[\bar{p}_{1}^{2}],
\end{align*}
where the second equality follows from \eqref{eq:p_2^3} and \eqref{eq:p^alpha}.

Note that $\mathbb{E}_{\alpha}[\bar{p}_{3}^{2}]=\frac{\int p^{\alpha-2}p_{3}^{2}dx}{\int p^{\alpha}dx}$,
thus
\begin{align*}
\frac{\partial\mathbb{E}_{\alpha}[\bar{p}_{3}^{2}]}{\partial t} & =\frac{(\int p^{\alpha-2}p_{3}^{2}dx)_{t}\int p^{\alpha}dx-\int p^{\alpha-2}p_{3}^{2}dx\cdot(\int p^{\alpha}dx)_{t}}{(\int p^{\alpha}dx)^{2}}\\
 & =\frac{\frac{(\alpha-2)(\alpha-3)}{2}\int p^{\alpha-4}p_{1}^{2}p_{3}^{2}dx+(\alpha-2)\int p^{\alpha-3}p_{2}p_{3}^{2}dx-\int p^{\alpha-2}p_{4}^{2}dx}{\int p^{\alpha}dx}+\frac{\frac{\alpha(\alpha-1)}{2}\int p^{\alpha-2}p_{1}^{2}dx\int p^{\alpha-2}p_{3}^{2}dx}{(\int p^{\alpha}dx)^{2}}\\
 & =\frac{(\alpha-2)(\alpha-3)}{2}\mathbb{E}_{\alpha}[\bar{p}_{1}^{2}\bar{p}_{3}^{2}]+(\alpha-2)\mathbb{E}_{\alpha}[\bar{p}_{2}\bar{p}_{3}^{2}]-\mathbb{E}_{\alpha}[\bar{p}_{4}^{2}]+\frac{\alpha(\alpha-1)}{2}\mathbb{E}_{\alpha}[\bar{p}_{1}^{2}]\mathbb{E}_{\alpha}[\bar{p}_{3}^{2}],
\end{align*}
where the second equality follows from \eqref{eq:p_3^2} and \eqref{eq:p^alpha}.
\begin{IEEEproof}[Proof of Lemma \ref{lem:4th derivative-1}]
Similar to Lemma \ref{lem:(p^alpha)_t} and Lemma \ref{lem:3rd derivative-1},
we have
\begin{align*}
\int p^{\alpha-7}p_{2}p_{1}^{6}dx & =\int p^{\alpha-7}p_{1}^{6}dp_{1}\\
 & =-\int(\alpha-7)p^{\alpha-8}p_{1}^{8}+p^{\alpha-7}\cdot6p_{1}^{6}p_{2}dx\\
 & =-(\alpha-7)\int p^{\alpha-8}p_{1}^{8}dx-6\int p^{\alpha-7}p_{1}^{6}p_{2}dx,
\end{align*}
thus we obtain \eqref{eq:p^alpha-7p_2p_1^6}.
\begin{align*}
\int p^{\alpha-6}p_{1}^{5}p_{3}dx & =\int p^{\alpha-6}p_{1}^{5}dp_{2}\\
 & =-\int(\alpha-6)p^{\alpha-7}p_{1}^{6}p_{2}+p^{\alpha-6}\cdot5p_{1}^{4}p_{2}^{2}dx\\
 & =\frac{(\alpha-6)(\alpha-7)}{7}\int p^{\alpha-8}p_{1}^{8}dx-5\int p^{\alpha-6}p_{1}^{4}p_{2}^{2}dx,
\end{align*}
where the second equality follows from \eqref{eq:p^alpha-7p_2p_1^6}.
\begin{align*}
\int p^{\alpha-5}p_{1}^{3}p_{2}p_{3}dx & =\frac{1}{2}\int p^{\alpha-5}p_{1}^{3}dp_{2}^{2}\\
 & =-\frac{1}{2}\int((\alpha-5)p^{\alpha-6}p_{1}^{4}p_{2}^{2}+p^{\alpha-5}\cdot3p_{1}^{2}p_{2}^{3})dx\\
 & =-\frac{\alpha-5}{2}\int p^{\alpha-6}p_{1}^{4}p_{2}^{2}dx-\frac{3}{2}\int p^{\alpha-5}p_{1}^{2}p_{2}^{3}dx
\end{align*}
\begin{align*}
\int p^{\alpha-4}p_{1}p_{2}^{2}p_{3}dx & =\frac{1}{3}\int p^{\alpha-4}p_{1}dp_{2}^{3}\\
 & =-\frac{1}{3}\int((\alpha-4)p^{\alpha-5}p_{1}^{2}p_{2}^{3}+p^{\alpha-4}p_{2}^{4})dx\\
 & =-\frac{\alpha-4}{3}\int p^{\alpha-5}p_{1}^{2}p_{2}^{3}dx-\frac{1}{3}\int p^{\alpha-4}p_{2}^{4}dx,
\end{align*}
\begin{align*}
\int p^{\alpha-4}p_{1}^{2}p_{2}p_{4}dx & =\int p^{\alpha-4}p_{1}^{2}p_{2}dp_{3}\\
 & =-\int((\alpha-4)p^{\alpha-5}p_{1}^{3}p_{2}p_{3}+2p_{1}p_{2}^{2}p^{\alpha-4}p_{3}+p^{\alpha-4}p_{1}^{2}p_{3}^{2})dx\\
 & =\frac{(\alpha-4)(\alpha-5)}{2}\int p^{\alpha-6}p_{1}^{4}p_{2}^{2}dx+\frac{3}{2}(\alpha-4)\int p^{\alpha-5}p_{1}^{2}p_{2}^{3}dx+\frac{2}{3}(\alpha-4)\int p^{\alpha-5}p_{1}^{2}p_{2}^{3}dx\\
 & +\frac{2}{3}\int p^{\alpha-4}p_{2}^{4}dx-\int p^{\alpha-4}p_{1}^{2}p_{3}^{2}dx\\
 & =\frac{(\alpha-4)(\alpha-5)}{2}\int p^{\alpha-6}p_{1}^{4}p_{2}^{2}dx+\frac{13}{6}(\alpha-4)\int p^{\alpha-5}p_{1}^{2}p_{2}^{3}dx+\frac{2}{3}\int p^{\alpha-4}p_{2}^{4}dx-\int p^{\alpha-4}p_{1}^{2}p_{3}^{2}dx,
\end{align*}
where the third equality follows from \eqref{eq:p^alpha-5p_1^3p_2p_3}
and \eqref{eq:p^alpha-4p_1p_2^2p_3}.
\begin{align*}
\int p^{\alpha-3}p_{2}^{2}p_{4}dx & =\int p^{\alpha-3}p_{2}^{2}dp_{3}\\
 & =-\int(\alpha-3)p^{\alpha-4}p_{1}p_{2}^{2}p_{3}+p^{\alpha-3}\cdot2p_{2}p_{3}^{2}dx\\
 & =\frac{(\alpha-3)(\alpha-4)}{3}\int p^{\alpha-5}p_{1}^{2}p_{2}^{3}dx+\frac{\alpha-3}{3}\int p^{\alpha-4}p_{2}^{4}dx-2\int p^{\alpha-3}p_{2}p_{3}^{2}dx,
\end{align*}
where the last equality follows from \eqref{eq:p^alpha-4p_1p_2^2p_3}.
\begin{align*}
\int p^{\alpha-3}p_{1}p_{3}p_{4}dx & =\frac{1}{2}\int p^{\alpha-3}p_{1}dp_{3}^{2}\\
 & =-\frac{1}{2}\int(\alpha-3)p^{\alpha-4}p_{1}^{2}p_{3}^{2}+p^{\alpha-3}p_{2}p_{3}^{2}dx\\
 & =-\frac{\alpha-3}{2}\int p^{\alpha-4}p_{1}^{2}p_{3}^{2}dx-\frac{1}{2}\int p^{\alpha-3}p_{2}p_{3}^{2}dx,
\end{align*}
\begin{align*}
\int p^{\alpha-2}p_{3}p_{5}dx & =\int p^{\alpha-2}p_{3}dp_{4}\\
 & =-\int(\alpha-2)p^{\alpha-3}p_{1}p_{3}p_{4}+p^{\alpha-2}p_{4}^{2}dx\\
 & =\frac{(\alpha-2)(\alpha-3)}{2}\int p^{\alpha-4}p_{1}^{2}p_{3}^{2}dx+\frac{\alpha-2}{2}\int p^{\alpha-3}p_{2}p_{3}^{2}dx-\int p^{\alpha-2}p_{4}^{2}dx,
\end{align*}
where the last equality follows from \eqref{eq:p^alpha-3p_1p_3p_4}.
\begin{align*}
\int p^{\alpha-5}p_{1}^{4}p_{4}dx & =\int p^{\alpha-5}p_{1}^{4}dp_{3}\\
 & =-(\alpha-5)\int p^{\alpha-6}p_{1}^{5}p_{3}dx-4\int p^{\alpha-5}p_{1}^{3}p_{2}p_{3}dx\\
 & =-\frac{(\alpha-5)(\alpha-6)(\alpha-7)}{7}\int p^{\alpha-8}p_{1}^{8}dx+7(\alpha-5)\int p^{\alpha-6}p_{1}^{4}p_{2}^{2}dx+6\int p^{\alpha-5}p_{1}^{2}p_{2}^{3}dx,
\end{align*}
where the last equality follows from \eqref{eq:p^alpha-6p_1^5p_3}and
\eqref{eq:p^alpha-5p_1^3p_2p_3}.
\end{IEEEproof}
\begin{IEEEproof}[Proof of Lemma \ref{lem:4th derivative-2}]
 We have
\begin{align*}
\frac{\partial\int p^{\alpha-6}p_{1}^{6}dx}{\partial t} & =\int(\alpha-6)p^{\alpha-7}\cdot\frac{1}{2}p_{2}p_{1}^{6}+p^{\alpha-6}\cdot6p_{1}^{5}p_{xt}dx\\
 & =\frac{\alpha-6}{2}\int p^{\alpha-7}p_{2}p_{1}^{6}dx+3\int p^{\alpha-6}p_{1}^{5}p_{3}dx\\
 & =\frac{\alpha-6}{2}(-\frac{\alpha-7}{7})\int p^{\alpha-8}p_{1}^{8}dx+\frac{3}{7}(\alpha-6)(\alpha-7)\int p^{\alpha-8}p_{1}^{8}dx-15\int p^{\alpha-6}p_{1}^{4}p_{2}^{2}dx\\
 & =\frac{5}{14}(\alpha-6)(\alpha-7)\int p^{\alpha-8}p_{1}^{8}dx-15\int p^{\alpha-6}p_{1}^{4}p_{2}^{2}dx,
\end{align*}
where the second equality follows from heat equation \eqref{eq:heat equation},
the third equality follows from \eqref{eq:p^alpha-7p_2p_1^6} and
\eqref{eq:p^alpha-6p_1^5p_3}.

\begin{align*}
\frac{\partial\int p^{\alpha-4}p_{1}^{2}p_{2}^{2}dx}{\partial t} & =\int(\alpha-4)p^{\alpha-5}\cdot\frac{1}{2}p_{2}p_{1}^{2}p_{2}^{2}+2p_{1}p_{xt}p^{\alpha-4}p_{2}^{2}+p^{\alpha-4}p_{1}^{2}\cdot2p_{2}p_{xxt}dx\\
 & =\frac{\alpha-4}{2}\int p^{\alpha-5}p_{1}^{2}p_{2}^{3}dx+\int p^{\alpha-4}p_{1}p_{2}^{2}p_{3}dx+\int p^{\alpha-4}p_{1}^{2}p_{2}p_{4}dx\\
 & =\frac{\alpha-4}{2}\int p^{\alpha-5}p_{1}^{2}p_{2}^{3}dx-\frac{\alpha-4}{3}\int p^{\alpha-5}p_{1}^{2}p_{2}^{3}dx-\frac{1}{3}\int p^{\alpha-4}p_{2}^{4}dx\\
 & +\frac{(\alpha-4)(\alpha-5)}{2}\int p^{\alpha-6}p_{1}^{4}p_{2}^{2}dx+\frac{13}{6}(\alpha-4)\int p^{\alpha-5}p_{1}^{2}p_{2}^{3}dx+\frac{2}{3}\int p^{\alpha-4}p_{2}^{4}dx-\int p^{\alpha-4}p_{1}^{2}p_{3}^{2}dx\\
 & =\frac{7}{3}(\alpha-4)\int p^{\alpha-5}p_{1}^{2}p_{2}^{3}dx+\frac{1}{3}\int p^{\alpha-4}p_{2}^{4}dx-\int p^{\alpha-4}p_{1}^{2}p_{3}^{2}dx+\frac{(\alpha-4)(\alpha-5)}{2}\int p^{\alpha-6}p_{1}^{4}p_{2}^{2}dx,
\end{align*}
where the second equality follows from heat equation \eqref{eq:heat equation},
the third equality follows from \eqref{eq:p^alpha-4p_1p_2^2p_3} and
\eqref{eq:p^alpha-4p_1^2p_2p_4}.
\begin{align*}
\frac{\partial\int p^{\alpha-3}p_{2}^{3}dx}{\partial t} & =\int(\alpha-3)p^{\alpha-4}\cdot\frac{1}{2}p_{2}^{4}+p^{\alpha-3}\cdot3p_{2}^{2}p_{xxt}dx\\
 & =\frac{\alpha-3}{2}\int p^{\alpha-4}p_{2}^{4}dx+\frac{3}{2}\int p^{\alpha-3}p_{2}^{2}p_{4}dx\\
 & =\frac{\alpha-3}{2}\int p^{\alpha-4}p_{2}^{4}dx+\frac{(\alpha-3)(\alpha-4)}{2}\int p^{\alpha-5}p_{1}^{2}p_{2}^{3}dx+\frac{\alpha-3}{2}\int p^{\alpha-4}p_{2}^{4}dx-3\int p^{\alpha-3}p_{2}p_{3}^{2}dx\\
 & =\frac{(\alpha-3)(\alpha-4)}{2}\int p^{\alpha-5}p_{1}^{2}p_{2}^{3}dx+(\alpha-3)\int p^{\alpha-4}p_{2}^{4}dx-3\int p^{\alpha-3}p_{2}p_{3}^{2}dx,
\end{align*}
where the second equality follows from heat equation \eqref{eq:heat equation},
the third equality follows from \eqref{eq:p^alpha-3p_2^2p_4}.
\begin{align*}
\frac{\partial\int p^{\alpha-2}p_{3}^{2}dx}{\partial t} & =\int(\alpha-2)p^{\alpha-3}\cdot\frac{1}{2}p_{2}p_{3}^{2}+p^{\alpha-2}\cdot2p_{3}p_{xxxt}dx\\
 & =\frac{\alpha-2}{2}\int p^{\alpha-3}p_{2}p_{3}^{2}dx+\int p^{\alpha-2}p_{3}p_{5}dx\\
 & =\frac{(\alpha-2)(\alpha-3)}{2}\int p^{\alpha-4}p_{1}^{2}p_{3}^{2}dx+(\alpha-2)\int p^{\alpha-3}p_{2}p_{3}^{2}dx-\int p^{\alpha-2}p_{4}^{2}dx,
\end{align*}
where the second equality follows from heat equation \eqref{eq:heat equation},
the last equality follows from \eqref{eq:p^alpha-2p_3p_5}.
\end{IEEEproof}
\bibliographystyle{plain}
\bibliography{ref}

\begin{thebibliography}{10}

\bibitem{barron1984monotonic}
A.~R. Barron.
\newblock Monotonic central limit theorem for densities.
\newblock {\em Department of Statistics, Stanford University, California, Tech.
  Rep}, 50, 1984.

\bibitem{bogachev2007measure1}
V.~I. Bogachev.
\newblock {\em Measure theory}, volume~1.
\newblock Springer Science \& Business Media, 2007.

\bibitem{borell1985geometric}
C.~Borell.
\newblock Geometric bounds on the {Ornstein--Uhlenbeck} velocity process.
\newblock {\em Probability Theory and Related Fields}, 70(1):1--13, 1985.

\bibitem{Cheng2015Higher}
F.~Cheng and Y.~Geng.
\newblock Higher {Order} {Derivatives} in {Costa's} {Entropy} {Power}
  {Inequality}.
\newblock {\em IEEE Transactions on Information Theory}, 61:5892--5905, 2015.

\bibitem{costa2003solutions}
J.~Costa, A.~Hero, and C.~Vignat.
\newblock On solutions to multivariate maximum $\alpha$-entropy problems.
\newblock In {\em International Workshop on Energy Minimization Methods in
  Computer Vision and Pattern Recognition}, pages 211--226. Springer, 2003.

\bibitem{costa1985new}
M.~Costa.
\newblock A new entropy power inequality.
\newblock {\em IEEE Transactions on Information Theory}, 31(6):751--760, 1985.

\bibitem{courtade2017concavity}
T.~A. Courtade.
\newblock Concavity of entropy power: {Equivalent} formulations and
  generalizations.
\newblock In {\em 2017 IEEE International Symposium on Information Theory
  (ISIT)}, pages 56--60. IEEE, 2017.

\bibitem{dembo1989simple}
A.~Dembo.
\newblock Simple proof of the concavity of the entropy power with respect to
  added {Gaussian} noise.
\newblock {\em IEEE Transactions on Information Theory}, 35(4):887--888, 1989.

\bibitem{eldan2015two}
R.~Eldan.
\newblock A two-sided estimate for the {Gaussian} noise stability deficit.
\newblock {\em Inventiones Mathematicae}, 201(2):561--624, 2015.

\bibitem{gallagerIT}
R.~G. Gallager.
\newblock {\em {Information Theory and Reliable Communication}}.
\newblock Wiley, New York, 1968.

\bibitem{gross1975logarithmic}
L.~Gross.
\newblock Logarithmic {Sobolev} inequalities.
\newblock {\em American Journal of Mathematics}, 97(4):1061--1083, 1975.

\bibitem{guo2021generalization}
L.~Guo, C.-M. Yuan, and X.-S. Gao.
\newblock A generalization of the concavity of {R{\'e}nyi} entropy power.
\newblock {\em Entropy}, 23(12):1593, 2021.

\bibitem{guo2022lower}
L.~Guo, C.-M. Yuan, and X.-S. Gao.
\newblock Lower bounds on multivariate higher order derivatives of differential
  entropy.
\newblock {\em Entropy}, 24(8):1155, 2022.

\bibitem{hung2022generalization}
L.-C. Hung.
\newblock Generalization of {Completely} {Monotone} {Conjecture} for {Tsallis}
  entropy.
\newblock {\em arXiv preprint arXiv:2212.09269}, 2022.

\bibitem{jungel2006algorithmic}
A.~J{\"u}ngel and D.~Matthes.
\newblock An algorithmic construction of entropies in higher-order nonlinear
  {PDEs}.
\newblock {\em Nonlinearity}, 19(3):633, 2006.

\bibitem{ledoux1994semigroup}
M.~Ledoux.
\newblock Semigroup proofs of the isoperimetric inequality in {Euclidean} and
  {Gauss} space.
\newblock {\em Bulletin des sciences math{\'e}matiques}, 118(6):485--510, 1994.

\bibitem{ledoux2014remarks}
M.~Ledoux.
\newblock Remarks on {Gaussian} noise stability, {Brascamp-Lieb} and {Slepian}
  inequalities.
\newblock In {\em Geometric Aspects of Functional Analysis: Israel Seminar
  (GAFA) 2011-2013}, pages 309--333. Springer, 2014.

\bibitem{ledoux2022differentials}
M.~Ledoux.
\newblock Differentials of entropy and {Fisher} information along heat flow: a
  brief review of some conjectures.
\newblock {\em
  https://perso.math.univ-toulouse.fr/ledoux/files/2023/03/Entropy-conjectures.pdf},
  2022.

\bibitem{ledoux2021log}
M.~Ledoux, C.~Nair, and Y.~N. Wang.
\newblock Log-convexity of {Fisher} information along heat flow.
\newblock {\em http://chandra.ie.cuhk.edu.hk/pub/papers/NIT/Log-cvx.pdf}, 2021.

\bibitem{li2020renyi}
S.~Li and X.-D. Li.
\newblock On the {R{\'e}nyi} entropy power and the
  {Gagliardo-Nirenberg-Sobolev} inequality on {Riemannian} manifolds.
\newblock {\em arXiv preprint arXiv:2001.11184}, 2020.

\bibitem{McKean1966}
H.~P. McKean.
\newblock Speed of approach to equilibrium for {Kac}'s caricature of a
  {Maxwellian} gas.
\newblock {\em Arch. Rational Mech. Anal.}, 21:343--367, 1966.

\bibitem{mossel2015robust}
E.~Mossel and J.~Neeman.
\newblock Robust optimality of {Gaussian} noise stability.
\newblock {\em Journal of the European Mathematical Society}, 17(2):433--482,
  2015.

\bibitem{nair2020signs}
C.~Nair.
\newblock On signs of derivatives of entropy along {Markov} semi-groups:{A}
  collection of some recent results.
\newblock {\em http://chandra.ie.cuhk.edu.hk/pub/papers/Misc/ITA-2023.pdf},
  2023.

\bibitem{ODonnell14analysisof}
R.~O'Donnell.
\newblock {\em Analysis of {Boolean} Functions}.
\newblock Cambridge University Press, 2014.

\bibitem{raginsky2013logarithmic}
M.~Raginsky.
\newblock Logarithmic {Sobolev} inequalities and strong data processing
  theorems for discrete channels.
\newblock In {\em IEEE International Symposium on Information Theory (ISIT)},
  pages 419--423, 2013.

\bibitem{Renyi1961OnMO}
A.~R{\'e}nyi.
\newblock On measures of entropy and information.
\newblock In {\em Proceedings of the Fourth Berkeley Symposium on Mathematical
  Statistics and Probability, Volume 1: Contributions to the Theory of
  Statistics}, pages 547--561, 1961.

\bibitem{savare2014concavity}
G.~Savar{\'e} and G.~Toscani.
\newblock The concavity of {R{\'e}nyi} entropy power.
\newblock {\em IEEE Transactions on Information theory}, 60(5):2687--2693,
  2014.

\bibitem{toscani2015concavity}
G.~Toscani.
\newblock A concavity property for the reciprocal of {Fisher} information and
  its consequences on {Costa’s} {EPI}.
\newblock {\em Physica A: Statistical Mechanics and its Applications},
  432:35--42, 2015.

\bibitem{Erven}
T.~{van Erven} and P.~Harremo\"es.
\newblock {R{\'e}nyi} divergence and {Kullback-Leibler} divergence.
\newblock {\em IEEE Transactions on Information Theory}, 60(7):3797--3820,
  2014.

\bibitem{villani2000short}
C.~Villani.
\newblock A short proof of the "concavity of entropy power".
\newblock {\em IEEE Transactions on Information Theory}, 46(4):1695--1696,
  2000.

\bibitem{wang2024entropy}
G.~Wang.
\newblock The entropy power conjecture implies the {McKean} conjecture.
\newblock {\em arXiv preprint arXiv:2408.07275}, 2024.

\bibitem{yu2023theentro}
L.~Yu.
\newblock {\em The Entropy Method}.
\newblock DOI: 10.13140/RG.2.2.26552.11527/1, 2023.

\bibitem{zhang2018gaussian}
X.~Zhang, V.~Anantharam, and Y.~Geng.
\newblock Gaussian optimality for derivatives of differential entropy using
  linear matrix inequalities.
\newblock {\em Entropy}, 20(3):182, 2018.

\end{thebibliography}

\end{document}